\documentclass[twocolumn]{aastex61} 

\usepackage{graphicx}           
\usepackage{latexsym}           
\usepackage{amsmath}            
\usepackage{nccmath}            
\usepackage{bm}                 
\usepackage[english]{babel}
\usepackage{color}              
\usepackage{booktabs}           
\usepackage{multirow}           
\usepackage{natbib}

\newcommand{\del}[1]{\relax}%
\newcommand{\DELETED}[1]{\relax}%
\newcommand{\DEL}[1]{\relax}%
{\relax}%

\definecolor{violet}  {rgb}{1.0,0.0,1.0}

\definecolor{dviolet} {rgb}{0.75,0.0,1.0}

\definecolor{blue}    {rgb}{0.0,0.7,1.0}

\definecolor{lblue}   {rgb}{0.5,1,1}

\definecolor{dblue}   {rgb}{0.0,0.0,1.0}

\definecolor{blgr}    {rgb}{0.70,0.80,1.00}

\definecolor{navy}    {rgb}{0.00,0.00,0.48}

\definecolor{green}   {rgb}{0.7,1.0,0.0}

\definecolor{dgreen}  {rgb}{0.0,1.0,0.0}

\definecolor{lgreen}  {rgb}{0.0,0.8,0.0}

\definecolor{dg}      {rgb}{0.0,0.6,0.0}

\definecolor{orange}  {rgb}{1.0,0.5,0.0}

\definecolor{dorange} {rgb}{1.0,0.6,0.0}

\definecolor{brown}   {rgb}{0.1,0.1,0.0}

\definecolor{lbrown}  {rgb}{0.7,0.5,0.0}

\definecolor{red}     {rgb}{1,0.0,0.0}

\definecolor{dred}    {rgb}{0.6,0.0,0.0}

\definecolor{grey}    {rgb}{0.1,0.1,0.1}

\definecolor{lgrey}   {rgb}{0.5,0.5,0.5}

\definecolor{black}   {rgb}{0.0,0.0,0.0}

\newcommand\n            {\noindent}

\newcommand\si           {\smallskip\indent}
\newcommand\bn           {\bigskip\noindent}
\newcommand\mn           {\medskip\noindent}
\newcommand\sn           {\smallskip\noindent}
\newcommand\cl           {\centerline}
\newcommand\ve           {\vfill\eject}

\newcommand\arcmpt       {{$\buildrel{\prime}      \over .$}}
\newcommand\arcspt       {{$\buildrel{\prime\prime}\over .$}}
\newcommand\degree       {{\ifmmode^\circ\else$^\circ$\fi}} 
\newcommand\arcm         {{\ifmmode {'\ }\else$'     $\fi}} 
\newcommand\arcs         {{\ifmmode{''\ }\else$''    $\fi}} 

\newcommand\bII          {{$b^{\rm II}$} }

\newcommand\cge          {{$\gtrsim$}}
\newcommand\cle          {{$\lesssim$}}

\newcommand\eg           {{\it e.g.},}
\newcommand\ie           {{\it i.e.},}

\newcommand\Ho           {{$H_{0}$} }

\newcommand\kms          {{km\ s$^{-1}$}}
\newcommand\kmsMpc       {{km\ s$^{-1}$\ Mpc$^{-1}$} }

\newcommand\lII          {{$l^{\rm II}$} }

\newcommand\Lya          {{Ly$\alpha$} }

\newcommand\mAB          {{$m_{\rm AB}$}}
\newcommand\MAB          {{$M_{\rm AB}$}}
\newcommand\Lo           {{$L_{\odot}$}}

\newcommand\Lstar        {{$L^{*}$} }
\newcommand\Laccr        {{$L_{\rm accr}$}}
\newcommand\Lbol         {{$L_{\rm bol}$}}

\newcommand\magarc       {{mag\ arcsec$^{-2}$}}
\newcommand\arcsecsq     {{arcsec$^{2}$}}
\newcommand\mum          {{\micron}} 
\newcommand\MBH          {{$M_{\rm BH}$} }
\newcommand\Mbol         {{$M_{\rm bol}$} }
\newcommand\MUV          {{$M_{\rm UV}$} }
\newcommand\Mo           {{M$_{\odot}$}}

\newcommand\nWsqmsr      {{nW\ m$^{-2}$\ sr$^{-1}$}}
\newcommand\nWsqmsrsq    {{nW$^{2}$\ m$^{-4}$\ sr$^{-2}$}}

\newcommand\Ro           {{R$_{\odot}$}}

\newcommand\rhl          {{$r_{\rm hl}$} }
\newcommand\Reff         {{$R_{\rm eff}$}}
\newcommand\Rs           {{$R_{\rm s}$}}
\newcommand\Raccr        {{$R_{\rm accr}$}}
\newcommand\Rmin         {{$R_{\rm min}$} }
\newcommand\RUV          {{$R_{\rm UV}$} }
\newcommand\RMS          {{$R_{\rm MS}$} }

\newcommand\taupreMS     {{$\tau_{\rm preMS}$}}
\newcommand\tauMS        {{$\tau_{\rm MS}$}}
\newcommand\tauGB        {{$\tau_{\rm GB}$}}

\newcommand\tauAGB       {{$\tau_{\rm AGB}$}}
\newcommand\Teff         {{$T_{\rm eff}$}}
\newcommand\Tmax         {{$T_{\rm max}$} }
\newcommand\vlos         {{$v_{\rm los}$}}
\newcommand\vT           {{$v_{T}$}}

\newcommand\VTs          {{$V_{T},s$}}
\newcommand\zreion       {{$z_{\rm reion}$} }
\newcommand\zmed         {{$z_{\rm med}$}}

\newcommand\Zo           {{$Z_{\odot}$}}

\defcitealias{driver_2016}{D16}
\defcitealias{windhorst_2011}{W11}

\def\ltsima{$\; \buildrel < \over \sim \;$}
\def\lsim{\lower.5ex\hbox{\ltsima}}
\def\gtsima{$\; \buildrel > \over \sim \;$}
\def\gsim{\lower.5ex\hbox{\gtsima}}

\newlength{\txw}
\newlength{\txh}

\begin{document}

\vspace*{-0.00cm}

\title{On the observability of individual Population III stars and their
stellar-mass black hole accretion disks through cluster caustic transits}

\author{Rogier A. Windhorst}
\affiliation{School of Earth and Space Exploration, Arizona State University,
Tempe, AZ 85287-1404}

\author{F. X. Timmes}
\affiliation{School of Earth and Space Exploration, Arizona State University,
Tempe, AZ 85287-1404}

\author{J. Stuart B. Wyithe}
\affiliation{University of Melbourne, Parkville, VIC 3010, Australia}

\author{Mehmet Alpaslan}
\affiliation{New York University, Department of Physics, 726 Broadway, Room 1005
New York, NY 10003, USA} 

\author{Stephen K. Andrews}
\affiliation{The University of Western Australia, 35 Stirling Highway, Crawley,
WA 6009, Australia}

\author{Daniel Coe}
\affiliation{Space Telescope Science Institute, 3700 San Martin Drive, 
Baltimore, MD 21218}

\author{Jose M. Diego}
\affiliation{IFCA, Instituto de Fisica de Cantabria (UC-CSIC), Avenida de Los 
Castros s/n, 39005 Santander, Spain}

\author{Mark Dijkstra}
\affiliation{Institute of Theoretical Astrophysics, University of Oslo, 0315
Oslo, Norway}

\author{Simon P. Driver}
\affiliation{The University of Western Australia, 35 Stirling Highway, Crawley,
WA 6009, Australia}

\author{Patrick L. Kelly}
\affiliation{University of California at Berkeley, Berkeley, CA 94720-3411}

\author{Duho Kim}
\affiliation{School of Earth and Space Exploration, Arizona State University,
Tempe, AZ 85287-1404}

\email{Rogier.Windhorst@asu.edu, Francis.Timmes@asu.edu, SWyithe@physics.unimelb.edu.au}

\begin{abstract}

We summarize panchromatic Extragalactic Background Light data to place upper
limits on the integrated near-infrared surface brightness (SB) that may come
from Population III stars and possible accretion disks around their stellar-mass
black holes (BHs) in the epoch of First Light, broadly taken from 
z$\simeq$7--17. Theoretical predictions and recent near-infrared power-spectra
provide tighter constraints on their sky-signal. We outline the physical
properties of zero metallicity Population III stars from \texttt{MESA} stellar
evolution models through helium-depletion and of BH accretion disks at
z$\gtrsim$7. We assume that second-generation non-zero metallicity stars can
form at higher multiplicity, so that BH accretion disks may be fed by
Roche-lobe overflow from lower-mass companions. We use these near-infrared SB
constraints to calculate the number of caustic transits behind lensing clusters
that the James Webb Space Telescope and the next generation ground-based
telescopes may observe for both Population III stars and their BH accretion
disks. Typical caustic magnifications can be $\mu$$\simeq$10$^4$--10$^5$, with
rise times of hours and decline times of $\lesssim$1 year for cluster
transverse velocities of \vT$\lesssim$1000 \kms. Microlensing by intracluster
medium objects can modify transit magnifications, but lengthen visibility
times. Depending on BH masses, accretion-disk radii and feeding efficiencies,
stellar-mass BH accretion-disk caustic transits could outnumber those from
Population III stars. To observe Population III caustic transits directly may
require to monitor 3--30 lensing clusters to AB$\lesssim$29 mag over a decade. 

\end{abstract}

\bn 

\bn \section{Introduction}
\label{sec1}

\sn In this paper we consider if the James Webb Space Telescope 
\citep[JWST;][]{gardner_2006, rieke_2005, beichman_2012, windhorst_2008} can
observe First Light objects directly. JWST's Near-InfraRed Camera (NIRCam) is
expected to reach medium-deep to deep (AB$\simeq$28.5--29 mag) flux limits
routinely, and in ultradeep surveys perhaps as faint as AB$\simeq$30--31 mag,
once JWST's on-orbit stray-light properties are mapped. 

Unlensed Population III (Pop III) stars or their stellar-mass black hole (BH)
accretion disks may have fluxes of AB$\simeq$35--43 mag at z$\simeq$7--25, and
therefore are not directly detectable by JWST, not even via ordinary
gravitational lensing targets \citep[\eg][]{rydberg_2013}, which typically have
magnification factors of $\mu$$\simeq$10\ or $\sim$2.5 mag
\citep[\eg][]{lotz_2017}. We use ``$\mu$'' throughout to indicate the lensing
magnification factor, and ``SB'' to indicate surface brightness.

However, cluster caustic transits, when a compact restframe UV-source transits
a caustic due to the cluster motion in the sky, or perhaps due to significant
velocity substructure in the cluster, have the great potential of magnifying
such compact objects temporarily by factors of $\mu$$\simeq$10$^3$--10$^5$
\citep[\eg][]{miralda_escude_1991, kelly_2017_a, kelly_2017_b, diego_2017,
rodney_2017, zackrisson_2015}. This could temporarily boost the brightness of 
a very compact object by $\mu$$\simeq$7.5--12.5 mag, which may render it 
observable by JWST. If Pop III stars --- and/or their resulting BH accretion
disks --- are numerous enough in the sky, it is therefore possible that
individual Pop III stars or their BH accretion disks are temporarily lensed by
foreground cluster caustics as the cluster transits across the background Pop
III target. This could render a AB$\simeq$35--41.5 mag Pop III star at
redshifts z$\simeq$7--17 temporarily visible to a medium-deep or deep
(AB$\simeq$28.5--29 mag), well time-sequenced set of JWST observations. 

The 2016 Planck results \citep{planck_XIII_2016_a, planck_XLVI_2016_b,
planck_XLVII_2016_d} reduced the polarization optical depth even further from 
earlier values --- and reduced its errors --- to
$\tau$$\simeq$0.058$\pm$0.012, thereby placing the redshift of reionization at
approximately \zreion$\simeq$7.8$\pm$0.9 {\it if} it had occurred 
instantaneously. \citet{sobral_2015} discovered an object at z$\simeq$6.7 with
both a clear \Lya 1216 \AA\ line and a possible He 1640 \AA\ line, which may
indicate a late, pristine stellar population dominated by very hot stars,
possibly Pop III stars. That is, the Pop III star epoch may have ended around
z$\simeq$7, and could have started very early, at z\cge 20--40
\citep{trenti_2009}. Of course at z$>>$30, the luminosity distance would be
very large, and render most Pop III stars fainter than \cge 43 mag. In the
hierarchical simulations of \citet{sarmento_2018}, most of the early
star-formation (SF) occurs between z$\simeq$20, when the star-forming
population consists predominantly of pristine Pop III stars, and z$\simeq$7,
when the population is predominantly polluted with metallicities of Z\cge
10$^{-4}$ \Zo. In this paper, we will therefore adopt a redshift range of
z$\simeq$12$\pm$5 where we may observe Pop III stars or their BH accretion
disks directly with JWST {\it if} they are sufficiently strongly lensed during a
cluster caustic transit. For brevity, we will take ``Pop III'' hereafter to 
include any objects at z\cge 7 that may have been already (slightly) polluted
by First Light objects. 

To discuss the possibilities of cluster caustic transits by Pop III objects,
we need to address four different main topics. In \S \ref{sec2}, we summarize
constraints to the possible sky-surface brightness (SB) from objects at z\cge
7, which is the foremost constraint that we must understand first before we can
predict a frequency of potential cluster caustic transits. In \S \ref{sec3},
we present the physical properties of Pop III stars from stellar evolution
models with HR-diagrams through the hydrogen-depletion and helium-depletion
stages, and from these derive their mass-luminosity (ML) relation, their
bolometric+IGM+K-corrections, and their relative contribution to the luminosity
density in faint star-forming objects. In \S \ref{sec4}, we evaluate limits to
the typical transverse velocities of massive lensing clusters, their typical
caustic lengths, the possible effects from microlensing, and estimate the
cluster caustic transit times and rates for the Pop III star parameters from \S
\ref{sec3}. In \S \ref{sec5}, we discuss the possible physical properties of
Pop III stellar-mass BH accretion disks, and under what conditions these may be
fed from early massive stellar binaries. In \S \ref{sec6}, we present estimates
of the cluster caustic transit rates that may result from BH accretion disks.
In \S \ref{sec7}, we discuss what a cluster caustic transit observing program
for Pop III objects with JWST might look like. In \S \ref{sec8}, we summarize
our conclusions.

Throughout we use Planck cosmology \citep{planck_XIII_2016_a}: \Ho~=~66.9 $\pm$
0.9 \kmsMpc, matter density parameter $\Omega_{m}$=0.32$\pm$0.03 and vacuum
energy density $\Omega_{\Lambda}$=0.68$\pm$0.03, resulting in a Hubble time of
13.8 Gyr. When quoting magnitudes, our fluxes are all in AB-magnitudes
(hereafter AB-mag), and our SB-values are in AB-\magarc\ \citep{okegunn_1983},
using $S_{\nu}$ = 10$^{-0.40({\rm AB} - 8.90\ {\rm mag})}$ in Jy.

\bn \section{Constraints to the Sky-Surface Brightness from Objects at z\cge 7}
\label{sec2}

\sn Before we can estimate the number of possible cluster caustic transits of
Pop III objects, we must estimate the maximum possible contribution of Pop III
stars and their stellar-mass BH accretion disks to the observed near--IR sky
surface brightness. In Fig.~\ref{fig:fig1} and \S \ref{sec21}--\ref{sec23}, we
therefore summarize the available data on the Extragalactic Background Light
(EBL) that are directly relevant to our caustic transit calculations in \S 
\ref{sec44} \& \ref{sec62}. Throughout, ``EBL'' will refer to the total
Extragalactic Background Light, including any diffuse EBL component, while 
``iEBL'' will refer to the {\it integrated} EBL extrapolated from the discrete
galaxy counts. 

\mn \subsection{Constraints from the Discrete Extragalactic Background Light}
\label{sec21}

\begin{figure*}[!hptb]
\vspace*{-0.500cm}
\n\cl{
\includegraphics[width=0.950\txw,angle=0]{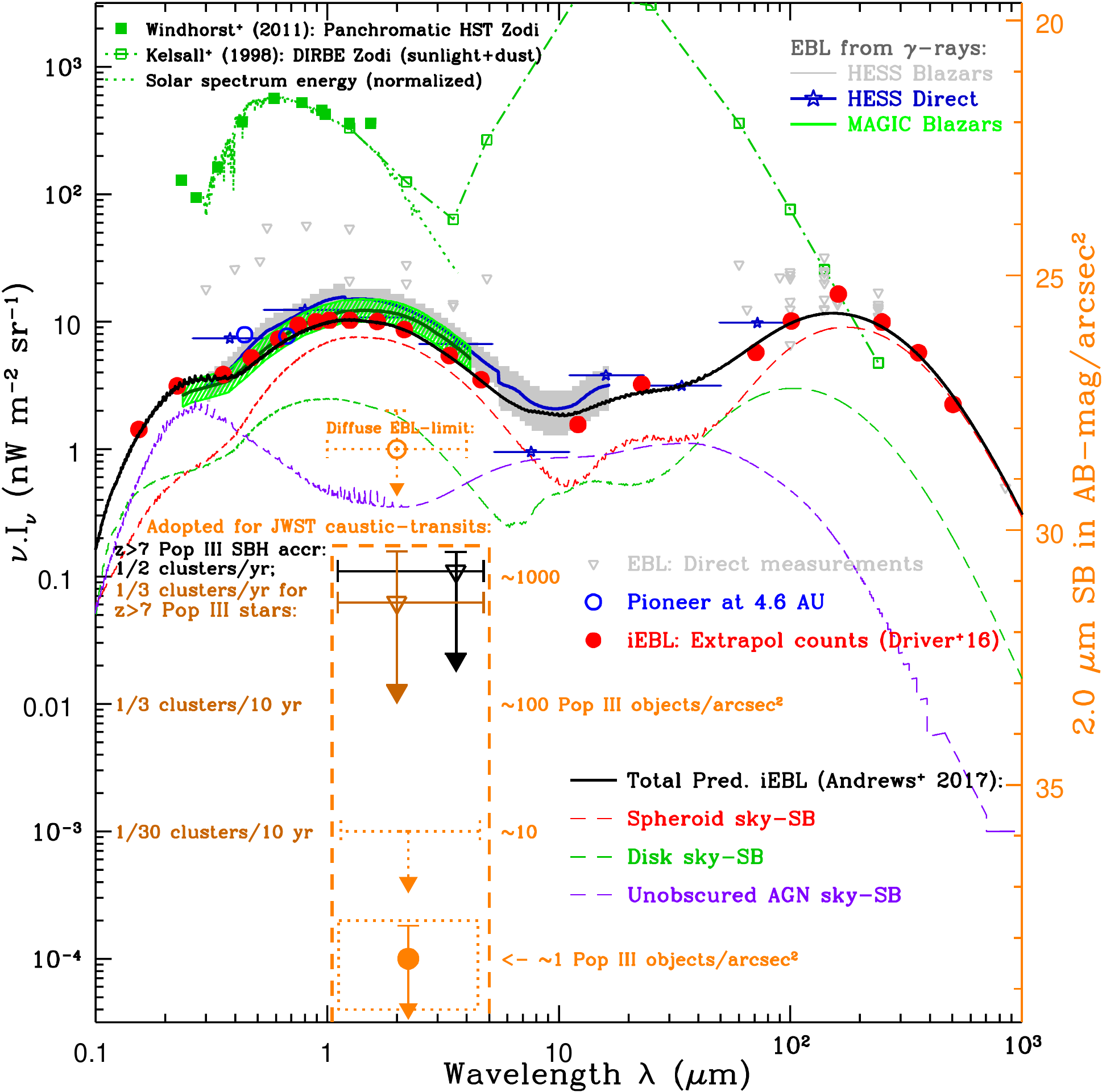}
}
\centering{}
\vspace*{-0.500cm}
\caption{
Summary of panchromatic backgrounds relevant for possible cluster caustic
transits of Pop III stars and their stellar-mass black hole accretion disks.
Green dot-dashed lines with open green squares indicate the scattered and
thermal Zodiacal foreground of \citet{kelsall_1998}. Filled green squares
indicate the panchromatic on-orbit Zodiacal (labeled ``Zodi'') foreground values
measured by HST \citep{windhorst_2011}. Light grey open triangles indicate {\it
direct} measurements of the Extragalactic Background Light from low-Earth orbit
or L2 \citep[for a review, see][]{dwek_2013}. Open blue circles indicate the
direct Pioneer spacecraft EBL values measured beyond most of the Zodiacal dust
at 4.6 AU. Red filled circles indicate the integrated and extrapolated (to AB
\cge 30 mag) panchromatic galaxy counts (iEBL) of \citet[][and references
therein]{driver_2016}. The dashed red, green, and purple lines are iEBL model
predictions for spheroids, disks, and unobscured AGN, respectively
\citep[][]{andrews_2017_b}. The solid black line is their total predicted
iEBL. EBL constraints from HESS $\gamma$-ray blazars are plotted as the light
grey shaded region plus its dark-blue best fit, and for MAGIC blazars as green
shaded region with its dark-green best fit. The orange open circle with dotted
range is our ``hard'' upper limit for the diffuse 1--4 \mum\ EBL, denoted as
``Diffuse EBL-limit''. The dashed orange box contains our adopted upper limits
on the 1--4 \mum\ near-IR sky-SB for Pop III stars at z\cge 7 (dark-orange) and
for their stellar-mass BH accretion disks at z\cge 7 (black). The possible
range in SB from Pop III objects is indicated at the level of $\sim$1, 10, 100,
and 1000 objects/\arcsecsq. The filled orange circle indicates the approximate
SB level of $\sim$1 Pop III star/\arcsecsq. Cluster caustic transit rates that
may be observed with JWST are listed in dark orange on the left for three SB
levels, ranging from $\sim$1 caustic transit per 3 clusters per year to $\sim$1
per 30 clusters {\it if} monitored over 10 years. This is the lowest rate JWST
could detect in a dedicated, large multi-year program. Details are given in \S 
\ref{sec2}--\ref{sec7}.} 
\label{fig:fig1} 
\end{figure*}

\sn In Fig.~\ref{fig:fig1}, the open green squares at 2--3 \mum\ indicate the
\citet{kelsall_1998} COBE DIRBE sky-SB from the Zodiacal light, which is
scattered sunlight. At 3--200 \mum, these COBE DIRBE points are dominated by the
$\sim$200 K thermal dust-component in the Zodiacal belt. Most of this dust is
piled up in the asteroid belt, and is clearly a limiting factor for
near--mid-IR observations, including for JWST observations at $\lambda$ \cge
3.5 \mum. Fig.~\ref{fig:fig1} plots as solid green points the Zodiacal
foreground as measured from low-Earth orbit using the panchromatic Hubble Space
Telescope (HST) Wide Field Camera 3 (WFC3) Early Release Science (ERS)
observations of \citet[]{windhorst_2011} [hereafter 
\citetalias{windhorst_2011}] and its precursor data from the Great Orbiting
Observatories Deep Survey (GOODS) Advanced Camera for Surveys (ACS) data
\citep{giavalisco_2004}. This includes the Zodiacal sky-measurements in the
Hubble Ultra Deep Field (HUDF) by \citet{hathi_2008}. The green dotted line is
the solar energy spectrum \citep{kurucz_2005} normalized to these HST data. 

All units in Fig.~\ref{fig:fig1} have been converted to $\nu$$I_{\nu}$ in units
of \nWsqmsr. For reference, 1.00 \nWsqmsr\ corresponds to 28.41\ \magarc\ at
2.00 \mum, which is indicated by the orange K-band SB-scale in AB-\magarc\ on
the right vertical axis of Fig.~\ref{fig:fig1}. At other near-IR wavelengths,
one can derive the SB-scale corresponding to the $\nu$$I_{\nu}$ scale on the
left by adding --2.5 log ($\lambda$/2.0 \mum) to the K-band scale on the right.

An important comment on the WFC3 ERS data of \citetalias{windhorst_2011} is in
order here. Fig. \ref{fig:fig1} shows that 8 of the 10 ERS filters have
sky-background measurements in line with the Zodiacal foreground at those
wavelengths. However, their bluest and reddest filters (WFC3/UVIS F225W and 
WFC3/IR F160W), have a sky-level significantly in excess of the Zodiacal
foreground for this Ecliptic latitude. This was expected for the F225W filter,
as this bluest WFC3 filter was intentionally scheduled at the end of each
available HST orbit, so that any Earthshine would add some sky level to the
highly readnoise-limited UV-images, since the Zodiacal sky is darkest at the
shortest HST wavelengths. Indeed, the resulting F225W background level was
significantly higher than that expected from the Zodiacal sky alone. In all
other ERS filters, every possible effort was made to avoid the Earth's limb,
but this was not fully successful for the WFC3/IR filter F160W, and its
resulting sky-background was $\sim$0.3 dex higher than expected, despite our
scheduling attempts to avoid this. In the remaining 8 ERS filters, the root
mean square (r.m.s.) variation from the best-fit normalized solar energy
spectrum is 10--20\%, illustrating that even in the case of requesting HST
``LOW-SKY'' observations --- and going at great length in the Astronomer's
Proposal Tool (APT) scheduling requests to make sure the sky background remains
close to the theoretical Zodiacal minimum --- some Earthshine may have
nonetheless leaked into the low-Earth orbit observations. 

The red dots in Fig.~\ref{fig:fig1} indicate the {\it integrated} EBL
measurements derived from the panchromatic (0.1--500 \mum) {\it discrete} galaxy
counts from GALEX, HST, ground-based, Spitzer, WISE and Herschel surveys, as
summarized in \citet[]{driver_2016} [hereafter \citetalias{driver_2016}],
which incorporated the panchromatic HST galaxy counts at
$\lambda$$\simeq$0.2--2 \mum\ to AB\cle 29--30 mag discussed in
\citetalias{windhorst_2011}. From 0.1--500 \mum, the discrete galaxy counts
converge well at almost all wavelengths, except for the less deep Spitzer/WISE
galaxy counts at 8--12 \mum, where the galaxy count extrapolation that yields
the iEBL integral is $\sim$40\% uncertain. Typically, the {\it normalized}
differential galaxy counts in \citetalias{driver_2016} reach a peak at
AB$\simeq$19--25 mag, where most of their iEBL energy is contained. At all
wavelengths except 8--12 \mum, the normalized differential counts converge ---
with a slope flatter than 0.4 dex/mag --- to a finite sky-integral that results
in a well-determined iEBL value for discrete objects to within 10--20\%,
including random errors, count-extrapolation errors, and cosmic variance that
were determined through Monte Carlo simulations. For clarity, error bars are
omitted from Fig.~\ref{fig:fig1}, but these can be found in
\citetalias{driver_2016}. The iEBL from discrete objects is thus well
determined to within \cle 20\% in general, as indicated by the small scatter
in the red dots in Fig.~\ref{fig:fig1} compared to the iEBL models of
\citet{andrews_2017_b}. 

The red, green, and purple dashed lines indicate the contributions that
spheroids, disks, and unobscured AGN at z\cle 6 may contribute to the EBL
energy, following \citet{andrews_2017_a, andrews_2017_b}. Obscured AGN in these
models are incorporated into the spheroidal galaxies, and not plotted
separately. The contribution from unobscured AGN to the discrete iEBL is
uncertain, but at their median redshift of \zmed$\simeq$2, AGN may produce
enough total restframe UV-radiation (at $\lambda_{rest}$\cge 912--1216 \AA) to
contribute significantly to the observed near-UV background
($\lambda_{obs}$\cle 0.4 \mum). Even below $\lambda_{rest}$$\simeq$912 \AA,
AGN at z$\simeq$2--3 may produce non-negligible LyC radiation (possibly made
visible through outflows) to the reionizing budget at these redshifts
\citep[\eg][]{haardt_2015, smith_2018}. As we will discuss below, these
discrete-object iEBL measurements are directly relevant to the possible sky-SB
contributed from unresolved objects, such as Pop III stars and their
stellar-mass BH accretion disks at z\cge 7. 

The light grey open downwards triangles in Fig.~\ref{fig:fig1} indicate the {\it
direct} measurements or limits to the EBL, which are in general {\it absolute}
measurements, and are summarized in detail in \citet{dwek_2013} and
\citetalias{driver_2016}. Most of these direct EBL estimates are a factor of
3--5$\times$ higher than the integrated and extrapolated discrete-objects
counts (the iEBL), and about \cge 2$\times$ higher in the far-IR, although the
latter is in general within the errors. Given that non-zodiacal foreground
light may enter into the low-Earth orbit observations at the \cge 10\% level as
discussed above, it is therefore possible that the true level of foreground
(Zodiacal+Earthshine and other straylight components) may have been
under-subtracted in some of the {\it direct} EBL measurements.


\vspace*{-0.00cm}
\n


\mn \subsection{Limits to the Diffuse Extragalactic Background Light}
\label{sec22}

\sn Here we summarize arguments that the {\it diffuse} EBL is likely smaller
than the iEBL that comes from {\it discrete} objects, especially in the 
near-IR. This will help us derive our first constraints to the diffuse EBL that
may be caused by Pop III stars and their stellar-mass BH accretion disks. Any
real diffuse EBL could be due to faint Inter-galaxy Halo Light
\citep[IHL;][]{cooray_2012}, IntraCluster Light (ICL), or IntraGroup Light
(IGL) not measured by Source Extractor-type algorithms \citep{bertin_2006} in
discrete object surveys, or by truly diffuse, unresolved populations, such as
Pop III stars and their BH accretion disks. Our reasoning that there may not be
a large amount of near-IR {\it diffuse} light hidden are:

\sn 1) Independent diffuse EBL estimates at 0.3--20 \mum\ come from $\gamma$-ray
blazar spectra and how much these are distorted from their original power-law
shape. When a $\gamma$-ray from the blazar hits an intervening EBL photon,
this can result in pair-production and energy loss in the power-law spectrum. 
This constrains the total EBL level that each of the low-redshift blazar
$\gamma$-ray photons are exposed to \citep{dwek_2013, lorentz_2015}.
Fig.~\ref{fig:fig1} indicates the resulting EBL constraints as a grey shaded
region+blue line and a green shaded region+dark-green line from the blazar
surveys with the High Energy Stereoscopic System
\citep[``HESS'';][]{abramowski_2013, abdalla_2017} and the Major Atmospheric
Gamma Imaging Cherenkov telescope \citep[MAGIC;][]{ahnen_2016}, respectively.
The MAGIC shaded region in Fig.~\ref{fig:fig1} is smaller than that of HESS,
and closer to the red iEBL points of \citetalias{driver_2016}. The extent to
which the $\gamma$-ray blazar spectra deviate from their intrinsic power-laws
constrains the amplitude and shape of the foreground component of the EBL
spectrum directly \citep{biteau_2015}, which is {\it completely independent}
from having to subtract the Zodiacal foreground. \citet{biteau_2015} found that
the amount of diffuse EBL at $\lambda$$\simeq$1--5 \mum\ is \cle 1--2 \nWsqmsr.
For a detailed discussion of these blazar data and their constraints to the
EBL, we refer the reader to \citet{dwek_2013} and \citetalias{driver_2016}. In
short, the allowed amount of total 1--5 \mum\ EBL from the $\gamma$-ray blazar
spectral-constraints is generally quite consistent with the integrated and
extrapolated {\it discrete} galaxy counts (red dots in Fig.~\ref{fig:fig1})
summarized in \citetalias{driver_2016}. 

At 0.45--0.65 \mum, the diffuse blazar EBL and the discrete iEBL measurements
are --- to within their errors --- also consistent with the {\it direct} Pioneer
spacecraft measurements \citep{matsuoka_2011}, which were made at a distance of
4.6 AU from the Sun, \ie\ well away from most of the Zodiacal foreground
brightness (blue open circles in Fig.~\ref{fig:fig1}). The direct R-band 
Pioneer EBL measurement was confirmed through the first measurement in a
broader R-band with the Long Range Reconnaissance Imager instrument onboard the
New Horizons spacecraft on its way to Pluto at $\sim$7--16 AU from the Sun
\citep{zemcov_2017}, albeit with a larger error bar, which will improve as
further New Horizons data are taken. At these very large distances from the
Sun, the uncertainties due to the Zodiacal foreground are much smaller than
from low-Earth orbit. Ground-based optical spectroscopy of dark clouds was done
by \citet{mattila_2017} to remove the Diffuse Galactic Light (DGL), suggesting
a diffuse EBL component at $\lambda$$\simeq$0.4--0.6 \mum\ possibly as high as
$\sim$4--6 \nWsqmsr. The good correspondence at $\lambda$\cle 1 \mum\ between
the iEBL from the discrete extrapolated galaxy counts \citepalias{driver_2016},
the direct Pioneer and New Horizons B+R-band observations at 4.6--16 AU, and
the independent constraints from the HESS/MAGIC blazar $\gamma$-ray spectra in
Fig.~\ref{fig:fig1}, suggests that a low-redshift, truly diffuse EBL component
at $\lambda$$\simeq$0.4--1 \mum\ may not exceed the iEBL component itself,
which is $\sim$4--10 \nWsqmsr. 

Despite uncertainties in the optical {\it diffuse} EBL, the 1--4 \mum\ iEBL
results are consistent with the blazar constraints on the diffuse EBL to within
their errors. Below, we will therefore adopt an upper limit to the {\it
diffuse} 1--4 \mum\ EBL based on the difference between the $\gamma$-ray
blazar constraints from HESS + MAGIC and the integrated plus extrapolated
galaxy counts of \citetalias{driver_2016}. If any diffuse 1--4 \mum\ EBL were
truly 3--5$\times$ higher than what the red dots in Fig.~\ref{fig:fig1}
indicate, such a high EBL level would have distorted the blazar spectra more
than what is observed in Fig.~\ref{fig:fig1}. Comparing the HESS and MAGIC
blazar constraints to the EBL from the discrete galaxy counts in
Fig.~\ref{fig:fig1} suggests that a {\it diffuse} 1--4 \mum\ EBL component
\citep{biteau_2015} may add $\sim$20\% to the iEBL from discrete objects
\citepalias{driver_2016}. 

\sn 2) The deepest ground-based surveys with large telescopes do not detect an
excessive amount of light in the outskirts of galaxies that have total fluxes
of AB$\simeq$20--23 mag. It is precisely in this flux range where most of the
iEBL is generated in the observed blue wavelength regime
\citepalias[see][]{driver_2016}. For instance, \citet{ashcraft_2017} present
32-hr LBT U-band images sorted as a function of image FWHM-value. The best 10\%
of their 320 images with the highest-resolution (FWHM\cle 0\arcspt 7) reach AB
\cle 27.0 mag for point source detection, while their best-depth 32-hr image
has FWHM\cle 1\arcspt 8, reaches AB\cle 28.0 mag for point sources, and has a
1-$\sigma$ SB-sensitivity of AB\cle 32\ \magarc. \citet{ashcraft_2017} then
compare the light-profiles of 220 galaxies with total fluxes of
AB$\simeq$20--23 mag in both their highest-resolution images and in their
best-depth LBT U-band image, and find that no more than 5--10\% of the total
galaxy flux is missing in the high-resolution images compared to the {\it
deeper} low-resolution images. That is, at least in U-band for galaxies
AB$\simeq$20--23 mag --- over which most of the iEBL is generated (see \S
\ref{sec21}) --- no more than 10\% of the light appears to be hidden in the
outskirts of these galaxies down to AB\cle 32\ \magarc. The integrated and
extrapolated U-band galaxy counts of \citetalias{windhorst_2011},
\citetalias{driver_2016} and \citet{ashcraft_2017} are consistent with the HESS
and MAGIC blazar constraints at 0.36 \mum, with little room to hide more than
10--20\% in diffuse EBL at 0.36 \mum. Longer wavelengths studies of this depth
have been done with the 10 meter Grand Canary Telescope \citep{trujillo_2016},
with similar results in the r-band. 

\sn 3) Combining the constraints from the previous two arguments, we derive the
following limit to the diffuse 1--4 \mum\ EBL: 1) A diffuse 1--4 \mum\ EBL
component can add $\sim$20\% to the iEBL from discrete objects; and 2) no 
more than 10--20\% in diffuse EBL seems to be hidden in galaxy outskirts to AB
\cle 32\ \magarc. More could come from low redshift ICL, but cluster galaxies
comprise a small fraction of the total galaxy population. Some could come from
IGL at low redshifts, since most galaxies reside in galaxy groups
\citep{robotham_2011}. Where the IGL has been measured, it does not appear to
dominate the total stellar light in galaxy groups \citep[\eg][]{robotham_2011}
or in galaxy clusters \citep[\eg][and references therein]{morishita_2017,
griffiths_2018}. In all, Fig.~\ref{fig:fig1} suggests that {\it diffuse} 1--4 
\mum\ EBL may well be as low as 20\% of the {\it discrete} iEBL, or \cle 1--2
\nWsqmsr\ at 2 \mum. We will use this level as a conservative or ``hard''
upper limit for any Pop III contribution to the near-IR EBL, as indicated by
the orange circle with its dotted 1--4 \mum\ range in Fig.~\ref{fig:fig1}. 

If the diffuse 1--4 \mum\ EBL from Pop III stars or accretion disks at z\cge 
7 was much larger than our hard upper limit of \cle 1--2 \nWsqmsr\ at 2 \mum,
it would exceed the known components from unobscured AGN and even galaxy disks
(blue and green dashed lines in Fig.~\ref{fig:fig1}) at z\cle 6, which would
be unheard of at any other wavelength in the electromagnetic spectrum. That is,
the diffuse 1--4 \mum\ EBL from Pop III stars and/or their accretion disks is
likely well below the level indicated by our hard upper limit at 1--2 \nWsqmsr\
in Fig.~\ref{fig:fig1}. In \S \ref{sec44}, we will estimate the Pop III caustic
transit rate for a range of possible diffuse 1--4 \mum\ EBL values, and
estimate which SB-levels may result in observable numbers of Pop III caustic
transits during JWST's lifetime. 

\mn \subsection{Diffuse EBL Limits Adopted for Pop III Stars and their 
Stellar Mass BH Accretion Disks}
\label{sec23}

\sn Next, we adopt tighter constraints to the sky-SB from Pop III stars from
recent theoretical and observational constraints, and from Pop III stellar-mass
BH accretion disks using recent near-IR--X-ray power-spectrum results. We need
both sky-SB constraints to estimate their cluster caustic transits in \S
\ref{sec44} \& \ref{sec62}, respectively. 

The thermal brightness of the Zodiacal belt rapidly increases at wavelengths
$\lambda$\cge 4 \mum\ (Fig.~\ref{fig:fig1}), and so in the calculations below
we do not anticipate to easily detect Pop III caustic transits with JWST at
wavelengths longer than 4 \mum. For Pop III objects at z\cge 7, the wavelength
range of interest is therefore $\lambda$$\simeq$1--4 \mum. The geometric
average of this wavelength range is $\lambda$=2.0 \mum, which is also equal to
the JWST diffraction limit \citep{rieke_2005}. JWST NIRCam will be most
sensitive over the wavelength range of 2--3.5 \mum, where the Zodical sky 
from L2 is darkest \citepalias[Fig.~\ref{fig:fig1} and][]{windhorst_2011}. 

\sn \subsubsection{Diffuse EBL Limits Adopted for Pop III Stars} 
\label{sec231}

\sn Based on metallicity arguments, \citet{madau_2005} provided a constraint
suggesting that Pop III stars must contribute less than a few \nWsqmsr\ to the
(1--4 \mum) InfraRed Background (IRB). This is consistent with our hard 
diffuse-EBL upper limit in \S \ref{sec22}. \citet{cooray_2012} provide a
detailed Pop III model for reionization, and estimate the Pop III flux to be 
\cle 0.04 \nWsqmsr\ (see their Fig.~4), which we confirm below. 
\citet{bovill_2016} suggests a Pop III star density of 0.1--10$^3$ stars per
\arcsecsq\ between z$\simeq$10--30, which we consider in more detail in \S 
\ref{sec35}. For their expected range in luminosities, Pop III stars could
have an observed flux of AB$\simeq$35--41.5 mag over the redshift range of
z$\simeq$7--17 (see \S \ref{sec3}). As an example, if there existed $\sim$1000
Pop III stars of 100 \Mo\ each per \arcsecsq, then their integrated 2.0 \mum\
sky-SB would be \cge 33\ \magarc\ or \cle 0.016 \nWsqmsr, which is comparable
to the \citet{cooray_2012} limit. 

To confirm these numbers, we will estimate the average sky-SB from star-forming
objects at z$\simeq$7--8 from the actual HUDF data corrected for 
incompleteness. For our caustic transit calculations, we need to estimate the
{\it maximum} possible SB from Pop III stars at z\cge 7 to use as most
conservative upper limit. This needs to take into account that the steep
faint-end of the galaxy luminosity function (LF) at z\cge 7 will contribute
additional flux from {\it unseen} Pop III objects beyond the detection limit of
the deepest HST and JWST images, {\it and} an estimate of the {\it maximum}
additional sky-SB from z$\simeq$9 to z$\simeq$17. We proceed with this
calculation in three steps: 

\sn {\it (a) The average sky-SB from star-forming objects at z$\simeq$7--8 from
the actual HUDF data corrected for incompleteness:}\ We use the actual HUDF
data at z$\simeq$7--8 \citep[Table A1 of][]{bouwens_2015} to estimate the
observed surface densities of star-forming objects at z$\simeq$7 and z$\simeq$8
to an average sky-SB. In the 4.7 arcmin$^2$ effective area of the WFC3/IR data,
there are 56 dropout candidates detected at z$\simeq$7 to the HUDF limit of
AB$\simeq$30.0 mag, while there are 28 dropout candidates at z$\simeq$8 to
AB\cle 30 mag. These can be directly converted to a total sky-SB, in this case
from {\it the objects detected to AB\cle 30 mag}. We need to correct these
observed surface densities by about a factor of 1.8, since in the deepest HUDF
WFC3/IR images, at least $\sim$45\% of the detector pixels are covered by the
wings foreground objects \citep{koekemoer_2013}. Our own insertion of
artificial objects into the HUDF WFC3/IR images confirms this correction
factor. 

\sn {\it (b) Maximum contribution from the steep faint-end of the galaxy
luminosity function down to the luminosity of single Pop III stars:}\ Next, we
correct this upper limit to the 2.0 \mum\ sky-SB that comes from z$\simeq$7--8
for the flux of objects that will have been missed {\it below} the current HUDF
object detection limit of AB$\simeq$30 mag. At z$\simeq$7 to z$\simeq$10, the
AB$\simeq$30 mag HUDF detection limits corresponds to absolute magnitudes of
\MAB$\simeq$--17.5 mag. According to the fits to the available galaxy LF data
in Fig.~5--6 of \citet{finkelstein_2016}, the faint-end slope of the galaxy LF
at z\cge 7 may become as steep as $\alpha$\cle --2.0 to --2.3, while the
characteristic Schechter luminosities (\Lstar or $M^{*}$) and space densities
($\Phi^{*}$) may well continue to get fainter and decline at z\cge 7,
respectively. High resolution hierarchical simulations of the faint-end galaxy
LF-slope evolution with redshift \citep[\eg][]{morgan_2015} suggested values of
$\alpha$$\simeq$--2.1 from z$\simeq$11 to z$\simeq$4. This is about as steep as
the Initial Mass Function (IMF) slope for more massive stars
\citep{coulter_2017}, and would occur if the luminosity density is dominated by
Pop III stars with M \cge 100 \Mo, for which L$\propto$M approximately holds.
We discuss this further in \S \ref{sec31} \& \ref{sec34}. 

We will adopt for simplicity in our extrapolation that $\alpha$$\simeq$--2.0,
so that each additional luminosity bin with objects at z\cge 7 that are
currently beyond the HST detection limit would contribute roughly equal amounts
of energy to the sky-SB. Since the M$^*$ values at z$\simeq$7--8 in the best
fits of \citet{finkelstein_2016} are about $M^{*}\simeq$--20.5 mag, the sky-SB
in the HUDF from objects that are currently {\it resolved} into galaxies comes
effectively from a $\sim$3 mag range in the {\it observed} LF. If we
extrapolate this LF with {\it a faint-end Schechter slope} $\alpha$=--2.0 from
\MAB$\simeq$--17.5 mag to \MAB$\simeq$--7 mag (\ie\ the luminosity of a 20
\Mo-star; see \S \ref{sec3}), then the integrated 2.0 \mum\ sky-SB will be
$\sim$3$\times$brighter than the estimate from {\it (a)} alone.

Integrating the maximum SB that can come from Pop III stars at z\cge 7 to 
\MAB$\simeq$--7 mag is meaningful and necessary, since at this luminosity a
faint star-forming ``object'' would simply consist of a {\it single} unresolved
Pop III star with M \cge 20 \Mo\ and \MAB$\simeq$--7 mag, which is the faintest
JWST could detect at z\cge 7 during a favorable caustic transit (see \S
\ref{sec3}). Given the homology relations in \S \ref{sec31}, the ML-relation
for such massive stars becomes approximately L$\propto$M, so that the very
faint-end slope of the object luminosity function may reflect the bright-end
slope of the stellar mass function at z\cge 8. 

If we integrate down to the limit of a 1.5 \Mo\ Pop III star luminosity of
\MAB$\simeq$+2 mag at z\cge 7 (see \S \ref{sec3}), then the maximum 2.0 \mum\
sky-SB will be $\sim$5$\times$brighter than the estimate from {\it (a)}. Since
these are the coolest stars that can contribute to reionization (\S
\ref{sec3}), we will use this multiplier to derive the most conservative upper
limit to the 2.0 \mum\ sky-SB that may come from z\cge 7. The {\it maximum 2.0
\mum\ sky-SB} we then obtain from the {\it entire} object LF to \MAB=+2 mag is
32.2 \magarc\ at z$\simeq$7 and 32.8 \magarc\ at z$\simeq$8.

\sn {\it (c) Maximum contribution from the cosmic star-formation history at
z\cge 8:}\ Last, we need to correct these limits for the maximum contribution
from the LF of star-forming objects at z$\simeq$9--17 that is not yet accounted
for. For this, we use a best fit of the cosmic star-formation history (SFH) data
summarized by \citet[][]{madau_2014} and \citet{finkelstein_2016}. Eq.~15 of
\citet{madau_2014} gives a best-fit to the cosmic SFR data over the {\it entire}
redshift range 0\cle z\cle 8: 

\begin{equation}
\psi(z) = 0.015\,{(1+z)^{2.7}\over 1+[(1+z)/2.9]^{5.6}}\,{\rm 
M_\odot\,yr^{-1}\,Mpc^{-3}}.
\label{eq:sfrd}
\end{equation}

\n Their best fit has its peak in the cosmic SFR at z$\simeq$1.9. The best-fit
power-law slope for z$>>$2 is approximately 2.7--5.6$\simeq$--2.9, so that at
z$\simeq$7 the SFR is $\sim$1.0 dex or 2.5 mag lower than at z$\simeq$1.9. This
decline is also seen in the more recent HST WFC3 data reviewed by
\citet{finkelstein_2016} and \citet{madau_2017}, who find a slightly steeper
decline of $\propto$(1+z)$^{-3.6}$ to (1+z)$^{-4.2}$ when {\it only} fitting
the data for z\cge 2. The difference in slope could be due to a truly steeper
decline in the cosmic SFR at z\cge 8, the smaller fitted redshift range used 
in these more recent papers, or larger incompleteness corrections for dropout
samples at z\cge 7, as discussed in {\it (a)}. 

To obtain {\it the most conservative upper limit} to the integrated sky-SB from
z=7 to z=17, we will use the {\it highest} predicted SFR at z\cge 8. Hence, we
will use the extrapolation of \citet{madau_2014} in Eq.~\ref{eq:sfrd}, since it
is $\sim$0.3 dex above the fits of \citet{finkelstein_2016} and
\citet{madau_2017} to the most recent WFC3 data at z$\simeq$8--10. The
extrapolation of Eq.~\ref{eq:sfrd} is also consistent with the hierarchical
model predictions of \citet{sarmento_2018} at 7\cle z\cle 20, which
approximately match the \citet{madau_2014} results at z$\simeq$7--8. The
extrapolation of Eq.~\ref{eq:sfrd} thus yields the {\it highest} observed 
sky-SB that may come from star-forming objects at z\cge 8, which is used in \S
\ref{sec44} to predict the highest level of caustic transits that could be
seen. That is, if the true Pop III star sky-SB is lower than what we predict
from Eq.~\ref{eq:sfrd} here, then the caustic transit rates will be
correspondingly smaller, as indicated in Fig.~\ref{fig:fig1} and discussed in
\S \ref{sec44}. 

With the \citet{madau_2014} extrapolation of Eq.~\ref{eq:sfrd}, more than half
of the sky-SB that comes from 7\cle z\cle 17 is already obtained from the
redshift shell at z$\simeq$7, while about 75\% comes from the two redshift
shells at z$\simeq$7 and z$\simeq$8 combined. Each redshift shell here is
assumed to have a width of $\Delta$$z$$\simeq$1. The contributions from the
redshift shells at z\cge 12 are negligibly small. Integrating the sky-SB
produced by each redshift shell by Eq.~\ref{eq:sfrd} from z$\simeq$7--17
produces thus approximately 1.33$\times$ the flux than from the z$\simeq$7--8
redshift shells alone, where we directly summed the observed sky-SB in {\it
(a)}. 

This then results in a {\it most conservative upper limit to the 2.0 \mum\
sky-SB} from star-forming objects at 7\cle z\cle 17 down to the luminosity of a
single Pop III star. This upper limit to the 2.0 \mum\ Pop III star sky-SB is
\cge 31.4 $\pm$0.6\ \magarc\ or \cle 0.06 \nWsqmsr, which is indicated by the
dark-orange open triangle and its error and wavelength range in
Fig.~\ref{fig:fig1}. 

\sn \subsubsection{Diffuse EBL Limits Adopted for Pop III Stellar Mass BH
Accretion Disks}
\label{sec232}

\sn \cite{kashlinsky_2012, kashlinsky_2015}, \citet{cappelluti_2013}, 
\citet{helgason_2016}, and \citet{mitchell-wynne_2016} provided estimates of
the object-free IR-power spectrum. After carefully subtracting all objects in
ultradeep Spitzer 3.6 and 4.5 \mum\ images in the CANDELS GOODS-South field
\citep{grogin_2011, koekemoer_2011}, these papers all found a consistent rather
uniform {\it power} in the power-spectrum on 100--1000\arcs\ scales with an
$r.m.s.$ (amplitude)$^2$ of \cle 0.004 \nWsqmsrsq, which is relatively flat on
the angular scales where it is well sampled, and is fairly similar between 3.6
and 4.5 \mum. While it is possible that residual, very low-level detector
systematics \citep{arendt_2016} or DGL \citep{cooray_2012} may have boosted
this signal, the 3.5 \mum\ power spectrum amplitude itself does provide an
upper limit to the diffuse 3.5 \mum\ sky-SB that may be generated by objects at
z\cge 7, as we will discuss below. 

\citet{cappelluti_2013} cross-correlated the object-subtracted ultradeep
Spitzer images with the deepest object-free 0.2--2 keV Chandra images in the
same CANDELS field, and found a similar power-spectrum signal on \cge 10\arcs\
scales. Their power spectra when cross-correlated with the object-free {\it
soft} (0.5--2 keV, or 1.2 keV in energy on average) Chandra images gave a
stronger signal than when cross-correlated with the {\it hard} (2--4.5 keV or
4.5--7 keV) Chandra images. \citet{cappelluti_2017} fit the 0.3--7 keV energy
spectrum of the X-ray background (XRB) with the redshifted X-ray spectra of
known populations, and constrain the fraction of the XRB that can come from
unresolved sources --- possibly early black holes at z\cge 6 --- as\ \cle 3\%
of the peak in the supermassive black hole (SMBH) growth-rate curve at
z$\simeq$1--2.\footnote{Throughout, ``SMBH'' indicates the rare supermassive
black holes, while ``BH'' indicates the much more numerous stellar-mass black
holes discussed in \S \ref{sec3}, \ref{sec5}, \& \ref{sec6} of this paper.}
\citet{mitchell-wynne_2016} cross-correlated the object-free Spitzer 3.6 and
4.5 \mum\ images with the deepest available object-free CANDELS HST ACS and
WFC3 images at 0.6, 0.7, 0.85, 1.25, and 1.60 \mum, and found {\it no}
correlation with the Spitzer images, or even an anti-correlation in these {\it
shorter} HST wavelength filters. 

This Spitzer--Chandra cross-correlation signal cannot be easily explained by
DGL alone \citep{mitchell-wynne_2016}. If this cross-correlation signal is
real, the implication is that some of it may come from First Light objects at
z\cge 7. Some of this signal may come from an {\it unresolved} AGN or hard
X-ray binary population in faint red bulge-dominated galaxies at {\it lower}
redshifts \citep{cooray_2012} --- from objects both below the Spitzer and
Chandra detection limits. But this signal has also been modeled with Primordial
Black Holes \citep[PBHs][]{kohri_2014}, Direct Collapse Black Holes
\citep[DCBHs][]{yue_2013}, or Obese Black Holes \citep[OBHs][]{natarajan_2017}
at z\cge 7--8. {\it If} part of this 3.6--4.5 \mum\ power-spectrum signal {\it
and} the Spitzer--Chandra cross-correlation signal truly came from z\cge 7,
then it must have an X-ray component that is much hotter than 10 keV in the
restframe (\ie\ T\cge 3$\times$10$^{7}$ K).

Regardless of its correct explanation, the near-IR power-spectrum signal 
provides an upper limit to the 3--4 \mum\ sky-SB that may come from Pop III BH
accretion disks, the inner regions of which may reach X-ray temperatures, as we
will discuss in \S \ref{sec552}. None of the evolutionary models for Pop III
stars that we discuss in \S \ref{sec31} reach temperatures much hotter than
T$\simeq$10$^{5}$ K, and so the redshifted spectral energy distribution (SED)
of Pop III {\it stars at z\cge 7 alone cannot} produce the Spitzer--Chandra
cross-correlation signal. 

Let us now consider the upper limit to the diffuse 3--4 \mum\ sky-SB that may
come from BH accretion disks at z\cge 7. Since the possible 3.6 (and 4.5) \mum\
sky-signal was derived from {\it power spectra} at $\theta$\cge 10-1000\arcs\
scales \citep{kashlinsky_2015, cappelluti_2013, mitchell-wynne_2016}, we must
first convert it to an upper limit to the {\it actual 3.6 \mum\ signal} in the
sky. For this we proceed as follows. The smallest angular scale $\theta$\cge
100\arcs\ at which the 3.6 \mum\ power-spectrum excess signal is seen
corresponds to 4.3--5.2 Mpc {\it physical} scales at z$\simeq$7--17 in our
adopted cosmology with an average of 4.4 Mpc at z$\simeq$8. (Note that the
physical scale needs to be used in this argument, not the co-moving scale). As
in \S \ref{sec231}, the redshift-shell 7\cle z\cle 8 contains about half of
the sky-SB that comes from 7\cle z\cle 17 {\it if} the source of this SB
intrinsically declines as $\propto$(1+z)$^{-2.9}$, or more steeply. At \cle 5
Mpc scales, the overdensities $\Delta\rho/\rho$ are about unity at z$\simeq$0
\citep{barkana_2001}. At redshift $z$, the physical overdensities
$\Delta\rho/\rho$ would thus have been (1+z)$\times$ lower, and so the
signal-amplitude itself (or the sky-SB of the signal) will scale with the
fluctuation in the signal as $\rho$$\simeq$(1+z)$\Delta\rho$. That is, if a
power-spectrum that came from z\cge 7 has an (amplitude)$^2$ at 100\arcs\
scales of \cle 0.004 \nWsqmsrsq, then its linear {\it flux amplitude} must be
less than (1+z)$\times$$\sqrt{(0.004)}$$\simeq$(1+z)$\times$0.06 \nWsqmsr, or
\cle 0.57 \nWsqmsr. 

From their Spitzer--Chandra cross-correlation, \citet{cappelluti_2013} suggest
that \cle 20\% of the large-scale power of the cosmic infrared fluctuations is
correlated with the spatial power spectrum of the X-ray fluctuations. Hence, we
will here adopt that no more than 0.2$\times$0.57 or 0.11 \nWsqmsr\ of the 3.6
\mum\ sky-SB may come from accreting sources at z\cge 7. In
Fig.~\ref{fig:fig1} we indicate this upper limit as the black open triangle
plus its error range in black. This limit is thus far only observationally
constrained at 3.6 and 4.5 \mum, but not yet at 2.0 \mum, although deep JWST
images of cirrus-free, low-extinction regions at the North Ecliptic Pole will
provide sky-SB constraints at 2.0 \mum\ as well \citep{jansen_2017}. At 3.6
\mum, this current SB-limit corresponds to \cge 30.2\ \magarc\ following the
wavelength-dependent conversion between \nWsqmsr\ and AB-\magarc\ in \S
\ref{sec21}. Since the Spitzer power-spectra and cross-correlation spectra with
Chandra of \citet{mitchell-wynne_2016} are fairly similar at both 3.6 and 4.5
\mum\ in units of \nWsqmsr, we will therefore adopt the equivalent sky-SB value
of \cge 30.8\ \magarc\ at 2.0 \mum\ as the upper limit for BH caustic transit
calculations, as indicated by the black upper limit in Fig.~\ref{fig:fig1}. 

In summary, \S \ref{sec231} and \ref{sec232} yield rather similar upper limits
to the 2.0 \mum\ sky-SB that may come from Pop III stars or their stellar-mass
BH accretion disks of \cge 31\ \magarc. In what follows, we will therefore do
the caustic transit calculations assuming that the {\it full}\ 2.0 \mum\ sky-SB
signal of \cge 31\ \magarc\ is {\it either} completely caused by Pop III stars
(\S \ref{sec44}), {\it or} by their BH accretion disks (\S \ref{sec62}). For
the plausible case where both Pop III stars {\it and} their BH accretion disks
{\it both} contribute to the 2.0 \mum\ sky-SB of \cge 31\ \magarc, one could
use a weighted sum of the caustic transit rates derived in \S \ref{sec44} and
\ref{sec62} for Pop III stars {\it and} their BH accretion disks, respectively.
Where appropriate, we give size, lifetime, and obscuration arguments regarding
the proportions of caustic transits Pop III stars and their BH accretion disks
that may be visible to JWST (\S \ref{sec31}, \ref{sec53}, \ref{sec62}). 

Note that for our caustic transit calculations it does not matter whether the
light that comes from z\cge 7 exists in {\it faint discrete objects} that have
already been detected down to the HUDF limit and contain Pop III stars and/or
stellar-mass BH accretion disks, or whether this light is fully {\it
unresolved} below the current HUDF object detection limit of AB$\simeq$30 mag.
Either way, the maximum 2.0 \mum\ SB of $\sim$31\ \magarc\ that can be produced
at z\cge 7 may be subject to cluster caustic transits.

\n \section{Parameters Adopted for Pop III Stars}
\label{sec3}

\sn In this section we present the physical properties of Pop III stars from
stellar evolution models with HR-diagrams through the hydrogen-depletion and
helium-depletion stages, and from these derive their mass-luminosity relation,
their bolometric+IGM+K-corrections, and their relative contribution to the 
luminosity density in a faint star-forming object. 

Simulations suggest that fragmentation of primordial gas allows central
concentrations to form in a mini-halo with a range of stellar masses,
depending on the dimensionality, spatial resolution, and local physics used in
the simulations. For instance, \citet{abel_2002} presented a 3D hydrodynamical
simulation to form the first stars, which resulted in a 100 \Mo\ star to form.
In a higher resolution simulation \citep{turk_2009}, a 50 \Mo\ clump breaks up
into two cores, each with a forming star that likely will become a binary
star. Radiation-hydrodynamic simulations of primordial clouds
\citep{hosokawa_2016} showed fragmentation into protostars with masses
M$\simeq$10--1000 \Mo, depending on the amount of UV-feedback that was
produced. \citet{stacy_2016} follow the formation of a mini-halo with gas
collapsing into central cores ranging from 20 \Mo\ to as many as $\sim$30 
stars with M\cle 1 \Mo. 

Strong radiative feedback from the most massive stars may initially prevent
lower mass stars --- and therefore binaries --- from forming in a mini-halo
\citep[\eg][]{abel_2002, trenti_2009}. We discuss the low near-IR sky-SB that
may result from this in \S \ref{sec35}. Given that more recent simulations
resulted in the formation of lower mass stars and binaries, we will also allow
for the possibility that slightly-polluted lower-mass stars --- and binaries
--- can form in the vicinity of previous more massive, zero metallicity Pop III
stars \citep[Z\cle 10$^{-4}$\Zo;][]{sarmento_2018}. We discuss this in more
detail in \S \ref{sec32} \& \ref{sec52}. When the distinction is relevant, we
refer to these slightly polluted stars as ``Pop II.5''. This paper will thus
consider stars of zero or very low metallicity that cover the mass range of 1
\Mo\cle M\cle 1000 \Mo. 

\mn \subsection{Pop III Star Physical Parameters from \texttt{MESA} Models}
\label{sec31}

\begin{figure*}[!hptb]
\vspace*{-0.700cm}
\n\cl{
\includegraphics[width=1.00\txw,angle=0]{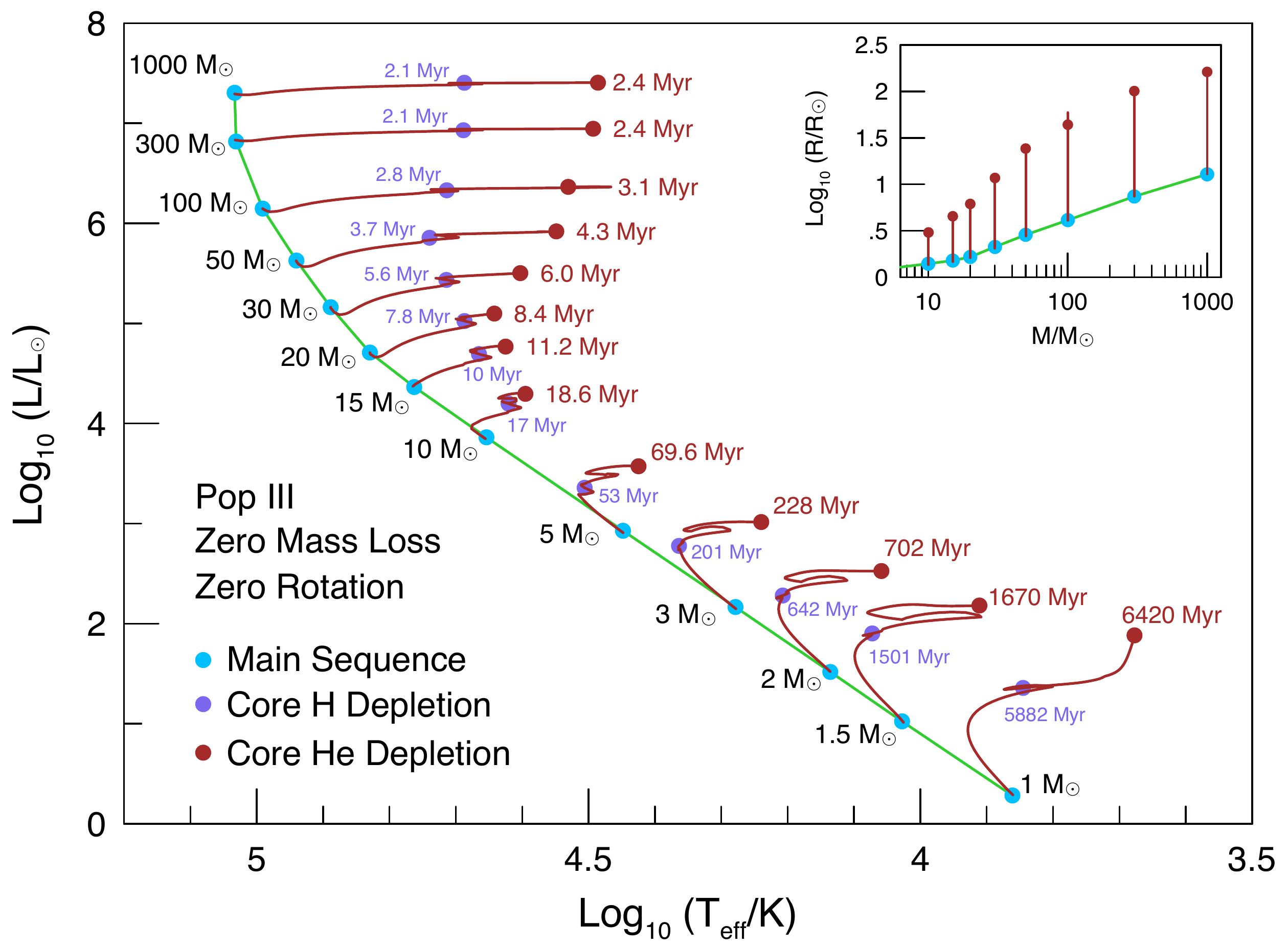}
}
\centering{}
\vspace*{-0.500cm}
\caption{
Loci of the zero age main-sequence in the HR diagram for non-rotating, zero
mass loss, Z = 0.00 \Zo\ \texttt{MESA} models. Evolutionary tracks to core
He-depletion are shown, with the final model marked by filled circles and
labeled by age. The actual model data are given in 
Tables~\ref{tab:tab1}--\ref{tab:tab4}. The inset plot shows the mass-radius
evolution, with filled circles marking the location of ZAMS and core 
He-depletion. \\
}\label{fig:fig2}
\end{figure*}

\sn We first need to outline the plausible physical parameter ranges for Pop
III stars. Fig.~\ref{fig:fig2} shows the zero age main-sequence (ZAMS) in an
HR diagram for stellar evolution models with Z = 0.00 \Zo, and the inset shows
their corresponding mass-radius relation. These non-rotating, zero metallicity,
zero mass-loss, single star 1--1000 \Mo\ models were calculated using the
\texttt{MESA} software instrument \citep{paxton_2011, paxton_2013, paxton_2015}
with physical and numerical parameters the same as those in \citet{farmer_2015},
\citet{fields_2016}, and \citet{farmer_2016}. We also calculated \texttt{MESA}
models for Z = 10$^{-8}$ \Zo, and their results were very similar to Z = 0.00
\Zo. This is because stars more massive than $\sim$2 \Mo\ make enough of their
own carbon in their cores to run the CNO cycle appropriate for their mass. In
other words, there is a floor metallicity, which --- if not provided by the
star's birth composition --- will be made by the star itself, and convective
episodes may bring part of these self-made metals to the stellar photosphere.
Here we adopt the set of Z = 0.00 \Zo\ \texttt{MESA} models for Pop III stars,
and will discuss the possible effects of metallicity in more detail below and
in \S \ref{sec52}. 

There may be model-dependent variations in the MS ages, depending on the age
definitions and on the chemical mixing algorithms used (\eg\ convection,
overshooting, etc). For details, we refer to \citet{paxton_2011, paxton_2013,
paxton_2015}. In our \texttt{MESA} models, the ZAMS by definition starts when
the nuclear luminosity reaches 90\% of the total stellar luminosity. The
Terminal-Age Main Sequence (TAMS) is defined when the central hydrogen mass
{\it fraction} drops to below 10$^{-6}$ of the star's core mass, which is when
the ZAMS ends. At this point, shell hydrogen burning dominates the energy
production, and can be taken as the ``beginning'' of the ``Giant Branch''. Core
Helium depletion is defined as the stage in the star's evolution when the
fraction of $^4$He drops to below 10$^{-6}$ of the core mass of the star. These
definitions are more precise than the common use of ``Red Giant Branch'' or
``Asymptotic Giant Branch'', but for the sake of brevity, we will henceforth
refer to these latter stages as the ``RGB'' and ``AGB'', respectively. The MS
age adopted here is defined as the time between the start of the ZAMS and the
start of the RGB (core-hydrogen depletion), while the ``Giant Branch'' (GB) age
is defined between the start of the RGB and the end of the AGB, when the star
has run out of $^4$He. 

Stars more massive than \cge 100 \Mo\ are radiation-pressure dominated. For
CNO burning, constant electron scattering, and radiative transport, the ZAMS
``homology'' relations for massive stars \citep{hoyle_1942, faulkner_1967,
pagel_1998, bromm_2001, portinari_2010} are: 

\begin{equation}
\begin{split}
\left ( \frac{R}{10 \ R_{\odot}} \right ) & \simeq \left ( \frac{Z}{10^{-7} \
Z_{\odot}} \right) ^{1/11} \left ( \frac{M}{400 \ M_{\odot}} \right) ^{5/11},\\
\left ( \frac{T_{eff}}{10^5 \ {\rm K}} \right ) & \simeq \left (
\frac{Z}{10^{-7} \ Z_{\odot}} \right ) ^{-1/20} \left ( \frac{M}{100 \
M_{\odot}} \right) ^{1/40},\\
\left( \frac{L}{L_{\odot}} \right )& \simeq \left( \frac{L_{\rm
edd}}{L_{\odot}} \right )\simeq 10^5 \left ( \frac{M}{M_{\odot}} \right ),
\end{split}
\label{eq:homologyrelations}
\end{equation}

\sn and approximate the trends shown by our detailed \texttt{MESA} models for
Z\cle 10$^{-8}$ \Zo\ (Fig.~\ref{fig:fig2} and Table~\ref{tab:tab1}). 

Metallicity affects the evolution of single stars in four distinct ways: it
sets their initial abundance, and it impacts their energy generation, opacity,
and their mass loss mechanism. For low metallicity stars, these homology
relations approximate the metallicity-dependence of their radii and
luminosities. For metallicities Z\cle 10$^{-4}$\ \Zo,
Eq.~\ref{eq:homologyrelations} suggest that the ZAMS tracks in
Fig.~\ref{fig:fig2} will shift systematically by a factor of $\sim$0.7 towards
lower effective temperatures almost independent of mass, while the ZAMS
luminosities would be nearly independent of the metallicity. The ZAMS radii
will correspondingly shift by a factor of $\sim$2 towards larger values. Given
the other much larger uncertainties in our Pop III star caustic transit
calculations in \S \ref{sec44}, we will adopt the physical parameter values of
zero metallicity Pop III stars (or Z = 0.00~\Zo) from our \texttt{MESA} modeling
runs.

Eq.~\ref{eq:homologyrelations} suggests that the bolometric luminosities of zero
metallicity Pop III stars --- as modeled in our \texttt{MESA} runs --- are to
first order {\it directly proportional} to their ZAMS mass, while the
mass-radius and mass-T$_{eff}$ relations have much shallower slopes. All three
parameters in Eq.\ref{eq:homologyrelations} need to be carefully traced as a
function of ZAMS mass for our caustic transit calculations in \S \ref{sec44}.

Closer inspection of Fig.~\ref{fig:fig2} suggests that
Eq.~\ref{eq:homologyrelations}c is only approximately correct for stars with
M\cge 100 \Mo. Over the mass range of 1\cle M\cle 1000 \Mo, the bolometric ZAMS
luminosities of Pop III stars in Fig.~\ref{fig:fig2} scale to a better
approximation with ZAMS mass $M$ as: 

\vspace{-0.30cm}
\hspace{-0.40cm}
\begin{eqnarray}
\nonumber 
\hspace{-0.40cm}
&&L \simeq L_{100}\ (M/100\ \mbox{M}_\odot)^{1.16},\ (100\lesssim M\lesssim
1000 \mbox{M}_\odot), \\
\nonumber 
\hspace{-0.00cm}
&\simeq& L_{100}\ (M/100\ \mbox{M}_\odot)^{2.06},\ \ \ \ \ \ (20\lesssim 
M\lesssim 100\ \mbox{M}_\odot), \\
\hspace{-0.00cm}
&\simeq& L_{20}\ (M/20\ \mbox{M}_\odot)^{3.20},\ \ \ \ \ \ \ (1\lesssim\ 
M\lesssim 20\ \mbox{M}_\odot),
\label{eq:masslumrelations}
\end{eqnarray}

\vspace*{-0.00cm}
\n where $L_{100}$ and $L_{20}$ are the luminosities of a 100 \Mo\ and a 20
\Mo\ star, respectively. The first two segments in this equation are rescaled
to the parameters of a 100 \Mo\ star, and the third to a 20 \Mo\ star. The
first segment is the nearly linear mass-luminosity relation for the most
massive (M\cge 100 \Mo) Pop III stars in Eq.~\ref{eq:homologyrelations}c, the
second segment is a good approximation for Pop III stars in the intermediate
mass range (20\cle M\cle 100\ \Mo), and the third segment is the ML-relation
for Pop III stars with M$\simeq$1--20 \Mo, which has a slope of $\sim$3.2,
similar to the slope of the ML-relation for lower-mass stars in our own Galaxy.
Our caustic transit calculations in \S \ref{sec44} are dependent on stellar
luminosity, and so we will propagate the segmented ML-relation of
Eq.~\ref{eq:masslumrelations} into the relevant equations
(Eqs.~\ref{eq:NM}--\ref{eq:Nlens}) of \S \ref{sec44}.

\sn Here we discuss in more detail the \texttt{MESA} Pop III star physical
parameters that are needed to estimate their resulting caustic transit rates
at z\cge 7:

\sn {\bf Masses:}\ The mass range for Pop III stars that have luminosities
bright enough for caustic transit detection by JWST at z\cge 7 is 30--1000
\Mo\ (see Table~\ref{tab:tab1} and \S \ref{sec44}). This corresponds to a
logarithmic mass range of M$\simeq$175 \Mo\ $\pm$0.75 dex, with a
corresponding bolometric absolute magnitude range of \MAB$\simeq$--10.8$\pm$2.5
mag. As discussed in \S \ref{sec51}, LIGO has detected several BHs at the
lower-end of this mass range, some of which may be the leftover of Pop III
stars. BH accretion disks in stellar binaries are therefore considered in \S
\ref{sec5}--\ref{sec6} as possible additional sources of caustic transits at
z\cge 7. When computing physical quantities below, we use a solar mass of
1.989$\times$10$^{30}$ kg \citep{mamajek_2015}. 

\sn {\bf Temperatures:}\ Effective temperatures were determined by integrating a
finely-zoned \texttt{MESA} photospheric model inward to an optical depth of
$\tau$=1. This radial location becomes the effective radius \Reff\ at which the
effective temperature is \Teff. Zero metallicity Pop III stars have ZAMS
photospheric temperatures ranging from \Teff$\simeq$7300--108,000 K for masses
in the range of M$\simeq$1--1000 \Mo, as summarized in Table~\ref{tab:tab1}.
During the RGB stage, Pop III stars with M$\simeq$1--1000 \Mo\ have lower
temperatures ranging from 7000--55,000 K, while during the AGB stage, their
temperatures are even lower, ranging from 6300--44,000 K (see
Fig.~\ref{fig:fig2}). For both post-MS stages, the photospheric temperatures
change non-monotonically with mass. This will affect their bolometric, IGM and
K-corrections in a non-linear way as a function of mass and stellar evolution
stage (see \S \ref{sec33}). When calculating their bolometric corrections
using black body curves below, we use as reference a solar photospheric
temperature of 5772 K \citep{mamajek_2015, prsa_2016}. 

\sn {\bf Radii:}\ Fig.~\ref{fig:fig2} and Table~\ref{tab:tab1} show that zero
metallicity Pop III stars have ZAMS effective radii ranging from 
\Reff$\simeq$0.9--13 \Ro. These are in line with previous predictions
\citep[\eg][]{woosley_2002, hirschi_2007, ohkubo_2009, yusof_2013}. In 2015,
the IAU adopted --- for stellar normalization purposes --- a value of the solar
radius of 1.00 \Ro\ $\equiv$ 695,700 km \citep{mamajek_2015, prsa_2016}, which
was guided by recent space-based measurements \citep[\eg][]{emilio_2012}. Hence,
the Pop III star ZAMS radii in Table~\ref{tab:tab1} range from $R_{\rm Pop III}$
= 6.05$\times$10$^{8}$--8.97$\times$10$^{9}$ m, which are the numbers we use for
Pop III star caustic transit rate predictions in \S \ref{sec44}. The Pop III
star radii are at most between 1.3--5.8$\times$larger during their RGB phase,
and 2.3--14$\times$ larger during their AGB phase (see Table~\ref{tab:tab1}).


\vspace*{-0.00cm}
\n



\begin{deluxetable*}{| c | c | ccc | cccc | cccc | c |}
\tablecolumns{14}
\tablewidth{1.0\linewidth}
\tablecaption{Adopted Pop III Star Physical Parameters from \texttt{MESA} models$^a$ 
\label{tab:tab1}}
\tablehead{
\colhead{Mass}                      $\vert$               & 
\colhead{Age}                       $\vert$               & 
\colhead{$T_{eff}$}                                       &
\colhead{$\log R$}                                        &
\colhead{$\log{L_{\rm bol}}$}        $\vert$              &
\colhead{$T_{eff}$}                                       &
\colhead{$\log R$}                                        &
\colhead{$\log{L_{\rm bol}}$}                             &
\colhead{Age}                       $\vert$               &
\colhead{$T_{\rm eff}$}                                   &
\colhead{$\log R$}                                        &
\colhead{$\log{L_{\rm bol}}$}                             &
\colhead{Age}                       $\vert$               &
\colhead{Time$^b$}                                        \\[-4pt]
\colhead{ }                         $\vert$               & 
\colhead{Pre-MS}                    $\vert$               & 
\multicolumn{3}{c}{--- at ZAMS ---} $\vert$               & 
\multicolumn{4}{c}{--- at Hydrogen-depletion ---} $\vert$ & 
\multicolumn{4}{c}{--- at Helium-depletion ---}   $\vert$ & 
\colhead{AGB-MS}                                          \\[-4pt]
\colhead{($M_{\odot})$}             $\vert$               &
\colhead{(Myr)}                     $\vert$               & 
\colhead{(K)}                                             & 
\colhead{($R_{\odot}$)}                                   & 
\colhead{($L_{\odot}$)}             $\vert$               & 
\colhead{(K)}                                             & 
\colhead{($R_{\odot}$)}                                   & 
\colhead{($L_{\odot}$)}                                   & 
\colhead{Myr}                     $\vert$                 & 
\colhead{(K)}                                             & 
\colhead{($R_{\odot}$)}                                   & 
\colhead{($L_{\odot}$)}                                   & 
\colhead{Myr}                     $\vert$                 & 
\colhead{(Myr)}                                           
}
\mn
\startdata
1.0  & 9.28 & 7.266e3 &--0.0581 & 0.2825 & 6.999e3 & 0.5119 & 1.3576 & 5882 & ---$^c$ & ---    & ---    & 6420 & 538 \\
1.5  & 6.11 & 1.065e4 &--0.0203 & 1.0227 & 1.181e4 & 0.3292 & 1.9015 & 1501 & 8.149e3 & 0.7913 & 2.1804 & 1670 & 169 \\
2.0  & 3.02 & 1.367e4 &  0.0108 & 1.5177 & 1.611e4 & 0.2498 & 2.2815 &  642 & 1.145e4 & 0.6685 & 2.5249 &  702 &  60 \\
3.0  & 1.38 & 1.899e4 &  0.0487 & 2.1654 & 2.311e4 & 0.1843 & 2.7770 &  201 & 1.736e4 & 0.5510 & 3.0138 &  228 &  27 \\
5.0  & 0.56 & 2.805e4 &  0.0911 & 2.9274 & 3.206e4 & 0.1903 & 3.3581 &   53 & 2.658e4 & 0.4608 & 3.5732 &   70 &  17 \\
 10  & 0.23 & 4.508e4 &  0.1462 & 3.8618 & 4.174e4 & 0.3807 & 4.1972 &   17 & 3.938e4 & 0.4811 & 4.2968 &   19 & 1.6 \\
 15  & 0.13 & 5.789e4 &  0.1803 & 4.3647 & 4.624e4 & 0.5401 & 4.6937 &   10 & 4.215e4 & 0.6581 & 4.7691 &   11 & 0.8 \\
 20  & 0.09 & 6.754e4 &  0.2183 & 4.7082 & 4.864e4 & 0.6612 & 5.0240 &  7.8 & 4.386e4 & 0.7879 & 5.0975 &  8.4 & 0.6 \\
 30  & 0.05 & 7.737e4 &  0.3270 & 5.1619 & 5.180e4 & 0.8120 & 5.4347 &  5.6 & 4.006e4 & 1.0688 & 5.5016 &  6.0 & 0.5 \\
 50  & 0.03 & 8.713e4 &  0.4570 & 5.6283 & 5.490e4 & 0.9722 & 5.8562 &  3.7 & 3.536e4 & 1.3862 & 5.9200 &  4.3 & 0.5 \\
100  & 0.02 & 9.796e4 &  0.6147 & 6.1470 & 5.173e4 & 1.2610 & 6.3303 &  2.8 & 3.392e4 & 1.6437 & 6.3627 &  3.1 & 0.3 \\
300  & 0.02 & 1.074e5 &  0.8697 & 6.8172 & 4.882e4 & 1.6111 & 6.9301 &  2.1 & 3.165e4 & 2.0041 & 6.9631 &  2.4 & 0.3 \\
1000 & 0.02 & 1.080e5 &  1.1090 & 7.3047 & 4.807e4 & 1.8740 & 7.4288 &  2.1 & 3.122e4 & 2.2119 & 7.3549 &  2.4 & 0.3 \\
\enddata
\mn
\tablenotetext{a}{All physical Pop III star parameters were calculated using
\texttt{MESA} models with zero initial metallicity (Z = 0.00 \Zo), zero mass
loss, zero rotation, and no stellar duplicity (\ie, no binaries--multiple
stars). Pop III star parameters are listed with a sufficient number of
significant digits to be able to integrate them assuming black body spectra,
which is needed in \S \ref{sec33}.}

\tablenotetext{b}{Ages are listed for the pre-main sequence collapse (pre-MS),
core hydrogen burning phase, shell hydrogen (H-depletion), and core+shell helium
burning phases (He-depletion), and the total giant branch lifetime (\ie\ the
AGB--MS age difference). The latter provides an upper limit to the BH feeding
times due to Roche-lobe overflow in non-zero metallicity massive-star binaries,
as discussed in \S \ref{sec5}.}

\tablenotetext{c}{The 1.0 \Mo, Z = 0.00 \Zo\ model did not ignite helium and may
thus turn directly into a helium white dwarf, so no AGB parameters are listed
here. (Its AGB ages and K-corrections in \S \ref{sec33} are those of the Z =
10$^{-8}$ $Z_{\odot}$ model, which did end in a white dwarf).} 
\end{deluxetable*}


\sn {\bf Luminosities:}\ The \texttt{MESA} models shown in Fig.~\ref{fig:fig2}
yield the bolometric absolute magnitudes of Pop III stars as a function of
their mass and for different stellar evolution ages. Zero metallicity Pop III
stars have ZAMS luminosities ranging from \Lbol$\simeq$1.9
\Lo--2.0$\times$10$^7$ \Lo\ for masses M$\simeq$1--1000 \Mo, respectively.
During the RGB stage, their luminosities range from \Lbol$\simeq$23
\Lo--2.7$\times$10$^7$ \Lo, while during the AGB stage, they are
\Lbol$\simeq$40 \Lo--2.4$\times$10$^7$ \Lo. These are the full {\it bolometric}
stellar luminosities as predicted by the \texttt{MESA} code. 

Several stellar atmosphere calculations have suggested that zero metallicity
Pop III stars in both the ZAMS, RGB, and AGB stages can be approximated as 
black body emitters \citep[\eg][]{bromm_2001} due to the lack of atomic
absorption features or line blanketing in their spectra. For our calculations
below, we therefore approximate the Pop III stars with black body curves of the
same photospheric temperatures and radii from the \texttt{MESA} models
summarized in Tables~\ref{tab:tab2}--\ref{tab:tab4}. We integrated these black
body curves in $I_{\nu}$ from hard X-ray to radio wavelengths, and use their
listed stellar radii (in km using \Ro\ above) to predict the theoretical
luminosities integrated under the full Planck curve for stars of that size. We
use these results to convert their bolometric luminosities to observed apparent
magnitudes in JWST's near-IR filters, which requires the distance modulus (DM)
as a function of redshift, their bolometric corrections, corrections for IGM
transmission, and their K-corrections (see \S \ref{sec33}). 

To normalize these calculations, we use the bolometric luminosity of the Sun,
\Lo=3.828$\times$10$^{26}$ W, which by definition is produced by a black body
with the effective temperature ($T_{eff}$=5772 K) and radius
(\Ro=6.957$\times$10$^{8}$ m) adopted for the Sun \citep{mamajek_2015}. This
corresponds to an absolute {\it bolometric} AB-magnitude of the Sun, which is by
definition \Mbol$({\odot}$)$\equiv$+4.74 mag \citep[][]{bessell_1998,
casagrande_2006}. Hence, all our Pop III star absolute magnitudes in
Tables~\ref{tab:tab2}--\ref{tab:tab4} are normalized to this \Mbol-value of the
Sun. 

This worked well for all Pop III stars in our \texttt{MESA} runs, except for
the 1.0 \Mo\ zero metallicity AGB model, whose \texttt{MESA} predictions were
2\% lower than a black body curve at its specified $T_{eff}$. Its
\texttt{MESA} model did not ignite helium, so the star may turn directly into a
helium white dwarf. Hence, no AGB parameters are listed in Table~\ref{tab:tab1}
for a 1.0 \Mo\ {\it zero} metallicity AGB star. To permit caustic calculations
for a 1.0 \Mo\ AGB star in any case, Table~\ref{tab:tab4} lists the parameters
for the Z = 10$^{-8}$ $Z_{\odot}$ \texttt{MESA} model, which {\it did} result in a
white dwarf. 

\sn {\bf Ages:}\ Pre-MS ages were estimated from the collapse time of a gas
cloud of the specified mass --- using atomic and molecular H-cooling --- to
the onset of ZAMS stage, and are not added to the other ages below.
Table~\ref{tab:tab1} lists the \texttt{MESA} ages for the (estimated) pre-main
sequence (pre-MS) collapse time, the core hydrogen burning phase (H-depletion),
the shell hydrogen and core+shell helium burning phases (He-depletion), and for
the total time spent on the giant branches. That is, (\tauAGB--\tauMS)
estimates the lifetime of the Red Giant Branch, Hot Horizontal Branch and
Asymptotic Giant Branch together (RGB+HHB+AGB). Note that for the most massive
Pop III stars, the HHB (core He-burning) is of very short duration, and the
stars essentially quickly transit from shell H-burning to shell He-burning in
one smooth, nearly horizontal giant branch towards cooler \Teff-values. We will
refer to the combined RGB+HHB+AGB phases as the giant branch (``GB''). For all
Pop III stars in Table~\ref{tab:tab1}, the time between core H-depletion and
core He-depletion is about 8--14\% of the time between ZAMS and core
H-depletion, with an average of $\sim$12\%. For our caustic transit
calculations in \S \ref{sec44}, we will take the approximate duration of the
Pop III RGB- and AGB-stages each to be about 6\% of their ZAMS lifetime. 

For Pop III stars, we will use the post-MS lifetimes in \S \ref{sec5} to
estimate the maximum time that a lower-mass He-burning star may be feeding the
accretion disk around a BH that was leftover from a more massive Pop III
companion star at z\cge 7. This assumes no major mass exchange during the 
{\it prior} stellar evolution stages, \ie\ we assume that stars in multiple
systems evolve in isolation during the ZAMS stage following the \texttt{MESA}
models.


\begin{deluxetable*}{| c | cccc | ccc | ccc | c | c |}
\tablecolumns{13}
\tablewidth{1.0\linewidth}
\tablecaption{Implied ZAMS Pop III Star Observational Parameters Relevant to Caustic Transit Calculations
\label{tab:tab2}}
\tablehead{
\colhead{Mass$^a$}                             $\vert$ &
\colhead{\Teff $^b$}                                   & 
\colhead{Radius $^c$}                                  & 
\colhead{\Lbol\ $^d$}                                  &                      
\colhead{M$_{\rm bol}$$^e$}                    $\vert$ & 
\multicolumn{3}{c}{Bolo+IGM+K-corr$^f$}        $\vert$ & 
\multicolumn{3}{c}{ZAMS m$_{\rm UV}$$^g$}      $\vert$ & 
\colhead{t$_{rise}$$^h$}                       $\vert$ & 
\colhead{transit$^i$}                                  \\[-4pt] 
\colhead{ZAMS}                                 $\vert$ &
\multicolumn{4}{c}{--- at ZAMS ---}            $\vert$ & 
\colhead{z=7}                                          & 
\colhead{z=12}                                         & 
\colhead{z=17}                                 $\vert$ & 
\colhead{z=7}                                          & 
\colhead{z=12}                                         & 
\colhead{z=17}                                 $\vert$ & 
\colhead{caust}                                $\vert$ & 
\colhead{rate}                                         \\[-4pt]
\colhead{($M_{\odot}$)}                        $\vert$ &
\colhead{(K)}                                          & 
\colhead{($R_{\odot}$)}                                & 
\colhead{($L_{\odot}$)}                                & 
\colhead{(AB)}                                 $\vert$ & 
\multicolumn{3}{c}{(AB-mag)}                   $\vert$ & 
\multicolumn{3}{c}{(AB-mag)}                   $\vert$ & 
\colhead{(hr)}                                 $\vert$ & 
\colhead{(/cl/yr)}         
}
\mn
\startdata
%
 1.0 & 7.266e3 &  0.87 &  1.92  &  +4.03 & +4.44 & +3.13 & +2.61 & 57.71 & 57.74 & 58.07  & 0.17  & 8$\times$10$^5$   \\ 
 1.5 & 1.065e4 &  0.95 &  10.5  &  +2.18 & +1.45 & +0.42 &--0.06 & 52.87 & 53.18 & 53.55  & 0.18  & 1.1$\times$10$^4$ \\ 
 2.0 & 1.367e4 &  1.03 &  32.9  &  +0.95 & +0.30 &--0.59 &--1.06 & 50.49 & 50.93 & 51.31  & 0.20  & 1.5$\times$10$^3$ \\ 
 3.0 & 1.899e4 &  1.12 &  146.  & --0.67 &--0.51 &--1.26 &--1.72 & 48.06 & 48.64 & 49.03  & 0.22  &  182.  \\ 
 5.0 & 2.805e4 &  1.23 &  846.  & --2.58 &--0.70 &--1.35 &--1.80 & 45.96 & 46.65 & 47.04  & 0.24  &  29.1  \\ 
  10 & 4.508e4 &  1.40 & 7.28e3 & --4.91 &--0.22 &--0.79 &--1.23 & 44.10 & 44.88 & 45.27  & 0.27  &  5.70  \\ 
  15 & 5.789e4 &  1.51 & 2.32e4 & --6.17 & +0.23 &--0.30 &--0.75 & 43.30 & 44.10 & 44.50  & 0.29  &  2.78  \\ 
  20 & 6.754e4 &  1.65 & 5.11e4 & --7.03 & +0.56 & +0.04 &--0.40 & 42.77 & 43.59 & 43.99  & 0.32  &  1.74  \\ 
  30 & 7.737e4 &  2.12 & 1.45e5 & --8.16 & +0.88 & +0.36 &--0.08 & 41.95 & 42.78 & 43.17  & 0.41? &  0.82? \\ 
  50 & 8.713e4 &  2.86 & 4.25e5 & --9.33 & +1.17 & +0.66 & +0.22 & 41.08 & 41.91 & 42.31  & 0.55* &  0.37* \\ 
 100 & 9.796e4 &  4.12 & 1.40e6 &--10.63 & +1.47 & +0.96 & +0.52 & 40.08 & 40.91 & 41.31  & 0.80* &  0.15* \\ 
 300 & 1.074e5 &  7.41 & 6.56e6 &--12.30 & +1.71 & +1.21 & +0.77 & 38.64 & 39.48 & 39.88  & 1.43* & 0.039* \\ 
1000 & 1.080e5 &  12.9 & 2.02e7 &--13.52 & +1.72 & +1.22 & +0.78 & 37.44 & 38.28 & 38.68  & 2.48* & 0.013* \\ 
\enddata
\mn
\tablenotetext{a}{Stellar mass in \Mo. The physical parameters listed are for
Pop III ZAMS stars in Table~\ref{tab:tab1}, as modeled with the \texttt{MESA}
code for zero initial metallicity, zero mass loss, no rotation, and no stellar
duplicity.}

\tablenotetext{b}{Pop III star photospheric temperature \Teff\ in K.}

\tablenotetext{c}{Pop III star radius \Reff\ at \Teff\ in \Ro.}

\tablenotetext{d}{Pop III star bolometric luminosity \Lbol\ in \Lo.}

\tablenotetext{e}{Pop III star bolometric absolute magnitude \Mbol in AB-mag.}

\tablenotetext{f}{Combined bolometric + IGM + K-correction to Pop III star
\Lbol\ at z=7, z=12, and z=17, respectively, from \S \ref{sec33}.} 

\tablenotetext{g}{Pop III star apparent restframe-UV AB-magnitudes at z=7,
z=12, and z=17 in 2016 Planck cosmology \citep{planck_XIII_2016_a}, using the
NIRCam filters that sample the restframe UV 1500 \AA, assuming K-corrections
as in Cols. 6--8 and no dust (for a discussion of dust, see \S \ref{sec61}).
Distance moduli used are DM = 49.24, 50.58, and 51.42 mag at z=7, z=12, and
z=17, respectively.} 

\tablenotetext{h}{Upper limits to caustic transit rise-time $t_{rise}$ (in
hours) as estimated in \S \ref{sec44}. Asterisks indicate Pop III star ZAMS
masses M\cge 50 \Mo, for which cluster caustic transit events are possibly
observable to the detection limits of JWST medium-deep to deep survey epochs
reaching AB\cle 28.5--29 mag, assuming that caustic transit magnifications of
$\mu$$\simeq$10$^{3}$--10$^{5}$ can elevate Pop III stars with AB\cle 35--41.5
mag temporarily above these JWST detection limits. Stars with M\cle 30 \Mo\
(labeled ``?'') likely remain undetectable even through caustic transits.} 

\tablenotetext{i}{The cluster caustic transit rate for Pop III stars (number of
events per cluster per year) at z=12 as estimated in \S \ref{sec44} by directly
using Eq.~\ref{eq:dNlensdt}. Appendices C--D summarize the uncertainties
relevant to caustic transits.}
\end{deluxetable*}



\begin{deluxetable*}{| c | cccc | ccc | ccc | c | c |}
\tablecolumns{13}
\tablewidth{1.0\linewidth}
\tablecaption{Implied Red Giant Branch Pop III Star Observational Parameters Relevant to Caustic Transit Calculations
\label{tab:tab3}}
\tablehead{
\colhead{Mass$^a$}                                $\vert$ &
\colhead{\Teff $^b$}                                      & 
\colhead{Radius $^c$}                                     & 
\colhead{\Lbol\ $^d$}                                     &                      
\colhead{M$_{\rm bol}$$^e$}                       $\vert$ & 
\multicolumn{3}{c}{Bolo+IGM+K-corr$^f$}           $\vert$ & 
\multicolumn{3}{c}{Giant Branch m$_{\rm UV}$$^g$} $\vert$ & 
\colhead{t$_{rise}$$^h$}                          $\vert$ & 
\colhead{transit$^i$}                                     \\[-4pt] 
\colhead{GB}                                      $\vert$ &
\multicolumn{4}{c}{--- at Hydrogen-depletion ---} $\vert$ & 
\colhead{z=7}                                             & 
\colhead{z=12}                                            & 
\colhead{z=17}                                    $\vert$ & 
\colhead{z=7}                                             & 
\colhead{z=12}                                            & 
\colhead{z=17}                                    $\vert$ & 
\colhead{caust}                                   $\vert$ & 
\colhead{rate}                                            \\[-4pt]
\colhead{($M_{\odot}$)}                           $\vert$ &
\colhead{(K)}                                             & 
\colhead{($R_{\odot}$)}                                   & 
\colhead{($L_{\odot}$)}                                   & 
\colhead{(AB)}                                    $\vert$ & 
\multicolumn{3}{c}{(AB-mag)}                      $\vert$ & 
\multicolumn{3}{c}{(AB-mag)}                      $\vert$ & 
\colhead{(hr)}                                    $\vert$ & 
\colhead{(/cl/yr)}         
}
\mn
\startdata
%
 1.0 & 6.999e3 &  3.25 &  22.8  &  +1.35 & +4.83 & +3.48 & +2.96 & 55.42 & 55.41 & 55.73 &  0.63  & 9$\times$10$^4$  \\
 1.5 & 1.181e4 &  2.13 &  79.7  & --0.01 & +0.91 &--0.06 &--0.53 & 50.13 & 50.51 & 50.88 &  0.41  &1.0$\times$10$^3$ \\
 2.0 & 1.611e4 &  1.78 &  191.  & --0.96 &--0.19 &--1.01 &--1.47 & 48.08 & 48.60 & 48.99 &  0.34  &  175.  \\
 3.0 & 2.311e4 &  1.53 &  598.  & --2.20 &--0.69 &--1.39 &--1.84 & 46.35 & 46.99 & 47.38 &  0.30  &  39.8  \\
 5.0 & 3.206e4 &  1.55 & 2.28e3 & --3.66 &--0.63 &--1.25 &--1.70 & 44.95 & 45.67 & 46.07 &  0.30  &  11.8  \\
  10 & 4.174e4 &  2.40 & 1.57e4 & --5.75 &--0.34 &--0.92 &--1.36 & 43.15 & 43.91 & 44.31 &  0.46  &  2.33  \\
  15 & 4.624e4 &  3.47 & 4.94e4 & --6.99 &--0.18 &--0.74 &--1.19 & 42.06 & 42.84 & 43.24 &  0.67? &  0.87? \\
  20 & 4.864e4 &  4.58 & 1.06e5 & --7.82 &--0.10 &--0.65 &--1.09 & 41.32 & 42.11 & 42.51 &  0.88* &  0.44* \\
  30 & 5.180e4 &  6.49 & 2.72e5 & --8.85 & +0.02 &--0.53 &--0.97 & 40.41 & 41.20 & 41.60 &  1.25* &  0.19* \\
  50 & 5.490e4 &  9.38 & 7.18e5 & --9.90 & +0.13 &--0.42 &--0.86 & 39.47 & 40.26 & 40.66 &  1.81* & 0.081* \\
 100 & 5.173e4 &  18.2 & 2.14e6 &--11.09 & +0.02 &--0.53 &--0.98 & 38.17 & 38.96 & 39.36 &  3.52* & 0.024* \\
 300 & 4.882e4 &  40.8 & 8.51e6 &--12.59 &--0.09 &--0.65 &--1.09 & 36.57 & 37.35 & 37.75 &  7.88* & 0.006* \\
1000 & 4.807e4 &  74.8 & 2.68e7 &--13.83 &--0.12 &--0.67 &--1.12 & 35.29 & 36.07 & 36.47 & 14.44* & 0.002* \\
\enddata
\mn
\tablenotetext{a}{Footnotes $^{a-i}$ are as in Table~\ref{tab:tab2}. All 
parameters in this Table are for Pop III stars at hydrogen-depletion (RGB).}
\end{deluxetable*}



\begin{deluxetable*}{| c | cccc | ccc | ccc | c | c |}
\tablecolumns{13}
\tablewidth{1.0\linewidth}
\tablecaption{Implied AGB Pop III Star Observational Parameters Relevant to Caustic Transit Calculations
\label{tab:tab4}}
\tablehead{
\colhead{Mass$^a$}                              $\vert$ &
\colhead{\Teff $^b$}                                    & 
\colhead{Radius $^c$}                                   & 
\colhead{\Lbol\ $^d$}                                   &                      
\colhead{M$_{\rm bol}$$^e$}                     $\vert$ & 
\multicolumn{3}{c}{Bolo+IGM+K-corr$^f$}         $\vert$ & 
\multicolumn{3}{c}{AGB m$_{\rm UV}$$^g$}        $\vert$ & 
\colhead{t$_{rise}$$^h$}                        $\vert$ & 
\colhead{transit$^i$}                                   \\[-4pt] 
\colhead{AGB}                                   $\vert$ &
\multicolumn{4}{c}{--- at Helium-depletion ---} $\vert$ & 
\colhead{z=7}                                           & 
\colhead{z=12}                                          & 
\colhead{z=17}                                  $\vert$ & 
\colhead{z=7}                                           & 
\colhead{z=12}                                          & 
\colhead{z=17}                                  $\vert$ & 
\colhead{caust}                                 $\vert$ & 
\colhead{rate}                                          \\[-4pt]
\colhead{($M_{\odot}$)}                         $\vert$ &
\colhead{(K)}                                           & 
\colhead{($R_{\odot}$)}                                 & 
\colhead{($L_{\odot}$)}                                 & 
\colhead{(AB)}                                  $\vert$ & 
\multicolumn{3}{c}{(AB-mag)}                    $\vert$ & 
\multicolumn{3}{c}{(AB-mag)}                    $\vert$ & 
\colhead{(hr)}                                  $\vert$ & 
\colhead{(/cl/yr)}         
}
\mn
\startdata
%
 1.0&6.312e3$^j$&5.23$^j$&39.8$^j$&+0.74& +6.01 & +4.57 & +4.03 & 55.99 & 55.89 & 56.19 &  1.01  &1.4$\times$10$^5$ \\
 1.5 & 8.149e3 & 6.18 &  151.  & --0.71 & +3.36 & +2.14 & +1.64 & 51.89 & 52.01 & 52.35 &  1.19  &4.0$\times$10$^3$ \\
 2.0 & 1.145e4 & 4.66 &  335.  & --1.57 & +1.06 & +0.07 &--0.40 & 48.73 & 49.08 & 49.45 &  0.90  &  273.  \\
 3.0 & 1.736e4 & 3.56 & 1.03e3 & --2.79 &--0.36 &--1.15 &--1.60 & 46.09 & 46.64 & 47.03 &  0.69  &  28.9  \\
 5.0 & 2.658e4 & 2.89 & 3.74e3 & --4.19 &--0.72 &--1.38 &--1.82 & 44.33 & 45.01 & 45.41 &  0.56  &  6.43  \\
  10 & 3.938e4 & 3.03 & 1.98e4 & --6.00 &--0.42 &--1.00 &--1.45 & 42.82 & 43.57 & 43.97 &  0.58  &  1.71  \\
  15 & 4.215e4 & 4.55 & 5.88e4 & --7.18 &--0.33 &--0.90 &--1.34 & 41.73 & 42.50 & 42.89 &  0.88? &  0.64? \\
  20 & 4.386e4 & 6.14 & 1.25e5 & --8.00 &--0.27 &--0.84 &--1.28 & 40.97 & 41.74 & 42.14 &  1.19* &  0.32* \\
  30 & 4.006e4 & 11.7 & 3.17e5 & --9.01 &--0.40 &--0.98 &--1.42 & 39.83 & 40.59 & 40.98 &  2.26* &  0.11* \\
  50 & 3.536e4 & 24.3 & 8.32e5 &--10.06 &--0.55 &--1.15 &--1.59 & 38.63 & 39.37 & 39.77 &  4.70* & 0.036* \\
 100 & 3.392e4 & 44.0 & 2.31e6 &--11.17 &--0.59 &--1.19 &--1.64 & 37.49 & 38.22 & 38.61 &  8.50* & 0.012* \\
 300 & 3.165e4 & 101. & 9.19e6 &--12.67 &--0.64 &--1.26 &--1.71 & 35.93 & 36.65 & 37.04 & 19.49* & 0.003* \\
1000 & 3.122e4 & 163. & 2.26e7 &--13.65 &--0.65 &--1.28 &--1.72 & 34.94 & 35.66 & 36.05 & 31.45* & 0.001* \\
\enddata
\mn
\tablenotetext{a}{Footnotes $^{a-i}$ are as in Table~\ref{tab:tab2}. All 
parameters in this Table are for Pop III stars at helium-depletion (AGB).}

\tablenotetext{j}{The 1.0 \Mo, Z = 0.00 \Zo\ \texttt{MESA} model did not ignite
helium and may thus become a helium white dwarf. To complete our calculations,
its AGB parameters listed are for the Z = 10$^{-8}$ $Z_{\odot}$ model, which
did end in a white dwarf.}
\end{deluxetable*}


For the most massive Pop III stars, their MS lifetime $\tau$ scales roughly as
mass/luminosity. Since for the highest masses (M\cge 100 \Mo), luminosities are
directly proportional to their ZAMS mass (see Eq.~\ref{eq:homologyrelations}),
the \texttt{MESA} models yield MS ages of 5.6--2.1 Myr that are only weakly
dependent on ZAMS mass for the mass range of 30--1000 \Mo, respectively (see
Table~\ref{tab:tab1}). The shortest MS lifetime possible is $\sim$2.07 Myr, which
happens when the star is radiating at the Eddington luminosity, and so its age
becomes nearly independent of mass and only a function of fundamental constants.
The \texttt{MESA} models in Table~\ref{tab:tab1} indeed approach this MS
age-limit for M$\simeq$300--1000 \Mo\ to within the modeling uncertainties.

\si In summary, Tables~\ref{tab:tab1}--\ref{tab:tab4} show that Pop III stars
in the mass range of M$\simeq$30--1000 \Mo\ have ZAMS photospheric temperatures
of 77,000--108,000 K, bolometric luminosities of
\Lbol$\simeq$10$^{5.2}$--10$^{7.3}$ \Lo, stellar radii of \RMS$\simeq$2--13
\Ro, and main sequence (MS) lifetimes of \tauMS$\simeq$2.1--5.6 Myr. They
may be therefore be bright enough for occasional caustic transit detections by
JWST, which is summarized in Col. 9--13 of
Tables~\ref{tab:tab2}--\ref{tab:tab4}, as calculated in \S \ref{sec44}. 

As discussed in \S \ref{sec23} \& \ref{sec44}, we only use {\it upper limits}
to the integrated 1--4 \mum\ sky-SB to estimate the maximum Pop III object
caustic transit rate. Hence, the actual Pop III star lifetimes do not directly
enter these calculations. However, plausible differences in Pop III star 
GB lifetimes as a function of ZAMS mass are relevant when estimating limits to
the caustic transit rates from Pop III stellar-mass BH accretion disks 
compared to those of Pop III stars. This is discussed in \S
\ref{sec5}--\ref{sec6}, where we consider the case that Pop III BH-feeding may
be done from lower-mass companion stars as soon as these can form in slightly
polluted environments. 

\mn \subsection{Multiplicity of Massive Stars}
\label{sec32} 

\sn The effect of binaries and stars of higher multiplicities is a complex
subject, that can have a significant effect on population synthesis models of
galaxies \citep[\eg][]{zhang_2010, stanway_2016, conroy_2013}. For our current
purpose, we must address the fact that the multiplicity factor $MF$ is nearly
unity for O-stars, at least in the local universe \citep[\eg][]{duchene_2013}:

\begin{equation}
MF = \frac{(B + T + Q)}{(S + B + T + Q)} \simeq 1.
\end{equation}

\n Here S is the number of single stars in a coeval stellar population, B the
number of binary stars, T the number of triples, and Q the number of quads,
etc., implying that one gets essentially a factor \cge 2 increase in the
luminosity from binary, triple, and quad stars together. 

\sn \subsubsection{Multiplicity --- Low-mass end}
\label{sec321} 

\sn At least 30\% of all lower mass stars in our own Galaxy occur in binaries
\citep{sana_2012, kiminki_2012, duchene_2013, mayer_2017}, but at z\cge 7 this
fraction is unknown. The exact ratio of Pop III stars to BH accretion disks
that are present will depend on the Pop III IMF slope, which is also unknown
\citep[\eg][]{greif_2011, guszejnov_2016, ishiyama_2016, susa_2014}. We
consider possible effects from the IMF slope in \S \ref{sec34}, \ref{sec44}, \&
\ref{sec5}. 

Table~\ref{tab:tab1} shows that the pre-MS lifetimes of Pop III stars with M
\cle 1.5--2 \Mo\ would be \cge 3--6 Myr, and thus generally exceed the \cle 4
Myr He-depletion age of 50--1000 \Mo\ stars. Hence, for {\it coeval} stellar
populations with a large number of stars and a sufficiently flat mass function
(\ie\ $dN/dM\propto M^{-\alpha}$ with $\alpha$$\simeq$2), a significant number
of \cge 50 \Mo\ Pop III stars may be present that will have polluted the
surrounding ISM with their AGB mass loss --- and supernovae in the right mass
range --- {\it before} stars with M\cle 1.5--2 \Mo\ can have finished forming
via their Hayashi tracks. Hence, most early low-mass stars with M \cle 1.5--2
\Mo\ may already have been polluted by coeval or precursor massive Pop III
stars (M\cge 50 \Mo), unless these low mass stars formed in very isolated
environments well away from the massive Pop III stars. The formation of
lower-mass Pop III stars may also have been prevented by the very strong UV 
radiation field of nearby more massive Pop III stars, as discussed in \S
\ref{sec35}. 

The very first stars likely did not form until z\cle 35 (age \cle 79 Myr), or
they would have left small-scale imprints in the Cosmic Microwave Background
(CMB). Even if \cle 1.5--2 \Mo\ Pop III stars had formed as early as
z$\simeq$35, Table~\ref{tab:tab1} shows that their MS ages will be \cge 640
Myr, so these low mass Pop III stars would not reach the giant branch until
well below z\cle 7.3 (cosmic age 720 Myr) when reionization has essentially
completed (\S \ref{sec1}). Therefore, Pop III stars with M\cle 1.5--2 \Mo\ ---
if they did manage to form as part of binary systems --- could not fill their
Roche lobes at z\cge 7, and would not be relevant to Pop III BH accretion disk
feeding at z\cge 7 if stellar binaries were their progenitors. In \S
\ref{sec5}, we will therefore not consider Pop III stars with masses M\cle 2
\Mo. With \Teff$\simeq$10$^4$ K (Table~\ref{tab:tab1}), low-mass Pop III stars
could, however, contribute some reionizing flux. Their fractional contribution
to the UV-luminosity density depends on the value of the Pop III IMF slope
(see \S \ref{sec34}). 

\sn \subsubsection{Multiplicity --- High-mass end}
\label{sec322} 

\sn Under the assumption that (slightly polluted) massive stars at z\cge 7 may
occur in binary or multiple systems, then for a \citet{salpeter_1955}-slope or
flatter IMF stars with M\cge 30 \Mo\ may have a lower mass companion with M\cle
30 \Mo. The last Column of Table~\ref{tab:tab1} shows that these lower mass
companion stars with M\cge 2 \Mo\ will be in their RGB--AGB stage for \tauGB\
\cle 30-60 Myr, \ie\ generally much longer than the plausible ages of a massive
Pop III star in the binary. They could thus be feeding the BH that was leftover
from the more massive Pop III star after 2.4--6 Myr. As long as the more
massive star --- during its short GB lifetime (\tauGB) --- does not transfer
the majority of its mass to the less massive companion star, the resulting
accretion timescale around the BHs leftover form M\cge 30 \Mo\ stars (\S
\ref{sec53}) would be mainly driven by the much longer \tauGB\ of the less
massive star, when it leaves the MS and fills its Roche lobe. 

A reasonable upper limit for the BH-feeding timescale by a lower-mass star in a
binary filling its Roche lobe is thus the \cle 60 Myr GB-age of a M\cge 2 \Mo\
star. Table~\ref{tab:tab1} suggests that the {\it lower} limit on the
timescale of BH accretion disk feeding from the more massive companion stars is
\cle 0.3 Myr. In \S \ref{sec53}, we will therefore assume a lifetime range for
BH accretion disk-feeding of 0.3--60 Myr. That is, Pop III BH accretion may
last up to \cle 10$\times$ longer than that of the Pop III stars themselves,
but it could also be $\sim$10$\times$ shorter. The Spitzer-Chandra
power-spectrum results may already hint at a BH contribution, as discussed in
\S \ref{sec232}. The \texttt{MESA} stellar evolution physics of zero
metallicity stars summarized in Table~\ref{tab:tab1} thus provides a
theoretical frame-work that allows massive Pop III stars to leave stellar-mass
BHs with accretion disks that may feed up to $\sim$10$\times$ longer than these
massive Pop III stars live themselves. We will discuss the implications of this
in \S \ref{sec5}-\ref{sec6}. 

\sn \subsubsection{Massive Star Multiplicity at Low Redshifts}
\label{sec323} 

O-stars in nearby surveys show significant multiplicity (\cge 80\%), and have a
rather flat mass-ratio distribution:

\begin{equation}
q \equiv M_{sec} / M_{pri}, 
\end{equation}

\n where $M_{pri}$ is the more massive star \citep{duchene_2013}. In theory,
the $q$-value distribution can be as steep as the \citet{salpeter_1955} mass 
function slope, \ie: 

\begin{equation}
N(q) \propto q^{-2.35}.
\label{eq:Nq} 
\end{equation}

\n The observed $q$-distribution of nearby O-stars seems to have a slope much
flatter than the IMF slope \citep{duchene_2013}. Since nearby surveys of
double/multiple stars may suffer from flux-bias, very faint low-mass stellar
companions around more massive stars are harder to find. In \S 
\ref{sec5}-\ref{sec6}, we will therefore assume that slightly-polluted early
massive stars have a mass-ratio no steeper than the IMF-slope, if they already
occur in binaries. 

Fig.~2 of \citet{duchene_2013} suggests that the majority of nearby binary
OB-stars have typical orbital periods in the range of $\sim$10--130 days and
typical orbital separations between 0.067--0.51 AU or 14\cle D\cle 110 \Ro. 
Larger and smaller separations can occur as well, some as small as a few \Ro.
Each of their Roche lobes will be about half that in effective radius, or 7\cle
R\cle 55\ \Ro. Following the \texttt{MESA} models of
Table~\ref{tab:tab1}--\ref{tab:tab4}, it is therefore possible that {\it if}
Pop III or II.5 stars exist in binaries at z\cge 7 like in OB-binary stars
today, their lower-mass companion stars with M\cge 2 \Mo\ will fill their Roche
lobes in the GB stage at z\cge 7 with sizes 2\cle R\cle 160 \Ro\ (see Col. 3 of
Table~\ref{tab:tab3}--\ref{tab:tab4}). This would feed the BH remnant from the
more massive Pop III star during their AGB--MS life time, which could last up
to \cle 60 Myr (\S \ref{sec321}). Their BH UV-accretion disk half-light radii
(\rhl) are estimated in \S \ref{sec55} to be in the range of 1\cle \rhl\cle 30
\Ro, and so in general will fit inside these Roche lobes when the companion star
in the binary reaches the AGB stage.

Since we do not include Pop III star multiplicity in the \texttt{MESA} models,
this will render the Pop III caustic transit rates of \S \ref{sec44} more
conservative. This is because caustic transit detections may be \cge 0.75 mag
brighter for binary stars --- and possibly multi-peaked in their detailed
time-sequence --- than we estimate for single Pop III stars in \S \ref{sec44}.
One example of a multiple caustic transit event has been suggested for a
possible massive binary star at z$\simeq$1.5 \citep{kelly_2017_b}. Future work
will need to model the evolutionary tracks of zero or very-low metallicity
stars that includes the evolution of massive binaries with possible mass
exchange, and how this may affect the BH feeding timescales. 

\mn \subsection{Bolometric Corrections after IGM Transmission, and
K-corrections}
\label{sec33}

\sn The luminosities and absolute magnitudes of Pop III stars summarized in \S~
\ref{sec31} and Table~\ref{tab:tab1}--\ref{tab:tab4} were calculated by the
\texttt{MESA} code in {\it bolometric} solar units {\it without} making 
bolometric corrections, corrections for IGM transmission, or K-corrections. We
thus need to correct the theoretical Pop III star luminosities for these
effects to predict the \mAB-values observed in the JWST filters at a given
redshift. The exact bolometric and K-corrections cannot be computed until the
actual object redshifts have been estimated from the 8-band JWST NIRCam
photometry, and/or measured with JWST NIRSpec or NIRISS spectra. For our
current photometric predictions, we will therefore proceed as outlined below. 

\sn \subsubsection{Pop III Star Bolometric + IGM Corrections}
\label{sec331}

\sn We use zero metallicity blackbody spectra for Pop III stars with ZAMS, RGB,
and AGB $T_{eff}$-values and radii from Tables~\ref{tab:tab2}--\ref{tab:tab4} to
estimate the bolometric correction (BC) and their IGM corrections as follows. Pop
III stars of 30\cle M\cle 1000 \Mo\ with photospheric temperatures of
T=77,400--108,000 K (Table~\ref{tab:tab1}) have SEDs with restframe wavelength
peaks in $I_{\nu}$ around $\lambda_{max}$$\simeq$620--444 \AA, respectively. 

The IGM at z\cge 7 is opaque for restframe $\lambda$\cle 1216 \AA\ due to the
significant fraction of neutral hydrogen in the immediate foreground of each
First Light object \citep{haardt_2012}. Hence, for our caustic transit
calculations in \S \ref{sec44} \& \ref{sec62}, it only matters that the Pop
III star or its stellar-mass BH accretion disk is UV-bright down to the 1216 
\AA\ Ly$\alpha$-limit, since the opaque neutral hydrogen forest at z\cge 7 will
certainly block the shorter hard-UV wavelengths from both the stars and BH
accretion disks. Therefore, while many Pop III objects may have unreddened UV
SED $\beta$-slopes \citep{calzetti_1994} much steeper than those corresponding
to T$\simeq$30,000 K, in what follows we will only consider their luminosities
and fluxes in the restframe UV-continuum at 1216--2000 \AA. At z$\simeq$7--17,
this is the only restframe wavelength range that JWST will sample to
AB$\simeq$28--30 mag over its most sensitive NIRCam wavelength range of 1--4
\mum. 

To predict their fluxes as observed in the JWST 1--5 \mum\ filters, we
integrated the assumed Pop III black body spectra in $I_{\nu}$, and computed the
fraction of flux longwards of $\lambda$=1216 \AA\ compared to the {\it total
bolometric flux} integrated from $\lambda$=0 to +$\infty$. The SEDs of Pop III
stars with M$\simeq$30--1000 \Mo\ as observed through the JWST NIRCam filters are
predicted to be about 9.5--23$\times$ fainter than their bolometric model
luminosities from Col. 5 of Table~\ref{tab:tab2}--\ref{tab:tab4}. After
accounting for the drop in IGM transmission to 0\% at z\cge 7, the actual BC
of Pop III stars would thus make them about 2.4--3.4 mag fainter in absolute
magnitude, and in the \mAB-magnitude to be observed with JWST. The average BC
that we need to apply is thus about +2.9 mag. This is much less than the formal
BC of $\sim$6 mag implied for a T$\simeq$10$^5$ K star \citep[][]{flower_1996}.
This is because the Pop III SED below \Lya 1216 \AA\ is completely blocked by
the IGM at z\cge 7, and so does not enter the BC. Note we derive the BC with the
opposite sign as defined in \citet{flower_1996}, since we need to go from
theoretical bolometric values to the predicted values to be observed with JWST. 

\sn \subsubsection{Pop III Star K-Corrections} 
\label{sec332}

\sn For each fixed JWST filter, we need to apply the K-correction to the Pop III
star flux at $\lambda$\cge 1216 \AA\ that makes it through the IGM.
\citet{hogg_1999} and \citet[][and references therein]{hogg_2002} define the
K-correction in $I_{\nu}$ units as following: 

\begin{equation}
K = -2.5\ log\ [ (1+z) L_{(1+z)\nu} / L_{\nu_e} ]\ \ (mag).
\label{eq:Kcorrection}
\end{equation}

\n This includes the effects of bandpass shifting due to the object's redshift,
and the change of the object's restframe SED with frequency or wavelength. The
factor of (1+z) accounts for the fact that the flux and luminosity are not
bolometric, but flux densities per unit frequency \citep{hogg_2002}. Due to
the complete IGM absorption for $\lambda$\cle 1216 \AA\ at z$\simeq$7--17, our
specific K-term needs to correct the flux predicted in the bluest available
JWST filter that is {\it completely longwards} of redshifted \Lya for these
effects. Hence, the luminosity ratio in Eq.~\ref{eq:Kcorrection} only needs to
account for the brighter flux shortward of this filter that still makes it
through the IGM. 

At z=7, the bluest available JWST filter that is above the Lyman-break (F115W
or short J-band) samples restframe $\lambda_{rest}$$\simeq$1440 \AA, while at
z=12 the F200W filter (short K-band) samples $\lambda_{rest}$$\simeq$1645\ \AA,
and at z=17 the F277W filter samples $\lambda_{rest}$$\simeq$1540 \AA. Since
the most massive Pop III ZAMS stars have nearly uniform temperatures of
T$\simeq$10$^5$ K (Table~\ref{tab:tab1}), all 1--5 \mum\ JWST filters sample
the Rayleigh-Jeans tail of their SEDs at z$\simeq$7--17, so that $I_{\nu}
\propto \nu^2$. The peak in their restframe ($I_{\nu}$) SEDs thus occurs only a
factor of 2--3.5 below the central restframe wavelength sampled by these
filters. We then compute this K-correction from the flux bluewards of each
filter down to the 1216 \AA\ IGM transmission cutoff. Using
Eq.~\ref{eq:Kcorrection}, the K-correction follows from these wavelengths
ratios as: 

\begin{equation}
K \simeq -2.5\ log\ [(1+z) . (1216/\lambda_c)^2]\ \ (mag). 
\label{eq:Kcorrectionb}
\end{equation}

\n Here, $\lambda_c$ is the central restframe wavelength of the bluest JWST
filter used for object detection at each redshift listed in
Tables~\ref{tab:tab2}--\ref{tab:tab4}. The wavelength ratio 
($\lambda_{Ly\alpha}$/$\lambda_c$)$^2$ reflects the fact that we can only
sample the $I_{\nu} \propto \nu^2$ tail of the very blue SEDs longwards of
Ly$\alpha$ at z\cge 7. For the extremely blue Pop III stars, this total K-term
amounts to about --1.9 mag brighter at z=7, --2.5 mag at z=12, and --2.9 mag
brighter at z=17. The K-correction gets somewhat brighter at the higher
redshifts, because the (1+z)-factor dominates the term from the wavelength or
frequency ratio. For all longer wavelength JWST filters, the K-corrections can
be computed similarly. 

\sn \subsubsection{The Combined Bolometric + IGM + K-corrections} 
\label{sec333}

\sn The combined bolometric+IGM+K-corrections (hereafter ``BIK''-corrections) 
were calculated in two independent ways. First, all integrations were done in
the {\it observed frame} while folding their black body SEDs with the JWST
NIRCam filters and integrating the bolometric+IGM-corrections as described in \S
\ref{sec331}, and the K-corrections as in \S \ref{sec332}. Second, the
calculations were done in the {\it restframe} after folding the black body SEDs
with the appropriately blueshifted JWST NIRCam filter curves, and integrating
between each filters' restframe FWHM wavelength cut-offs. The NIRCam
interference filters were designed to resemble block-functions
\citep{rieke_2005}, so this approximation is valid. Both methods gave similar
\mAB-fluxes results to within \cle 0.05--0.20 mag, with the second method
producing \mAB-fluxes that are on average only 0.09 mag brighter than the first
method. The \mAB-values listed in Tables~\ref{tab:tab2}--\ref{tab:tab4} are
therefore the slightly brighter values from the second method, since our
caustic transit calculations use in all cases upper limits in predicted fluxes. 

These combined BIK-corrections need to be added to the \MAB-values from
Tables~\ref{tab:tab2}--\ref{tab:tab4} to yield the predicted \mAB-values that
JWST would observe:

\begin{equation}
m_{AB}(z) = M_{AB} + DM(z) + (BC+IGM+K)(z),
\label{eq:DM}
\end{equation}

\n where the distance moduli DM at redshift $z$ are listed in the footnotes of
Table~\ref{tab:tab2}. To first order, the bolometric+IGM and K-corrections are
comparable in magnitude, but opposite in sign for the Pop III stars in
Table~\ref{tab:tab1}. The combined BIK-corrections are therefore in general
modest, but can on average result in making objects $\sim$1--2 mag fainter than
what the intrinsic bolometric luminosities from the \texttt{MESA} models would
yield. 

We list the combined corrections for z=7, z=12, and z=17 in Cols. 6--8 of
Tables~\ref{tab:tab2}--\ref{tab:tab4} for the ZAMS, RGB, and AGB, respectively.
Cols. 9--11 list the resulting apparent restframe UV AB-magnitudes for these
redshifts that result from the absolute bolometric magnitudes in Col. 5 and
these combined corrections, respectively. These are directly used in the
calculation of our caustic transit rise-times and rates in \S \ref{sec44},
which are listed in Col. 12--13, respectively. 

For the lowest mass stars that have \Teff\cle 10$^4$ K in
Tables~\ref{tab:tab2}--\ref{tab:tab4}, the combined BIK-corrections are
significantly positive, since for such cool stars most of their blackbody-like
SED is redshifted to well longwards of the bluest JWST filter that samples above
the 1216 \AA-break at the anticipated object redshift. These cool, low-mass Pop
III stars are thus predicted to be always very faint (AB\cge 50 mag), and
permanently out of reach for JWST caustic transits. At the intermediate
temperatures of Pop III RGB and AGB stars of $T_{eff}$$\simeq$50,000--30,000 K,
respectively (see Fig.~\ref{fig:fig2} and \S \ref{sec31}), the combined
BIK-corrections are in general negative but no brighter than --1 to --2 mag
(Tables~\ref{tab:tab3}--\ref{tab:tab4}), because the peak of their blackbody
SED falls between the \Lya\ 1216\AA\ IGM cutoff and the restframe
wavelength-range covered by the bluest NIRCam filter that JWST will use to
detect the Pop III object at z$\simeq$7--17, so that the full benefit of the
K-correction for a very blue SED is achieved. For much hotter ZAMS Pop III star
temperatures of $T_{eff}$$\simeq$10$^5$ K, the BIK-corrections are
generally positive but no dimmer than +1 to +2 mag (Table~\ref{tab:tab2}), 
because most of the energy in their blackbody SED now falls well below the \Lya\
1216\AA\ IGM cutoff. {\it Hence, the combined BIK-corrections are more
advantageous for detecting the cooler Pop III RGB and AGB phases than for the
hotter Pop III ZAMS stars.} This can be seen in Cols. 6--8 of
Tables~\ref{tab:tab2}--\ref{tab:tab4}, and will be folded into the caustic
transit rate calculations in \S \ref{sec44}.


\vspace*{-0.00cm}
\n\begin{figure}[!hptb]
\vspace*{-0.00cm}
\n\cl{
\includegraphics[width=0.4850\txw,angle=0]{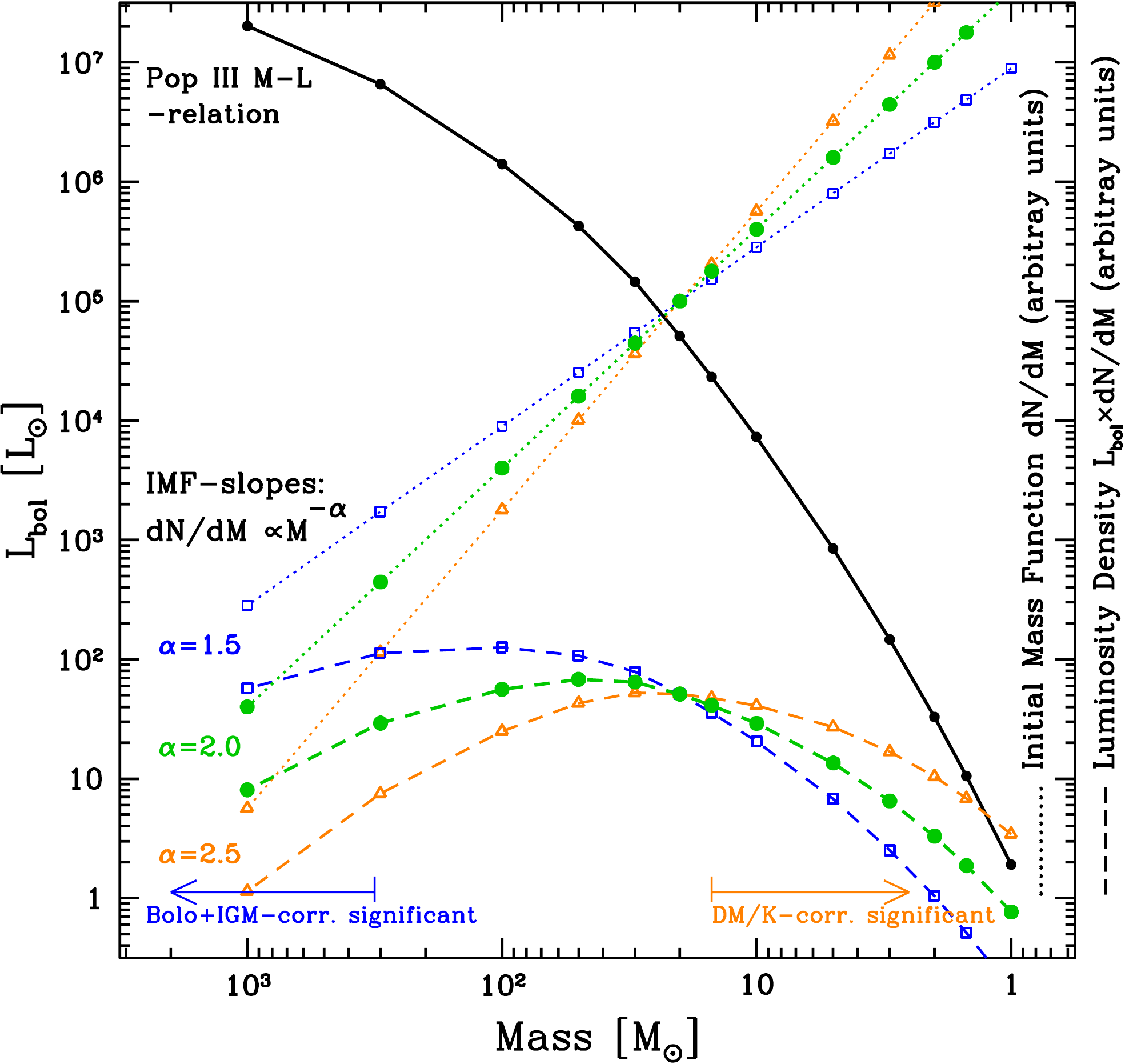}
}
\centering{}
\vspace*{-0.400cm}
\caption{
The luminosity density (dashed curves) for early star-forming objects inferred
from the ZAMS Pop III mass-luminosity relation (solid black line) from 
Table~\ref{tab:tab1} in \S \ref{sec31}. The ZAMS Pop III ML-relation is 
folded with three different IMF slopes (dotted lines), ranging from
$\alpha$=1.5 (top heavy; blue), $\alpha$=2.0 (normal; green), and $\alpha$=2.5
(steep IMF; orange). For a Pop III IMF slope of $\alpha$=2.0, the luminosity
density peaks around 30 \Mo, while most of the population's luminosity density
is produced between 10--100 \Mo. 
}\label{fig:fig3}
\end{figure}


\vspace*{-0.00cm}
\mn \subsection{Luminosity Density from Mass-Luminosity Relation and Initial 
Mass Function}
\label{sec34}

\sn The ZAMS Pop III mass-luminosity relation discussed in \S \ref{sec31} and
Table~\ref{tab:tab2} has important implications for the mass range that
dominates the luminosity density of a faint star-forming object at z\cge 7.
This is indicated in Fig.~\ref{fig:fig3}, where the ZAMS ML-relation from 
Table~\ref{tab:tab2} is indicated by the solid black line:

\begin{equation}
L_{bol} \propto M^{\lambda},
\label{eq:masslumlambda}
\end{equation}

\n with mass-dependent slope $\lambda$ from Table~\ref{tab:tab2}, as 
approximated by the segmented power-laws in Eq.~\ref{eq:masslumrelations}. 

In our caustic transit calculations of \S \ref{sec44}, we assume that small
early star-forming objects exist at z\cge 7. These will be mostly fainter than
the HST or JWST detection limits, and contribute a total 1--4 \mum\ sky-SB
whose upper limits were discussed in \S \ref{sec231}. In this context, it is
necessary to consider which stars will dominate the luminosity density of these
faint star-forming objects, which is defined as the number of stars per unit
area on the sky. We consider their IMF to be a power law:

\begin{equation}
dN/dM \propto M^{-\alpha}
\label{eq:IMF}
\end{equation}

\n with three different IMF slopes in Fig.~\ref{fig:fig3} (dotted curves),
ranging from ``top-heavy'' ($\alpha$=1.5; blue), ``intermediate''
($\alpha$=2.0; green), and ``steep'' ($\alpha$=2.5; orange) that bracket a
range of plausible IMFs \citep[dotted curves; \eg][]{bastian_2010,
coulter_2017, scalo_1986}. The ZAMS Pop III ML-relation is folded with these
three IMF slopes in Fig.~\ref{fig:fig3} to yield the luminosity density:

\begin{equation}
\hspace*{-0.20cm}
L_{bol}(dN/dM) \propto M^{\lambda} M^{-\alpha}\propto M^{\delta},
\label{eq:lumdensity}
\end{equation}

\n where $\delta$=($\lambda$--$\alpha$) is the slope of the luminosity-density
vs. mass relation. For an IMF slope $\alpha$$\simeq$2.0 and the mass-dependent
slope of the ML-relation in Eq.~\ref{eq:masslumrelations}, we infer a strongly
positive slope of the luminosity-density vs. mass relation (dashed green line): 
$\delta$$\simeq$+1.2 for M$\simeq$1--20 \Mo, a nearly zero slope
($\delta$$\simeq$0.1) for M$\simeq$20--100 \Mo, and a negative slope
($\delta$$\simeq$--0.8) for M$\simeq$ 100--1000 \Mo. 

For an IMF-slope of $\alpha$$\simeq$2.0, most of the Pop III ZAMS {\it
bolometric energy} from faint star-forming objects at z\cge 7 is thus produced
by stars with masses between 10--100 \Mo, with a somewhat smaller contribution
from stars with M$\simeq$100--1000 \Mo, and a much smaller contribution from
M$\simeq$1--10 \Mo, which is compounded by the significant K-correction for the
lowest mass stars (\S \ref{sec33}). For an IMF slope of $\alpha$$\simeq$2.0,
the Pop III luminosity density peaks around 30 \Mo\ with a broad plateau
(green dashed curve in Fig.~\ref{fig:fig3}). These are the Pop III stars with
the most advantageous bolometric+IGM+K-correction values
(Tables~\ref{tab:tab2}--\ref{tab:tab4}), and are within reach for JWST,
assuming caustic transits can occur as described in \S \ref{sec44}. 

For a top-heavy IMF, most of the luminosity density would be produced by stars
with M$\simeq$100 \Mo\ (blue dashed lines in Fig.~\ref{fig:fig3}), while for a 
very steep Salpeter-like IMF, most energy is still produced by stars as massive
as M$\simeq$20 \Mo\ (orange dashed lines). Hence, irrespective of any
reasonable IMF-slope at z\cge 7, the Pop III star mass-luminosity relation
implies that the highest near-IR sky-SB will be produced by stars with 20\cle
M\cle 100 \Mo. These are precisely the stars that are most likely to become
visible during caustic transits, as discussed in \S \ref{sec44}. 

\mn \subsection{Estimating the Surface Brightness from Massive Pop III Stars}
\label{sec35} 

\sn \citet{mas-ribas_2016} give the number of Lyman-Werner (LW) photons
produced by ZAMS Pop III stars as a function of their mass. A 300 \Mo\ Pop III
star emits $\dot{N}=3.1\times 10^{49}$ photons s$^{-1}$ in the LW-band, which
spans the energy range of E=11.2--13.6 eV. From these numbers, we find a flux
of $m_{\rm LW}$$\simeq$38.6 mag at z$\simeq$12. The ZAMS Pop III stars from the
\texttt{MESA} runs in Table~\ref{tab:tab1}--\ref{tab:tab2} are about 0.3 mag
brighter in their {\it bolometric} \MAB-magnitude. 

We estimate the surface density of Pop III stars using the models of
\citet{sarmento_2018} in which the Pop III star-formation rate density
reaches $\sim$$10^{-3.5}$ \Mo\ yr$^{-1}$ Mpc$^{-3}$ at z=12 (see their Fig.~2).
If we assume that these consist of M=300 \Mo\ stars that each live $t\simeq
2\times 10^6$ yr \citep[][see Table~\ref{tab:tab1} here]{schaerer_2002}, then
the total number of Pop III stars per \arcsecsq\ is:

\begin{equation} 
\approx 0.03 \left(\frac{{\rm SFR}_{\rm Pop III}}{10^{-3.5}\hspace{1mm}M_{\odot}
/{\rm yr}}\right)
\left(\frac{M_*}{300\ M_{\odot}}\right)^{-1}
\frac{{\rm stars}}{{\rm ('')}^{2}\Delta z}.
\end{equation}

\n If each of these stars has $m_{\rm AB}$\cge 38.6 mag (see Table
\ref{tab:tab2}), then the total surface brightness\deleted{\footnote{{\it
Footnote:}\ The contribution to the surface brightness (in erg s$^{-1}$
Hz$^{-1}$ cm$^{-2}$ sr$^{-1}$) at some wavelength $\lambda$ by stars within the
redshift range z$\pm$dz/2 equals:
\begin{equation}
\frac{dS_{\nu}(\lambda)}{dz}dz=\frac{c \epsilon^{\rm prop}_{\rm
Pop III}(z,\lambda)}{4\pi H(z)(1+z)^4}dz=\frac{c \epsilon^{\rm com}_{\rm Pop
III}(z,\lambda)}{4\pi H(z)(1+z)}dz,
\end{equation}
where $c$ denotes the speed of light, $H(z)$ the Hubble parameter at redshift
z, $\epsilon^{\rm prop}_{\rm Pop III}$ the proper volume emissivity density
(total luminosity density from all Pop III stars within a proper volume) at
wavelength $\lambda/(1+z)$, and $\epsilon^{\rm com}_{\rm Pop III}$ the comoving
volume emissivity.}} in Pop III stars could be fainter than $\sim$36.0 \magarc. 

Theoretically, it appears unlikely that Pop III stars {\it alone} can fully
account for an IR background with SB$\simeq$31 \magarc. In order to reach
SB$\simeq$31 \magarc, we need $\sim$10$^3$ massive Pop III stars per
\arcsecsq, each with AB$\simeq$38.5 mag. Most of these must be at lower
redshift (z \cle 12), because of the strong redshift-dependence of $dS/dz$, as
discussed in \S \ref{sec231}. This would require \cge 100 massive Pop III
stars per \arcsecsq\ at z$\simeq$12, with a weak z-dependence. This is much
larger than the numbers we calculate above, although close to the ``no LW''
case of \citet{trenti_2009}. 

Strong LW radiation from massive Pop III stars can significantly suppress the
formation of subsequent lower-mass stars in their immediate environment,
resulting in a possibly much dimmer total Pop III sky-SB of $\sim$36 \magarc.
This SB-level is also indicated in orange in Fig.~\ref{fig:fig1}. In \S 
\ref{sec45} and \ref{sec7}--\ref{sec8}, we will therefore consider a {\it
range} of Pop III near-IR SB-levels and the scope of observing programs needed
for JWST to detect their resulting caustic transits in each case. 

\bn \section{Estimates of Cluster Caustic Transits for Pop III stars}
\label{sec4}

\sn The question that we address in this section is: under what conditions could
JWST detect the individually lensed Pop III stars of \S \ref{sec3} at very high
magnification, identified as a sudden onset of an AB$\simeq$28.5 mag point
source, corresponding to a caustic crossing in which two additional critical
images\footnote{The critical images are two formally infinite magnification
images of a point source, which form on the critical curve for a source at the
location of the caustic line.} are formed? 

\mn \subsection{Selection of Lensing Clusters for JWST Caustic Transit
Observations of Pop III Objects} 
\label{sec41}

\sn To address the caustic transit rate and duration for Pop III stars, we 
first need to evaluate the plausible limits to the transverse velocities of
massive lensing clusters, their typical caustic lengths, and the possible
effects from microlensing. From this, we will in \S \ref{sec44} estimate the
cluster caustic transit rates for the Pop III stars of \S \ref{sec3}. 

A Pop III caustic-transit observing program with JWST would likely select the
best lensing clusters that also have matching deep HST images in previous
epochs --- including WFC3 IR data --- such as the Hubble Frontier Field
clusters \citep[HFF; \eg][]{lotz_2017, kawamata_2016, lagattuta_2017,
acebron_2017, mahler_2018} or the CLASH clusters \citep[\eg][]{postman_2012,
rydberg_2015}. 

The HFF clusters were chosen to have the capability for significant lensing
magnification. Many are highly elongated, and could well have significant
internal velocities between cluster sub-components, and/or a significant space
velocity compared to the nearby cosmic web, as discussed in \S \ref{sec42} and
Appendix A. Indeed, in two of these clusters (MACS J0416-2403 and MACS
J1149.5+2223) to date, possible caustic transits have been identified at lower
redshifts \citep[z$\simeq$1.0--1.5;][]{kelly_2017_b, diego_2017, rodney_2017}. 

A JWST lensing cluster program should select the best lensing clusters with
redshifts 0.3\cle z\cle 0.5. This is because a combination of the following
two factors. First, the SED of the 5--8 Gyr old stellar population in these
clusters at z$\simeq$0.4 peaks at $\lambda$$\simeq$1.6 \mum\ in the restframe
\citep[\eg][]{kim_2017}. This includes the SED of the substantial ICL that is
present in massive virialized clusters. Also, the Zodiacal foreground in JWST's
second Lagrange-point (L2) orbit strongly declines between 1--3.5 \mum\ (see
Fig.~\ref{fig:fig1}), so the best JWST sensitivity per unit time is obtained in
the observed wavelength range of 2--3.5 \mum. This is {\it the} critical
wavelength range for detecting First Light objects at z\cle 17. Hence, ideally
the redshift of lensing clusters to be observed with JWST should be kept at
z\cle 0.5. This is so that the restframe peak SED of the cluster galaxies and
the ICL does not redshift as much into the most sensitive 2.5--3.5 \mum\ NIRCam
filters, and thus compromise the ability to make First Light object detections,
including Pop III caustic transits. 

Higher redshift clusters will of course have lower ICL and cluster galaxy 
brightness, because of the stronger cosmological (1+z)$^4$ SB-dimming. They
are also less massive by selection, and may therefore not always be the most
optimal gravitational lenses. Because of their younger ages, they may also be
less virialized. There exist exceptional clusters at z\cge 0.5, of course,
that could be used for lensing studies with JWST, such as, \eg\ El Gordo at
z$\simeq$0.87 \citep{zitrin_2013}. 

In \S \ref{sec45}, we suggest that 3--30 clusters need to be monitored by JWST
for several years for Pop III caustic transit studies. In the end, practical
arguments, such as available HST images --- especially at shorter wavelengths
($\lambda$\cle 0.6 \mum) than those that JWST can observe --- the quality of
available lensing models, ancillary data such as available ground-based
spectra and X-ray images, and the ability to schedule JWST observations for at
least half a year during each JWST Cycle will likely determine which cluster
sample is best suited for Pop III caustic transit observations with JWST. 

\mn \subsection{Maximum Plausible Transverse Velocity of Lensing Clusters}
\label{sec42}

\sn In this section we consider the possible maximum transverse (or tangential)
velocity, \vT, of a massive cluster, which has visible substructure in both
its measured redshift/velocity-distribution, as well as in its spatial extent
on the sky (see Fig.~\ref{fig:fig4} and Appendix A). 

\sn \subsubsection{General Limits to Transverse Velocities of Nearby Clusters}
\label{sec421}

\sn First, we will summarize what typical space velocities are seen for clusters
locally, to get an idea what \vT-values to expect for massive clusters at z
\cge 0.3. The best determined CMB dipole value is 3364.5$\pm$2.0$\mu$K
\citep{fixsen_1996, hinshaw_2009, planck_II_2016_c}. Compared to the best
determined current CMB temperature values of T=2.72548$\pm$0.00057
\citep{fixsen_2009, planck_II_2016_c}, this corresponds to the average velocity 
of the solar system of 370.1$\pm$0.2 \kms\ in the direction of \lII =
264.00$\pm$0.03\degree\ and \bII = 48.24$\pm$0.20\degree\ in Galactic
coordinates \citep{hinshaw_2009, planck_XXVII_2014, planck_XLVII_2016_d}. 

The Local Group is falling into the Virgo Cluster at $\sim$250 \kms\
\citep[\eg][]{dressler_1991}. More recent studies suggest that the bulk
velocity of Virgo {\it plus} the Local Group towards the CMB is 631$\pm$20 \kms\
\citep[\eg][]{hoffman_2015, watkins_2015_a, watkins_2015_b, hoffman_2017}. This
bulk motion may be as much due to gravitational pull from a Great Attractor in
the direction of the Shapley {\it overdensity}, as well as a push from large
local {\it underdensities or voids} roughly in the opposite direction
\citep[\eg][]{hoffman_2017}. That is, the net space velocity of the solar system
moving with the Local Group {\it and} the Virgo Cluster towards the Great
Attractor would correspond to a one-dimensional velocity of $\sim$631/$\sqrt{3}$
$\simeq$364 \kms\ when viewed from a random direction. This would be close to
its \vT-value when viewed from a random point in space. To calculate the net
\vT-value below that includes the solar system motion, we will use the actual
solar system velocity of 370 \kms\ towards the CMB from the
\citet{planck_XLVII_2016_d} results above, which is similar in value. 

From a large sample of clusters, \citet{bahcall_1996} suggest a \cle 5\%
probability of finding clusters with one-dimensional peculiar velocities greater
than 600 \kms, while the one-dimensional cluster peculiar velocity ranges
between 300--600 \kms\ (r.m.s.) for various cosmological models, in line with
the bulk velocity implied for the Virgo cluster above. It thus seems reasonable
for nearby massive clusters to have transverse space velocities of 300\cle
\vT\cle 600 \kms.

These velocities are substantially smaller than the relative velocity of the
sub-cluster components seen in the Bullet cluster at z$\simeq$0.296
\citep[\eg][]{tucker_1998, clowe_2006}. The relative transverse velocity of
the two Bullet sub-cluster components may be as high as 3000\cle \vT\cle 4500
\kms\ \citep{molnar_2013}. Based on cosmological simulations with the largest
possible volumes, \citet{thompson_2012} and \citet{watson_2014} emphasize that
the probability of merging sub-clusters with masses exceeding $10^{14}$ \Mo\
and velocities this high are rare, except perhaps in non-standard models
\citep{angus_2008}. Based on hydrodynamical models, \citet{springel_2007}
suggest a more modest transverse velocity for the Bullet cluster sub-components
of $\sim$2700 \kms. The question then arises: what are reasonable values for
\vT\ for the massive (M\cge 10$^{15}$ \Mo) and best lensing clusters at 0.3\cle
z\cle 0.5, to be selected for observations of the First Light epoch with JWST?

\sn \subsubsection{Maximum Transverse Velocities Adopted for Lensing Clusters
at 0.3\cle z\cle 0.5}
\label{sec422}

\sn Throughout, we take (\VTs) to mean the net transverse velocity, accounting
for the transverse motion of both the observer, the lens, and the source
planes. The effective transverse velocity \VTs\ when observing a source at
z$_s$$\simeq$7--17 lensed by a cluster at z$_d$$\simeq$0.4 is computed as
following, starting with the sum of the relevant velocity {\it vectors} scaled
with the appropriate angular diameter distances, using Eq. B9 of
\citet{kayser_1986}:

\begin{eqnarray}
\nonumber
\vec{V_T,s}\ &=&\ \vec{v_s}/(1+z_s)\ +\ \vec{v_{obs}} (D_{ds}/D_d)/(1+z_d)\\ 
&-&\ \vec{v_T} (D_s/D_d)/(1+z_d).
\label{eq:vT}
\end{eqnarray}

\n Here, the first term due to the source motion at z$\simeq$7--17 is
negligible at $v_s$\cle 30/(1+z$_s$)\cle 2--4 \kms\ \citep{barkana_2002},
using the expected small velocity dispersion of the low mass halos they likely 
reside in at high redshifts. The unknown bulk motion of the halo at z\cge 7
could increase this to several 100/(1+z) \kms, or \cle 40 \kms. In either
case, this first term is much smaller than the last two terms. The second term
is due to the velocity of the solar system (moving in the Galaxy, the Local
Group {\it and} with the Virgo cluster) towards the CMB of $v_{obs}$$\simeq$370
\kms\ from \S \ref{sec421} --- modulated by the Earth's motion around the Sun
of $\sim$30 \kms\ --- and scales with the ratio of the angular diameter distance
between deflector-to-source, $D_{ds}$, and the angular diameter distance to the
cluster deflector, $D_d$. The third term \vT\ is due to the transverse cluster
motion itself, and scales with the angular diameter distance ratio of
source-to-deflector, $D_s/D_d$. We ignore here the intrinsic velocities of 
any microlenses in the lens plane, since these are demagnified in the source 
plane by large factors, as shown by \citet{kayser_1986} in the
high-magnification regime of interest.

To assess the transverse velocity \vT\ for lensing clusters at 0.3\cle z\cle
0.5, we perturbed the observed redshift distribution of three promising HFF
clusters with a random {\it space velocity}, and determine how much of the {\it
projected} space velocity can be added to the {\it transverse} direction,
before the other projected component added to the line-of-sight velocities
disturbs the observed cluster redshift distribution too much. Details of this
simulation are given in Appendix A and its figure. These show that adding space
velocities with projected {\it transverse} components much larger than \vT
$\simeq$1000 \kms\ imply projected components of this space velocity added
{\it along the line-of-sight} that are not consistent with the available
redshift data in the cluster core, although somewhat smaller values are
certainly allowed. This results in an upper limit of \vT\cle 1000 \kms\ for
the maximum transverse velocity of these clusters at 0.3\cle z\cle 0.5 in the
plane of the sky. For some substructures in each cluster, the \vT-values may
well be as high as 1000 \kms. Since we cannot currently distinguish if the
whole cluster --- or only several of its sub-clumps --- are moving transversely
at \vT\cle 1000 \kms, the integral caustic length that we defined in \S 
\ref{sec43} and use in \S \ref{sec44} to calculate the caustic transit rate is
therefore also an upper limit. 

For the HFF cluster sub-sample discussed in Appendix A at least, it seems
appropriate to use \vT\cle 1000 \kms\ for the maximum transverse velocity of
these clusters in the plane of the sky, as projected from their space velocity.
\citet{kelly_2017_b} find that for cluster MACS J0416-2403 --- behind which
their caustic-transiting star at z$\simeq$1.5 was identified --- the
transverse velocity is about 1000 km/s. From large N-body simulations,
\citet{watson_2014} find that most pairwise halo velocities are \cle 3000
\kms\ with a median of $\sim$1000 \kms. 

The maximum effect for caustic transits in Eq.~\ref{eq:vT} is obtained when the
solar system velocity vector, $\vec{v_{obs}}$, towards the CMB as projected on
the sky at the cluster location is exactly anti-aligned with the transverse
cluster velocity vector, \vT, which is captured by the minus-sign in
Eq.~\ref{eq:vT}. If both lens and observer are going in the same direction, 
one would obtain the smallest value for this velocity difference. The actual
transverse vector sum will be different for each cluster by an unknown amount,
depending on how the unknown direction of the transverse vector of each cluster,
$\vec{v_T}$, aligns with the velocity vector, $\vec{v_{obs}}$, of the solar
system towards the CMB. Because both projected transverse velocities are
vectors, the typical expected value of the velocities is the $r.m.s.$, not the
velocity difference. Hence, we add both in quadrature, so that for
z$_d\simeq$0.4 and z$_s\simeq$12 we obtain: 

\begin{equation}
\vert{V_T,s}\vert \simeq \sqrt{ {[370 \sqrt{2/3}\times 0.40 ]}^2\ +\
{[1000 \times 0.48]}^2},
\end{equation}

\sn using the velocities discussed above and the appropriate angular diameter
distance ratios for our adopted cosmology. This amounts to \VTs$\simeq$495
\kms. When exactly anti-aligned, the two velocity components would just add
without the transverse projection factor $\sqrt{2/3}$ of the solar system
velocity at the location of the lensing cluster, so that \VTs\cle 502 \kms.
Given the significant differences in the allowed \vT-values between the three
HFF clusters discussed in Appendix A, we will adopt for our caustic transit
calculations in \S \ref{sec44} an upper limit of \VTs\cle 1000 \kms.

\mn \subsection{Estimates of Cluster Caustic Lengths for Pop III Caustic
Transit Calculations}
\label{sec43}

\begin{figure*}[!hptb] 
\n\cl{ 
\vspace*{-0.00cm} 
\includegraphics[height=0.500\txw,angle=0]{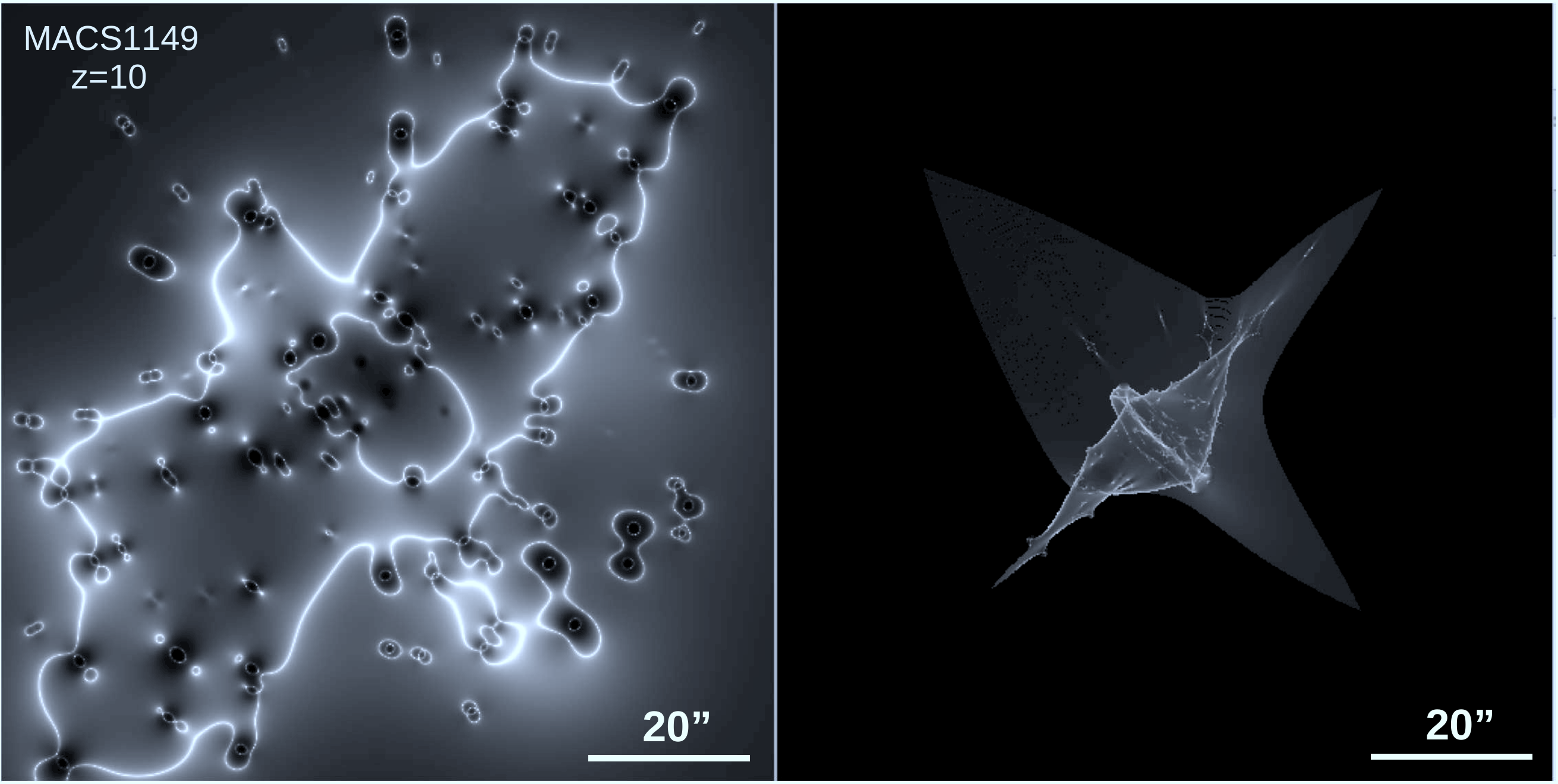} 
} 
\centering{}
\vspace*{-0.300cm} 
\caption{
[LEFT] Example of the lensing magnification map for galaxy cluster MACS
J1149.5+2223 at z$\simeq$0.4 and a background source at z=10 \citep[\eg][and
references therein]{lotz_2017}. Light from the cluster galaxies is not shown to
avoid overcrowding, but can be found in these papers. The white areas mark the
critical curves, where maximum lensing magnification is observed from this
cluster for a background source with half-light radius \rhl\cle 0\arcspt 5 at
$z=10$. The lightest regions have the highest magnification ($\mu$\cge 10--20),
while the darkest regions are areas of low magnification ($\mu\simeq$1 or even
$\mu$\cle 1) around the cluster member galaxies. 
[RIGHT] Example of the caustic map produced by the cluster mass model for a
background source at z=10. This is the location where a point source at z=10
produces maximum magnification. The total length of the cluster caustics can be
as large as L$\simeq$100\arcs, which we adopt as upper limit to the typical
caustic length in our caustic transits calculations.} 
\label{fig:fig4}
\end{figure*}

\sn Gravitational lensing modeling will result in lensing maps in the plane of
the cluster. The clusters selected for JWST are assumed to be in the redshift
range 0.3\cle z\cle 0.5, following the arguments of \S \ref{sec41}. An
example of a lensing map is shown in Fig.~\ref{fig:fig4}a for sources at z=10
behind the HFF cluster MACS J1149.5+2223 at z$\simeq$0.4
\citep[\eg][]{lotz_2017}. Detailed lensing maps of the HFF clusters have been
made by \eg\ \citet[][]{jauzac_2014}, \citet{jauzac_2015}, \citet{lam_2014},
\citet{diego_2015_a}, \citet{diego_2015_b}, \citet{diego_2016_a},
\citet{diego_2016_b}, and \citet{caminha_2017}. An example of the caustic maps 
for a source at z=10 behind MACS J1149.5+2223 is given in Fig.~\ref{fig:fig4}b. 
These are similar for Pop III sources at 7\cle z\cle 17, which may be
observable to JWST via caustic transits. 

Making lensing maps is a complex process that introduces its own uncertainties,
which depend, \eg\ on the detailed input cluster mass distribution, the number
and redshift distribution of available sources with multiple images through
which the lensing model gets refined, the point-spread-function (PSF), quality,
and depth of the images at HST resolution used to reconstruct the lensed
sources, and a number of other factors such as the actual amount of cluster
sub-structure present. For details, we refer to \citep[\eg][]{meneghetti_2017},
where errors in the reconstruction of HFF-like clusters are discussed based on
simulated data. 

In essence, the more numerous the input redshifts and the better the input
imaging data are, the more reliable the lensing model will become. Ideally,
the entire gravitational field of the cluster with all its substructure, dwarf
galaxies, detailed ICL distribution and stellar microlenses would have been
modeled. For exact caustic transit modeling of a known source at lower 
redshift \citep[\eg][]{kelly_2017_b, diego_2017}, detailed lensing modeling is
necessary, since a caustic transit may have been observed at {\it one}
location, where the local gravitational lensing model then exactly matters for
the correct interpretation of the observed data. This is, \eg\ the case when a
known background galaxy provides the stellar object that transits the cluster
caustic at a specific location. 

For the current work, detailed lensing models are not required, since Pop III
stars at z\cge 7 may be present {\it everywhere} at average sky-SB levels no
brighter than the upper limits adopted in \S \ref{sec23}, where we calculated
that most of their diffuse flux will come from objects that are well beyond the
HST and JWST point-source detection limits. For the current caustic transit
calculations, we will assume that the integrated near-IR sky-SB of these
``unresolved'' objects at z\cge 7 is rather uniform (see \S \ref{sec231}). That
is, we do not need to know the exact lensing model at each location along the
caustic, since Pop III caustic transits at z\cge 7 can happen {\it anywhere} at
unpredictable locations along the caustic lines in the cluster. 

Instead, we need the general properties of the caustics to estimate the rate of
transits, and an estimate of the maximum magnification around the caustic as a
function of source angular size. We are interested in the
statistics/probability of seeing these rare events. Hence, typical global
properties of the lenses, such as the area in the background plane with
magnifications above a given threshold, or the statistical presence of
microlenses that can modify these properties in the high-magnification regime
are the only quantities that are relevant here. 

For the current purpose of order-of-magnitude estimates of Pop III object
caustic transits at z\cge 7, we will thus assume average caustic lengths $L$
and geometry. Line integration of the lensing models in clusters like
Fig.~\ref{fig:fig4}b shows that their typical {\it total} caustic length is
$L_{\rm caust}$\cle 100\arcs, which we will use as upper limit in for the 
caustic transit calculations \S \ref{sec44} \& \ref{sec62}.


\vspace*{-0.00cm} 
\n \\


\vspace*{-0.50cm}
\n \subsubsection{Details of Lensing Magnification near the Cluster Caustics}
\label{sec431}

\sn Although usually represented in the source plane for convenience, caustics
actually form in the observer plane, and it is the relative motion with
respect to these caustics that produce the peaks in the observed light curves
(see \S \ref{sec422}). For most clusters, caustics tend to adopt a
diamond-like shape aligned in the same direction as the main symmetry axis of
the ellipsoid that encloses the cluster (see Fig.~\ref{fig:fig4}ab). Since
these three planes are uncorrelated, the vector of the relative transverse
velocity can point in any direction with respect to the caustic pattern. As
discussed in \S \ref{sec422}, we should consider all possible directions when
estimating caustic transit rates. In detail, the velocity of the caustics is
complex, because the shape of the network changes in addition to the transverse
movement. Given the other larger uncertainties in cluster geometry, \vT, and 
$L_{\rm caust}$, we will henceforth ignore the projection effects from the
angle, $i$, between the cluster's unknown main velocity vector and the main
direction of the caustic at each location, which will average out to
$<$sin(i)$>$$\simeq$1/2. 

When a background star crosses a cluster caustic it can be magnified by a
factor of up to $\mu$$\simeq$$10^5$--$10^6$ for a short period of time (few
weeks--months; see \S \ref{sec44}), depending on the strength of the caustic and
the stellar radius. This magnification can thus boost the apparent brightness
of the star by $\sim$12.5--15 mag. Possible modifications from microlensing are
discussed in \S \ref{sec432} and Appendix B1--B2. Fainter Pop III stars with
AB$\simeq$41--43 mag could then be observed with JWST to AB\cle 28.5--29 mag 
during one of these caustic crossing events. At larger distances from the
caustic ($\simeq1$ pc) the magnification is more moderate ($\mu\simeq$10$^3$),
and only the brightest stars with AB\cle 36 mag could be observed via a caustic
transit, but they could remain visible for many years because they remain
visible further away from the caustic. Microlensing can reduce these
magnifications and spread the microlensed events over a larger area still,
which lengthens their visibility in time. This is discussed further in Appendix
B1--B2. 

For the more ubiquitous fold caustics, the magnification near a caustic varies
with the distance to the caustic, $d$, as:

\begin{equation}
\mu = B_o / \sqrt{d} 
\label{eq:muvsd1}
\end{equation}

\n where $B_o$ is a constant that depends on the derivatives of the 
gravitational potential. For clusters like the HFFs, $B_o$ is normally in the
range 10--20, while $d$ is expressed in arcseconds \citep[see \eg][for a
detailed discussion]{miralda_escude_1991, diego_2017}. Hence, for a background
Pop III star at z \cge 7, magnifications of order $\mu$$\simeq$10$^3$ can be
attained once the background star is $\simeq$1 pc away from the caustic (or
$d$$\simeq$0\arcspt 001). For an HFF-like cluster with ${\rm L_{\rm
caust}}$$\simeq$100\arcs\ at z$\simeq$12, this implies that an area of
$\sim$0.1 \arcsecsq\ in the source plane can magnify background stars by more
than a factor $\mu$$\simeq$10$^3$, so that any star brighter than AB$\simeq$36
mag can be lensed to above the detection limit of JWST and produce a double
image separated by less than 0\arcspt 5 around the critical curve. When we get
closer to the critical curve, the double lensed image that would appear on each
side of the critical curve will be unresolved at JWST's near-IR resolution of
$\sim$0\arcspt 08 FWHM {\it if} the separation between the two images is
smaller than $\sim$25 milliarcsec. At these separations, the total
magnification would be $\mu$$\simeq$10$^4$, and any star brighter than
AB$\simeq$38 mag could be lensed to above the detection limit of JWST. This
corresponds to an area of $\simeq$3$\times$10$^{-4}$ \arcsecsq\ in the source
plane. 

At even smaller distances to the true caustic, fainter and smaller stars would
become visible, but the probability of magnifying a star in this narrower
region would be smaller. Clearly, there is a trade-off between the luminosity
function slope of the background stars and the probability of being magnified
above a certain value, which is given by the area in the source plane,
$A(>\mu)$, that has a magnification larger than $\mu$. This magnification area
seems to follow a power law for the more ubiquitous fold caustics: $A(>\mu) =
B_o/\mu^2$, where $B_o$ will vary somewhat from cluster to cluster. 

Owing to this scaling of the area with $\mu^2$, it is easy to see
\citep{kelly_2017_b} that the optimal trade-off between luminosity function and
$A(>\mu)$ happens when the stellar LF slope is close to $\alpha$$\simeq$--2,
where $dN/dL$$\propto$$L^{\alpha}$ is the luminosity function of the background
stars. One possible complication when observing a lensed bright star at
moderate magnifications (\ie\ a star brighter than AB$\simeq$36 mag and with
$\mu$\cle 10$^3$), is that the timescale for the flux variation could be very
long (\cge hundreds of years). This makes it a challenging task to distinguish
between a lensed Pop III star and a larger unlensed substructure, such as a
globular cluster in the background galaxy. Spectroscopy of brighter caustic
transits will be necessary to help reveal their nature, as discussed in
\S~\ref{sec7}. Microlensing fluctuations will likely make the light curve of a
caustic transit more variable and spread over a longer period of time, as
discussed in \S \ref{sec432} \& Appendix B1--B2.

\n \subsubsection{Possible Role of Microlenses during Caustic Transits}
\label{sec432}

\sn For completeness, we will consider here also the case where the caustics are
disrupted by microlenses, as discussed in \citet{kelly_2017_b} and
\citet{diego_2017}. A way to distinguish a Pop III star from a small background
substructure would be through microlensing by low-mass stars and stellar
remnants in the intra cluster medium \citep[\eg][]{lewis_2000}. First, the
timescale for microlensing would be on the order of days to weeks instead of
years. Second, microlensing by star-like objects in the lens plane would affect
only very compact objects in the background, such as Pop III stars and their
stellar-mass BH accretion disks. Larger objects magnified by factors of 
$\mu$$\simeq$10$^3$ would be much larger than the Einstein radius of the
microlenses, resulting in microlensing being irrelevant for such objects.
Third, microlensing events would take place around the critical curve, at
separations of $\sim$0\arcspt 1 on either side of the critical curve, instead
of just at the location of the critical curve. Finally, if microlenses are
ubiquitous in the lens plane, a single bright star in the background can be
responsible for multiple peaks, all of them with exactly the same spectrum,
which would increase the rate of observed caustic transit events. These would
have to be monitored over the long term (see \S \ref{sec73}) and modeled in
detail for proper interpretation.

As discussed in \citet{diego_2017}, the relatively high magnifications near the
critical curve of a cluster amplify not only the background object, but also the
lensing distortion produced by otherwise negligible microlenses from the
intracluster medium. In the magnification regime of $\mu$$\simeq$10$^3$ (about
1 pc from the caustic at z$\simeq$12), a microlens with M$\simeq$1 \Mo\ behaves
as a microlens with an effective mass of hundreds of \Mo. These large effective
masses can magnify a Pop III star by $\mu$\cge 10$^4$, instead of by the
expected factor of $\mu$$\simeq$10$^3$ that would occur without microlenses.
This translates into a temporary boost on timescales of days to weeks of
$\sim$3 mag with respect to the case without microlensing, depending on the
microlens mass and the radius of the Pop III star. Multiple microlens
magnification events can occur for many years {\it before or after} the Pop
III star aligns with the cluster caustic, thereby greatly increasing the chance
of seeing these stars. On the other hand, a large number of microlenses around
the critical curve may disrupt the caustic in such a way that extreme
magnifications of order $\mu$$\simeq$10$^5$ are no longer possible. Hence, 
only the brightest Pop III stars may be observed this way, thereby reducing the
pool of background objects that can be observed by JWST. For more details on
caustic transits in the presence of microlensing, we refer to the discussion in
\citet{diego_2017}. 

To compute the caustic transit rates in this paper, we consider the two cases:
with and without microlenses. The case {\it without} microlenses is more
straightforward, since it involves only the properties of the caustic, the
sky-SB of background Pop III stars, and the relative velocity between the
caustic and the stars. As mentioned above, all Pop III stars brighter than 
AB$\simeq$35--41.5 mag could potentially be observed during a caustic crossing
($\mu_{max}$\cle 10$^3$--10$^5$). The caustic transit rate for this particular
case is discussed in \S \ref{sec44}. The case {\it with} microlenses renders
similar results, but is more uncertain since it depends on the actual IMF of
faint microlensing stars in the foreground cluster ICL. We refer to Appendix
B1--B2 for further details of possible Pop III object caustic transits in the
presence of microlenses. 

\mn \subsection{Implied Estimates of Cluster Caustic Transits for Pop III
Stars without Microlensing}
\label{sec44}

\sn The question we will address in this section is: {\it if} a fraction of 
the diffuse near-IR background {\it is} generated by Pop III stars --- with a
conservative upper limit to their near-IR sky-SB of \cge 31\ \magarc\ (\S 
\ref{sec23}) --- then what is the probability that JWST will catch one of these
Pop III stars being lensed by a cluster caustic transit? 

For our calculations, we start with the premise that this maximum 1--4 \mum\ 
sky-SB is made up of ZAMS Pop III stars with AB\cge 37.5 mag at z\cge 7
(Table~\ref{tab:tab2}). During their RGB and AGB stages, these Pop III stars
may be as ``bright'' as AB\cge 35 mag at z\cge 7
(Tables~\ref{tab:tab3}--\ref{tab:tab4}). Pop III stars in the mass range of
30\cle M\cle 1000 \Mo\ are the most likely to be detected by JWST at z\cge 7 
at AB\cle 28.5--29 mag {\it if} the caustic magnifications reach $\mu$\cge 
10$^4$--10$^5$. We will assume that the geometrical optics approximation still
holds in this very small source regime. To reference our calculations
following Eq.~\ref{eq:masslumrelations}, we define the apparent magnitude of a
100 \Mo\ star with luminosity $L_{100}$ at $z$=12 as $m_{100}$, and that of a
20 \Mo\ star with luminosity $L_{20}$ at z=12 as $m_{20}$. The AB-magnitudes
at other redshifts scale with the DM and BIK-corrections in
Tables~\ref{tab:tab2}--\ref{tab:tab4}. 

Our caustic transit calculations depend on stellar luminosity, which depends on
ZAMS mass following the Pop III star mass-luminosity relation of
Eq.~\ref{eq:masslumrelations}. To generalize our caustic transit calculations
in the relevant equations below, we will propagate the three different
mass-dependent power-law slopes in Eq.~\ref{eq:masslumrelations} over the
entire Pop III ZAMS mass range of 1\cle M\cle 1000 \Mo. For 10\cle M\cle 1000 
\Mo, the ZAMS radii scale as:

\begin{equation} 
R = R_{100}(M/100\ \mbox{M}_\odot)^{0.45},
\label{eq:RM}
\end{equation}

\n following Eq.~\ref{eq:homologyrelations}a in \S \ref{sec31}. 

Given a population of Pop III stars with luminosity $L$, the number density
$N(L)$ required to make up a surface brightness of AB$\simeq$31\ \magarc\
follows from: 

\begin{eqnarray} 
\nonumber (m_{100}-31) = 2.5\ log_{10} (N(L)\ \times\
L_{100}),\ or:\\ N(L)\ \times\ L_{100} = 10^{(m_{100}-31)/2.5}.\ \ \ \
\end{eqnarray}

\vspace*{-0.10cm}
\n Given the segmented ML-relation in Eq.~\ref{eq:masslumrelations}, we can
generalize this as a function of mass $M$ as follows: 

\vspace*{+0.10cm}
\begin{eqnarray}
\nonumber 
\hspace{-0.00cm}
\nonumber
&&N(M)= 10^\frac{(m_{100}-2.5\ log\ (L/L_{100})-31)}{2.5} \\ 
\nonumber 
\hspace{-0.00cm}
&=& \left(\frac{L}{L_{100}}\right)^{-1}10^\frac{(m_{100}-31)}{2.5}\\ 
\nonumber 
\hspace{-0.00cm}
&\simeq& \left(\frac{M}{100}\right)^{-1.16}10^\frac{(m_{100}-31)}{2.5},\ \ 
for\ \ M\gtrsim\ 100\ \mbox{M}_\odot,\\
\nonumber 
\hspace{-0.00cm}
&\simeq& \left(\frac{M}{100}\right)^{-2.06}10^\frac{(m_{100}-31)}{2.5},\ \ 
for\ \ 20\lesssim M\lesssim 100\ \mbox{M}_\odot,\\
\hspace{-0.00cm}
&\simeq& 2.33 \left(\frac{M}{20}\right)^{-3.20}10^\frac{(m_{20}-31)}{2.5},\ \
for\ \ M\lesssim 20\ \mbox{M}_\odot\ 
\label{eq:NM} 
\end{eqnarray}

\vspace*{-0.00cm}
\n in units of $\mbox{arcsec}^{-2}$, while all masses are in \Mo. The extra
constant in the last line of Eqs.~\ref{eq:NM}--\ref{eq:Nlens} reflects the
change in normalization at 100 \Mo\ in the first two mass ranges to 20 \Mo\ 
in the last mass range. 

To be observed with JWST in a single epoch at a flux limit of AB\cle 28.5 mag,
a star of mass $M$ would need a lensing magnification of:

\begin{eqnarray}
\nonumber 
\hspace{-0.00cm}
&&\mu(M) = 10^\frac{(m_{100}-2.5\ log\ (L/L_{100})-28.5)}{2.5}\\ 
\nonumber 
\hspace{-0.00cm}
&=& \left(\frac{L}{L_{100}}\right)^{-1}10^\frac{(m_{100}-28.5)}{2.5}\\ 
\nonumber 
\hspace{-0.00cm}
&\simeq& \left(\frac{M}{100}\right)^{-1.16}10^\frac{(m_{100}-28.5)}{2.5},\ \ 
for\ \ M\gtrsim 100\ \mbox{M}_\odot,\\
\nonumber 
\hspace{-0.00cm}
&\simeq& \left(\frac{M}{100}\right)^{-2.06}10^\frac{(m_{100}-28.5)}{2.5},\ \
for\ \ 20\lesssim M\lesssim 100\ \mbox{M}_\odot,\\
\hspace{-0.80cm}
&\simeq& 2.33\ \left(\frac{M}{20}\right)^{-3.20}10^\frac{(m_{20}-28.5)}{2.5},\ 
\ for\ M\lesssim 20\ \mbox{M}_\odot.\ \ 
\label{eq:muM}
\end{eqnarray}

\vspace*{-0.00cm}
\n Now the typical magnification at a distance $d_{\mu}$$\simeq$1\arcs\ from a
caustic is $\mu$(1\arcs)=10 (see \S \ref{sec43} and Fig.~\ref{fig:fig4}b), 
using the conservative lower value of $B_o$$\simeq$10 in \S \ref{sec431}, so
that:

\begin{equation}
\mu\ \simeq\ 10 \frac{1}{\sqrt{d_{\mu}}}.
\label{eq:muvsd}
\end{equation}

\vspace*{-0.10cm}
\n The angular distance (in \mbox{arcsec}) from the true caustic maximum
corresponding to a magnification $\mu$ is then: 

\vspace{-0.10cm}
\begin{eqnarray}
\nonumber
\hspace{-0.40cm}
&& d_{\mu}(M) = 100\ \mu^{-2}\\
\nonumber
\hspace{-0.00cm}
&=& 100\left(\frac{L}{L_{100}}\right)^{2}10^\frac{-(m_{100}-28.5)}{1.25}
\mbox{arcsec}\\
\nonumber
\hspace{-0.00cm}
&\simeq& 100\left(\frac{M}{100}\right)^{2.32}10^\frac{-(m_{100}-28.5)}{1.25},\ \ 
for\ \ M\gtrsim 100\ \mbox{M}_\odot,\\
\nonumber
\hspace{-0.00cm}
&\simeq& 100\left(\frac{M}{100}\right)^{4.12}10^\frac{-(m_{100}-28.5)}{1.25},\ 
20\lesssim M\lesssim 100\ \mbox{M}_\odot,\\
\hspace{-0.00cm}
&\simeq& 18.4 \left(\frac{M}{20}\right)^{6.4}10^\frac{-(m_{20}-28.5)}{1.25},\ 
1\lesssim M\lesssim 20\mbox{M}_\odot.\ \ 
\label{eq:dmuM}
\end{eqnarray}

\vspace*{-0.00cm}
\n To estimate the relevant timescales, we need to know the crossing time for
a caustic passing over a distance $d_{\mu}$, and the crossing time for a
typical radius of a ZAMS Pop III star with M$\simeq$30--1000 \Mo. Our adopted
cosmology yields 3740 pc per arcsec at z$\simeq$12. Using the 2012 IAU value
for one Astronomical Unit (AU) of 149.6 $\times$10$^{9}$ m \citep{prsa_2016},
then 1.0 \Ro\ (or 695,700 km; \S \ref{sec31}) corresponds to
$\sim$6.03$\times$10$^{-12}$ arcsec at z$\simeq$12. Hence, the ZAMS Pop III
stars in Table~\ref{tab:tab1} are
$\sim$5.2$\times$10$^{-12}$--7.78$\times$10$^{-11}$ arcsec across at
z$\simeq$12, and at most between $\sim$1.3--14$\times$larger during their 
RGB--AGB phases, which together last $\sim$8$\times$ shorter than the ZAMS 
(\S \ref{sec31}). Pop III RGB--AGB star caustic transits will thus be more
rare, although according to Table~\ref{tab:tab3}--\ref{tab:tab4} also
$\sim$1.5--2.5 mag brighter than those of ZAMS Pop III stars
(Table~\ref{tab:tab2}). To obtain lower limits to the caustic transit
rise-times and upper limits to their caustic transit rates, we will therefore
use the Pop III star ZAMS parameters in Table~\ref{tab:tab1}--\ref{tab:tab2}. 

The upper limit of \VTs\cle 1000 \kms\ (\S \ref{sec42} and Appendix A)
corresponds to an angular speed of $d\theta/dt$$\simeq$$1.83\times 10^{-7}$
arcsec/yr for a galaxy cluster at z$\simeq$0.4. At this redshift, there are
$\sim$5590 pc/arcsec in Planck cosmology. Using Eq.~\ref{eq:RM}, the crossing
times for the Pop III star radius $R(M)$ across the $\mu>10^4$ magnification
region $\theta_{\mu}(M)$ --- needed to make stars of mass $M$ detectable to
JWST --- leads then to a mass-dependent Pop III star caustic transit timescale
of:\ 

\begin{equation} 
t_R(M) = \frac{R}{v_{\rm T,s}} = \frac{R}{R_{100}}\frac{R_{100}}{v_{\rm T,s}}
\simeq \frac{R_{100}}{v_{\rm T,s}}\left(\frac{M}{100}\right)^{0.45},
\label{eq:tRMb}
\end{equation}

\n and:

\begin{eqnarray}
\nonumber
\hspace{-0.20cm}
&& \theta_\mu(M) = (\frac{d\theta}{dt})^{-1}\frac{100''}{\mu^2(M)}\\ 
\nonumber
\hspace{-0.00cm}
&\simeq& \frac{100}{d\theta/dt}\left(\frac{M}{100}\right)^{2.32}
10^\frac{-(m_{100}-28.5)}{1.25},\ for\ M\gtrsim 100\ \mbox{M}_\odot,\\
\nonumber
\hspace{-0.00cm}
&\simeq& \frac{100}{d\theta/dt}\left(\frac{M}{100}\right)^{4.12}
10^\frac{-(m_{100}-28.5)}{1.25},\ \ 20\lesssim M\lesssim 100\ \mbox{M}_\odot,\\
\hspace{-0.00cm}
&\simeq& \frac{18.4}{d\theta/dt}\left(\frac{M}{20}\right)^{6.4}
10^\frac{-(m_{20}-28.5)}{1.25},\ for\ M\lesssim 20\ \mbox{M}_\odot.
\label{eq:thetamuM}
\end{eqnarray}

\vspace*{-0.00cm}
\n This implies that the brightening time --- defined as the time for the
magnification to go from zero to its maximum value --- for a Pop III star is
very short ($\sim$0.5--3 hours) when it transits the caustic starting at the
``highest-magnification edge''. The star would then stay bright for several
months to a year, with brightness decaying as 1/$\sqrt{t-t_o}$, where ($t-t_o$)
is the time since the stellar disk started the caustic crossing at time $t_o$. 

Also, this entire process is reversible, so one could witness a very slow rise
of an object's flux as 1/$\sqrt{t_f-t}$ when it moves {\it towards} the caustic
starting from the low-magnification end, followed by an abrupt disappearance
once it crosses the caustic at some future time $t_f$. We will discuss below
and in \S \ref{sec7} how JWST may detect each of these cases. 

To calculate how often we expect such a brightening, we assume that the cluster
has a length $L_{\rm caust}$$\simeq$100\arcs\ of caustic (see \S \ref{sec43}).
To calculate the rate at which Pop III lensing occurs, we need the area crossed
by the caustics per unit time. This change in area is given by: 

\begin{eqnarray} 
\nonumber 
&&\frac{dA}{dt} = L_{\rm caust} \times v_{\rm T} = 100 \times 2.4 \times
10^{-14}\ \mbox{arcsec}^2 / \mbox{sec} \\ 
\noindent 
&\simeq& 1.6\times 10^{-5}\left(\frac{L_{\rm caust}}{100''}\right)
\left(\frac{v_{\rm T}}{1000\ \mbox{km/s}}\right)\mbox{('')}^2/yr.\ \ \ 
\label{eq:dAdt} 
\end{eqnarray}

\n The surface areas referred to here are all in the source plane, and so there
is no depletion correction for magnification. The number of events therefore
follows from the surface density of Pop III stars $N$, yielding: 

\begin{eqnarray} 
\nonumber 
\frac{dN_{\rm lens}}{dt} &=& N(M)\times \frac{dA}{dt} \\ 
&=& \frac{dA}{dt} \left(\frac{L}{L_{100}}\right)^{-1}
10^\frac{(m_{100}-31)}{2.5}\hspace{1mm}\mbox{per yr}. 
\label{eq:dNlensdt1} 
\end{eqnarray}

\n To quantify the values for $N$, $t_\mu$, and $\frac{dN_{\rm lens}}{dt}$, we
base our numbers on the discussion of the physical parameters of Pop III stars
in \S \ref{sec31}. From Table~\ref{tab:tab2}, the luminosity of a 100 \Mo\
star is $\sim$1.40$\times$10$^6$ \Lo, giving an absolute magnitude of --10.63
AB-mag, using M$\equiv$$+4.74$ mag for the absolute magnitude of the Sun (see
\S \ref{sec31}). Including the bolometric+IGM+K-corrections of
Table~\ref{tab:tab2}, the corresponding apparent magnitude $m_{100}$ at z$=$12
is then $m_{100}$=40.91 mag, assuming no extinction. For a discussion of dust,
see \S \ref{sec61}. From Table~\ref{tab:tab1}, the radius of a 100 \Mo\ star
is $R_{100}$ = 4.12 \Ro. We then find the following caustic crossing rate for
lensed Pop III stars: 

\hspace{-0.30cm}
\begin{eqnarray}
\nonumber 
\hspace{-0.00cm}
&& \frac{dN_{\rm lens}}{dt} = N(M)\times \frac{dA}{dt} = \\
\nonumber
\hspace{-0.00cm}
\noindent 
&=& 0.064\ \frac{L_{\rm caust}}{100''}\ \frac{v_{\rm T}}{1000}
\left(\frac{M}{100}\right)^{-1.16},\ for\ M\gtrsim 100\ \mbox{M}_\odot,\\
\nonumber
\hspace{-0.00cm}
\noindent 
&=& 0.064\ \frac{L_{\rm caust}}{100''}\ \frac{v_{\rm T}}{1000}
\left(\frac{M}{100}\right)^{-2.06},\ 20\lesssim M\lesssim 100\ 
\mbox{M}_\odot,\\
\hspace{-0.00cm}
\noindent 
&=& 1.76\ \frac{L_{\rm caust}}{100''}\ \frac{v_{\rm T}}{1000}
\left(\frac{M}{20}\right)^{-3.20},\ 1\lesssim M\lesssim 20\ \mbox{M}_\odot\ 
\label{eq:dNlensdt} 
\end{eqnarray}

\vspace*{-0.00cm}
\n per year. The duration of a brightening time is then:

\begin{equation}
t_R(M) = 9.1\times10^{-5} \left(\frac{M}{100}\right)^{0.45} \mbox{yr},
\label{eq:tRM}
\end{equation}

\n which for a 100 \Mo\ ZAMS star is about 0.80 hours for \vT\cle 1000 \kms.
The range in rise-times in Table~\ref{tab:tab2} is $\sim$0.4--2.5 hrs for
M$\simeq$30--1000 \Mo\ Pop III stars. 

The time spent above the detection limit is:

\vspace{-0.20cm}
\hspace{-0.50cm}
\begin{eqnarray}
\nonumber 
\hspace{-0.50cm}
&& t_\mu(M) \simeq 0.4 \left(\frac{v_{\rm T}}{1000}\right)^{-1} 
\left(\frac{M}{100}\right)^{2.32},\ for\ M\gtrsim 100\ \mbox{M}_\odot,\\
\nonumber 
\hspace{-0.00cm}
&\simeq&0.4 \left(\frac{v_{\rm T}}{1000}\right)^{-1} 
\left(\frac{M}{100}\right)^{4.12},\ \ \ \ for\ \ 20\lesssim M\lesssim 100\ 
\mbox{M}_\odot,\\
\hspace{-0.00cm}
&\simeq&5.3\times10^{-4} \left(\frac{v_{\rm T}}{1000}\right)^{-1}
\left(\frac{M}{20}\right)^{6.4},\ for\ M\lesssim 20\ \mbox{M}_\odot
\label{eq:tmuM}
\end{eqnarray}

\vspace*{-0.00cm}
\n in units of years. This assumes that after the flux has peaked upon caustic
crossing, the flux declines as 1/$\sqrt{t}$ following Eq.~\ref{eq:muvsd},
assuming constant velocity \vT. 

The number of lensed Pop III stars visible at a given time is then:

\vspace{-0.20cm}
\hspace{-0.50cm}
\begin{eqnarray}
\nonumber 
\hspace{-0.50cm}
&& N_{\rm lens} = t_\mu(M)\frac{dN_{\rm lens}}{dt}\\
\nonumber 
&\simeq& 0.026\ \frac{L_{\rm caust}}{100''}
\left(\frac{M}{100}\right)^{1.16},\ for\ M\gtrsim 100\ \mbox{M}_\odot,\\
\nonumber 
\hspace{-0.00cm}
&\simeq& 0.026\ \frac{L_{\rm caust}}{100''} \left(\frac{M}{100}\right)^{2.06},\ 
for\ 20\lesssim M\lesssim 100\ \mbox{M}_\odot,\\
\hspace{-0.00cm}
&\simeq& 9.4\times10^{-4} \frac{L_{\rm caust}}{100''}
\left(\frac{M}{20}\right)^{3.20},\ 1\lesssim M\lesssim 20\ \mbox{M}_\odot.\ 
\label{eq:Nlens}
\end{eqnarray}

\vspace*{-0.00cm}
\n Note that while $t_\mu(M)$ and $\frac{dN_{\rm lens}}{dt}$ are sensitive to
the transverse velocity \vT, the visible number of events is not. {\it
Eq.~\ref{eq:dNlensdt}--\ref{eq:Nlens} contain the key relations of this paper
for calculating Pop III star caustic transits.}

For an IR background of \cge 31 \magarc\ (\S \ref{sec23}) made up of
AB$\simeq$41 mag Pop III stars with M$\simeq$100 \Mo, we estimate that one
lensing event can be observed above a flux limit of AB$\simeq$28.5 mag per
cluster per $\sim$2.7 years, or one event when monitoring $\sim$3 clusters 
during a year. Because these events should stay detectable at $\mu>\mu(M=100\
M_{\odot})$ for $t_{\mu}\simeq 0.4$ years, this implies that $\sim0.15$ such
lensed Pop III sources per cluster would be observed above the flux limit at
any given time. 

These results are sensitive to the luminosity of the Pop III stars. For example,
let us instead try the extreme case where the Pop III stars are 1000 \Mo, and
so have AB$\simeq$38.3 mag at z$\simeq$12 rather than AB$\simeq$40.9 mag. This
implies that the source needs to be magnified less, and so it can be observed
while further from the caustic, with a visible time of $t_{\mu} = 40$ yr.
However the rate is lower with $dN_{\rm lens}/dt \simeq$0.013 per year per
cluster, giving one event per cluster per 75 years. In this case of brighter,
more massive stars we find that $\sim$0.5 events per cluster would be visible 
at any given time. 

Thus, for 100 \Mo\ Pop III stars, about 6 clusters observed twice about 6
months apart would make the likelihood of observing a lensed Pop III star of
order unity, while for more massive stars, detecting a new lensing event (with
a time baseline limited to 1 year) would require observation of a larger number
of clusters in proportion to the mass $M$. For lower-mass stars, fewer clusters
would need to be observed, as long as they can appear magnified above the
detection thresholds of Tables~\ref{tab:tab2}--\ref{tab:tab4}. 

The observed rate of events will thus also depend on the mass function of Pop
III stars. A mass-function weighted average over Eq.~\ref{eq:dNlensdt} is:

\begin{eqnarray}
\nonumber
\hspace{-0.00cm}
&& \left\langle \frac{dN_{\rm lens}}{dt} \right\rangle = 0.064\left(\frac{L_{\rm
caust}}{100''}\right) \left(\frac{v_{\rm T}}{1000\ \mbox{km/s}}\right) \\ 
\hspace{-0.00cm}
\noindent 
&\times& \int_{M_{\rm min}}^{M_{\rm max}} dM
\left(\frac{M}{100}\right)^{-1} \frac{dP}{dM}dM\hspace{1mm}\mbox{\bigg/yr}. 
\label{eq:avgdNlensdt} 
\end{eqnarray}

\vspace*{-0.00cm}
\n Here, $dP/dM$ is the normalized {\em instantaneous} mass function of Pop III
stars, \ie\ the population of stars available to be lensed, {\it not} the
entire IMF. We assume a power-law mass function $dP/dM\propto M^{-\alpha}$ with
slope $|\alpha|>1$ in the range $M_{\rm min}<M<M_{\rm max}$, leading to:

\begin{eqnarray}
\nonumber
\hspace{-0.00cm}
&& \left\langle \frac{dN_{\rm lens}}{dt} \right\rangle = 0.064\left(\frac{L_{\rm
caust}}{100''}\right) \left(\frac{v_{\rm T}}{1000\mbox{km/s}}\right) 
\left(\frac{M_{\rm min}}{100}\right)^{-1} \\ 
\noindent 
&\times& \left(\frac{\alpha-1}{\alpha}\right)\left(\frac{1-(M_{\rm min}/
M_{\rm max})^\alpha}{1-(M_{\rm min}/M_{\rm max})^{\alpha-1}}\right) 
\mbox{\bigg/yr}.
\label{eq:dNlensdtalpha}
\end{eqnarray}

\vspace*{-0.00cm}
\n For $M_{\rm max}\gg M_{\rm min}$, the last term is close to unity, while
for steep mass functions with $\alpha\gg1$ the integral converges to the value
of $\langle \frac{dN_{\rm lens}}{dt} \rangle \rightarrow \frac{dN_{\rm
lens}}{dt}(M_{\rm min})$. The choice of the mass function slope
$\alpha$$\simeq$2.0 was discussed in \S \ref{sec231} \& \ref{sec34}, and is used
below. 

We chose the lower mass boundary here at 30 \Mo, since such stars may be visible
through caustic transits to JWST (Table~\ref{tab:tab2}--\ref{tab:tab4}), and 
such stars may produce BH leftovers in the mass range already observed by LIGO
at M\cge 14 \Mo\ \citep[\eg][see also \S \ref{sec51}]{abbott_2016_e}. Lower Pop
III stellar masses would render the stars too faint to be reasonably observed 
at AB\cle 28.5 mag through caustic transits in a single JWST epoch
(Table~\ref{tab:tab2}), except for perhaps RGB--AGB Pop III stars with M\cge
15--20 \Mo. The latter may be visible because their K-corrections are more
advantageous (Tables~\ref{tab:tab3}--\ref{tab:tab4}) than for the much hotter
ZAMS Pop III stars (\S \ref{sec333}). 

For the lowest Pop III star mass considered here ($M_{\rm min}$=30 \Mo), its
physical parameters of \S \ref{sec31}, and adopting an IMF slope of
$\alpha$$\simeq$2.0, we get the following upper limits for L$\simeq$100\arcs\
and \VTs\cle 1000 \kms: 

\begin{eqnarray}
&& \left\langle \frac{dN_{\rm lens}}{dt} \right\rangle 
\le 0.064 \left(\frac{30}{100}\right)^{-1}
\left(\frac{1.0}{2.0}\right)\mbox{\bigg/yr}.
\label{eq:avgdNlensdtalpha}
\end{eqnarray}

\n For a ZAMS Pop III star mass function slope of $\alpha$$\simeq$2, the
weights for each mass bin in Table~\ref{tab:tab2} are very similar at
0.23--0.17 following Eq.~\ref{eq:dNlensdtalpha}. 

The resulting {\it total} transit rates for stars with M\cge 30 \Mo\ that are
in principle observable with JWST across the caustics are then predicted to be
$\frac{dN_{\rm lens}}{dt}$\cle 0.30 events per cluster per year. These caustic
transits that may be visible to JWST are marked with an asterisk in Cols.
12--13 of Table~\ref{tab:tab2}--\ref{tab:tab4}. 

To this we need to add the caustic transit rates expected for the RGB from
Table~\ref{tab:tab3} and the AGB from Table~\ref{tab:tab4}. These must be
weighted with their approximate lifetimes compared to the ZAMS, which are
$\sim$6\% of the ZAMS lifetime (\S \ref{sec31}) for each of the RGB and AGB
phases detectable by JWST. This amounts to an additional 0.01 transits per
cluster for each of the RGB and AGB phases. Hence, the weighted total number of
caustic transits for ZAMS, RGB and AGB Pop III stars in
Tables~\ref{tab:tab2}--\ref{tab:tab4} are $\sim$0.32 per cluster per year. 

\mn \subsection{Observing Strategies for JWST to Detect Pop III Stars via
Caustic Transits}
\label{sec45}

\sn From \S \ref{sec44}, it follows that in order to see one Pop III star
caustic transit per year at the top of the Pop III star mass function (M\cge 
15--30 \Mo), one would need to observe about 3 clusters at least two times per
year about 6 months apart in one--two successive JWST Cycles. Observing more
often when scheduling allows for clusters at higher Zodiacal latitude would, of
course, be preferred. The first exposure pair would be needed to identify a
potential Pop III star caustic transit events, and the last pair is needed to
monitor its expected decay on a timescale less than one year. Imaging in all 8
broad-band NIRCam filters is essential to identify the high-redshift dropout
nature of a potential caustic transit event and to identify foreground
interloping events, which will be more numerous but interesting in their own
right. For the brighter caustic transit events, follow-up spectroscopy should
be attempted to confirm the nature of the transit, as described in 
\S~\ref{sec7}. 

The caustic transit rate of 0.32 per cluster per year is indicated by the
orange upper limit in Fig.~\ref{fig:fig1}. If the actual 2.0 \mum\ SB of Pop
III stars is dimmer than $\sim$31 \magarc, then their caustic transit rate
would be correspondingly lower. This is indicated in Fig.~\ref{fig:fig1} for SB
levels (in light orange) that are 10, 100, and 1000$\times$ dimmer than
$\sim$31.0 \magarc, with the corresponding caustic transit rates indicated in
dark orange. The minimum number of caustic transits JWST could reasonably see
--- in a large monitoring program spread over many years --- is a Pop III SB of
$\sim$36 \magarc, which would require monitoring 30 clusters at least twice
every year over 10 years. Such a large JWST observing program could reach the
level of $\sim$10 Pop III objects per \arcsecsq, and would need to be a
dedicated multi-year community effort. 

To reach levels of only a few Pop III objects per \arcsecsq\ (SB\cge 37
\magarc\ in Fig.~\ref{fig:fig1}) through JWST caustic transits would either
require to observe $\sim$100 clusters per year for 10 years --- prohibitive in
terms of JWST time --- or the existence of stellar-mass BH accretion disks that
are feeding much longer than massive Pop III stars live on average. This is
discussed in \S \ref{sec53}. 

Appendix C discusses the uncertainty estimates in the main parameters that
determine the caustic {\it transit rates} and {\it rise times} of Pop III stars
at z\cge 7. The combined uncertainty in their caustic transit rates follows
from the multiplicative sources of error in Eqs.~\ref{eq:NM}, \ref{eq:dAdt},
and \ref{eq:dNlensdt1}. These are the adopted effective caustic length
$L_{caust}$ (with $\sim$0.3 dex uncertainty), the cluster transverse velocity
\vT\ ($\sim$0.3 dex), the Pop III stellar luminosity $L$ at z\cge 7 ($\sim$0.2 
dex), the uncertainty from the presence of microlensing in the ICL (\cge 0.5
dex), and the uncertainty in the 1--4 \mum\ sky-SB from Pop III stars (\cge 0.5
dex). Further details are given in Appendix C.

These five main parameters that determine the Pop III star caustic transit rates
are independent. Therefore, the combined uncertainty in the Pop III star {\it
caustic transit rates} follows from taking these factors in quadrature, and is
estimated to be at least 0.7 dex, which is indicated by the vertical (dark
orange) error range in Fig.~\ref{fig:fig1}. For this reason, a JWST survey to
find caustic transits at z\cge 7 will need to be prepared to cover at least this
factor of 5 uncertainty in Pop III star caustic transit rates. Since these
uncertainty factors can be larger, JWST may need to observe at least 3--30
clusters per year during the first couple years of its lifetime. Such a survey
would need to be maintained until a sufficient number of Pop III star caustic
transits have been detected, at which point the actual Pop III star caustic
transit rate can be better estimated, and the survey strategy updated
accordingly. 

If Pop III stars at z\cge 7 are weakly clustered, their SB may be fairly
uniform compared to the size of the caustics (see \S \ref{sec43} and
Fig.~\ref{fig:fig4}b). Therefore, one could instead monitor fewer clusters for
a correspondingly longer period of time. That is, for an anticipated 5--10 year
lifetime of JWST (see \S \ref{sec7}), one could instead monitor a number of
well understood lensing clusters at high Zodiacal latitude every few months
during JWST's lifetime. Any of these possibilities would constitute a minimum
observing program to potentially identify Pop III star caustic transits during
the lifetime of JWST. The program could then be adjusted after the number of
caustic transits at z\cge 7 is known after the first couple of years when
monitoring a number of clusters. The presence of microlensing will likely also
require to observe the clusters more frequently to catch caustic transits at
shorter timescales, as discussed in \S \ref{sec432} and Appendix B. 

\bn \section{Parameters Adopted for Pop III Star Black Hole Accretion Disks}
\label{sec5}

\sn The question that we address in this section is: under what conditions could
JWST see the UV accretion disks of Pop III stellar-mass black holes lensed
individually through cluster caustic transits at very high magnification? To
address this, we first need to discuss the plausible range in physical
properties of Pop III stellar-mass BH accretion disks at z\cge 7, and under
what conditions these may be fed from early massive stellar binaries for the
expected range in IMF-slope (\S \ref{sec34}) and metallicity evolution (\S
\ref{sec52}). We refer the reader to recent work on PBHs \citep{kohri_2014},
DCBHs \citep{yue_2013}, or OBHs \citep{natarajan_2017} for other direct BH
feeding mechanisms. Their surface density and accretion rates are uncertain,
but if these could be estimated from theory, one could use the same formalism
as in \S \ref{sec44} \& \ref{sec62} here to estimate their caustic transit
rates. 

\mn \subsection{Range in Pop III Stellar Black Hole Masses}
\label{sec51}

\sn The mass of the final Pop III star end-product is more nuanced than just the
black hole mass. For example, theoretical models predict that stars in the
general mass range of 100 \Mo\ $\lesssim M \lesssim$ 200 \Mo\ do not lose much
mass, and that they may undergo an $e^+-e^-$ pair-creation instability
\citep{barkat_1967, fraley_1968, wheeler_1977, sugimoto_1980, bond_1984,
fryer_2001, woosley_2002, kozyreva_2017, woosley_2017}. Such stars may undergo
thermonuclear explosions that completely disrupts the star without forming a
stellar-mass BH and eject a large amount of iron-group elements, especially
$^{56}$Ni \citep[\eg][]{smith_2007, kozyreva_2015}. Theoretical models predict
that stars with 260 \Mo\ $\lesssim M \lesssim$ 5$\times$10$^5$ \Mo\ enter the
pair-instability region, but are too massive to be disrupted. They undergo
standard core-collapse, and form intermediate-mass black holes
\citep[IMBH,][]{fryer_2001, chatzopoulos_2013, belczynski_2016}. Fig.~12 of
\citet{woosley_2002} offers a map of the Pop III initial--final mass relation
for massive stars from stellar evolution theory. For the Pop III ZAMS mass
range in our \texttt{MESA} models, we adopt similar end-products. Their
end-product mass and the BH Schwarzschild radii \Rs, are listed in 
Table~\ref{tab:tab5}, which are used in our caustic transit calculations for
stellar-mass BH accretion disks. 

In this context, we briefly consider possible constraints from the recent LIGO
detections on stellar-mass BHs at z\cle 0.1 \citep{abbott_2016_a,
abbott_2016_c}. These are very plausibly examples of merging black hole pairs
with M$\simeq$29--36 \Mo, 14--21 \Mo, and 19--31 \Mo, respectively, about 1--3
Gyr ago \citep{abbott_2016_b, abbott_2016_d, abbott_2016_e, abbott_2017_a}.
\citet{de-mink_2016} suggest that these BHs are possibly leftover from later
(Pop II) starbursts about 5--12 Gyr before z$\simeq$0.1, with a median age of
$\sim$7 Gyr for these mass pairs, which in 2016 Planck cosmology corresponds to
a range in their formation redshift of z$_f$$\simeq$0.7--10 with a median of
z$_f$$\simeq$1.1. If true, such BHs may not have had significant accretion 
rates since their progenitor-star Supernovae (SNe) went off 5--12 Gyr before
their binary merger produced gravitational waves at their detection distance of
z\cle 0.1. In each LIGO case, a pair of massive stars formed of somewhat
unequal mass, and so their evolutionary scenarios may have resulted in
accretion onto the black holes left by the more massive parent stars with M\cge
30--80 \Mo\ after it produced a SN. 

Another issue that we need to consider in this section is the {\it lowest}\ 
ZAMS mass that can with some fidelity produce a BH, also for Pop III stars at
z\cge 7. This is a very active topic of research where different groups get
different results \citep{sukhbold_2014, sukhbold_2016, petermann_2018}.
Depending on the models used (1D, 2D or 3D, with or without rotation), the
``compactness'' of the end-product is rather uncertain in the mass range of 10
\cle M\cle 30 \Mo. Rotation and binary interaction can produce different
initial--final mass landscapes \citep[\eg][]{yoon_2008}. For 10\cle M\cle 30
\Mo\ not all models get a clean explosion. On the other hand, at M\cge 30 \Mo\
nature can produce SNe with BH remnants, since LIGO has already seen 14--36
\Mo\ BHs at z\cle 0.1--0.2. Hence, Pop III stars with 10\cle M\cle 30 \Mo\ {\it
may} yield BHs, while for M\cge 30 \Mo\ they most likely do. 

For our calculations of Pop III BH accretion disk caustic transits, we will
assume that Pop III stars with M\cge 30 \Mo\ --- with the exception of the mass
range of 100\cle M\cle 200 \Mo\ --- can and will produce BHs of roughly 
15--70\% of the ZAMS Pop III stellar mass, or M$\simeq$5--720 \Mo\ 
\citep[deduced from Fig.~12 of][see Col. 2 of Table~\ref{tab:tab5}
here]{woosley_2002}. A full treatment of the evolution of Pop III binary or
multiple stars, their end-products, and their impact on Pop III BH accretion
disks is beyond the scope of this study, and needs to be the focus of more
detailed modeling in future work. 

The actual resulting BH masses themselves are not as relevant for our caustic
transit calculations. It only matters that such BHs exist --- and for M\cge 
14 \Mo\ LIGO has clearly shown that they do --- {\it and} that they accrete while
producing a sufficiently high UV-luminosity to be detected by JWST during a
caustic transit. Any accretion \citep{frank_2002} would have to be maintained
for \cge 0.1 year in the restframe at z\cge 7 (\ie\ $\sim$ 1 year in the
observed frame) with \Lbol-values\cge 10$^{5}$\Lo\
(Tables~\ref{tab:tab2}--\ref{tab:tab4}) to be possibly seen transiting across a
cluster caustic by JWST and decay for about a year or less above the JWST
detection threshold (\S \ref{sec44}).


\begin{deluxetable*}{| c | c | c | ccc | ccc | ccc | c | c |}
\tablecolumns{14}
\tablewidth{1.0\linewidth}
\tablecaption{Pop III Stellar Mass Black Hole Accretion Disk Parameters Adopted 
for Caustic Transit Calculations 
\label{tab:tab5}}
\tablehead{
\colhead{Mass$^a$}                                    $\vert$ &
\colhead{M$_{compact}$$^b$}                           $\vert$ & 
\colhead{R$_{s}$$^c$}                                 $\vert$ & 
\colhead{Radius$^d$}                                          & 
\colhead{L$_{bol}$$^e$}                                       & 
\colhead{M$_{bol}$$^f$}                               $\vert$ & 
\multicolumn{3}{c}{bolo+IGM+K-corr$^g$}               $\vert$ & 
\multicolumn{3}{c}{\mAB-limits at$^h$}                $\vert$ &
\colhead{t$_{rise}$$^i$}                              $\vert$ & 
\colhead{Transit$^j$}                                         \\[-4pt]
\colhead{ZAMS}                                        $\vert$ &
\colhead{ }                                           $\vert$ & 
\colhead{BH}                                          $\vert$ & 
\multicolumn{3}{c}{--- of the UV accretion disk --- } $\vert$ & 
\colhead{z=7}                                                 & 
\colhead{z=12}                                                & 
\colhead{z=17}                                        $\vert$ & 
\colhead{z=7}                                                 & 
\colhead{z=12}                                                & 
\colhead{z=17}                                        $\vert$ & 
\colhead{(z=12)}                                      $\vert$ & 
\colhead{rate}                                                \\[-4pt]
\colhead{(M$_{\odot}$)}                               $\vert$ &
\colhead{(M$_{\odot}$)}                               $\vert$ & 
\colhead{(km)}                                        $\vert$ & 
\colhead{(R$_{\odot}$)}                                       & 
\colhead{(L$_{\odot}$)}                                       & 
\colhead{AB-mag}                                      $\vert$ & 
\multicolumn{3}{c}{(AB-mag)}                          $\vert$ &
\multicolumn{3}{c}{(AB-mag)}                          $\vert$ &
\colhead{(hr)}                                        $\vert$ & 
\colhead{(/cl/yr)}                                            
}
\startdata
\multicolumn{14}{c}{BH accretion-disk bolometric luminosities and UV half-light radii scaling from microlensed quasars \citep{blackburne_2011}}\\
 30      & $\sim$5.0 BH &   15  &  1.4 &\cle 4.2$\times$10$^4$ &\cge  --6.8 & --0.6 & --1.4 & --1.7 &\cge 41.8 &\cge 42.4 &\cge 42.9 & 0.27? &\cge 0.58? \\[-2pt]
 50      & $\sim$24  BH &   72  &  3.0 &\cle 2.0$\times$10$^5$ &\cge  --8.5 & --0.4 & --1.2 & --1.5 &\cge 40.3 &\cge 40.9 &\cge 41.4 & 0.58* &\cge 0.15* \\[-2pt]
 100     & $\sim$65  BH &  195  &  4.9 &\cle 5.4$\times$10$^5$ &\cge  --9.6 & --0.2 & --0.9 & --1.3 &\cge 39.4 &\cge 40.0 &\cge 40.5 & 0.95* &\cge 0.06* \\[-2pt]
 300     & $\sim$230 BH &  690  &  9.2 &\cle 1.9$\times$10$^6$ &\cge --11.0 & --0.2 & --1.0 & --1.3 &\cge 38.1 &\cge 38.6 &\cge 39.2 &  1.8* &\cge 0.02* \\[-2pt]
1000     & $\sim$720 BH & 2160  & 16.3 &\cle 6.0$\times$10$^6$ &\cge --12.2 & --0.2 & --0.9 & --1.3 &\cge 36.8 &\cge 37.5 &\cge 37.9 &  3.2* &\cge 0.01* \\[0.05in]
\hline
\multicolumn{14}{c}{BH accretion-disk bolometric luminosities and UV half-light radii estimated from multi-color thin-disk model }\\
 30      & $\sim$5.0 BH &   15  &  1.9 &\cle 3.1$\times$10$^4$ &\cge  --6.5 & --0.6 & --1.4 & --1.7 &\cge 42.1 &\cge 42.8 &\cge 43.2 & 0.37? &\cge 0.84? \\[-2pt]
 50      & $\sim$24  BH &   72  &  4.5 &\cle 1.8$\times$10$^5$ &\cge  --8.4 & --0.4 & --1.2 & --1.5 &\cge 40.4 &\cge 41.1 &\cge 41.5 & 0.87* &\cge 0.18* \\[-2pt]
 100     & $\sim$65  BH &  195  &  7.8 &\cle 5.9$\times$10$^5$ &\cge  --9.7 & --0.2 & --0.9 & --1.3 &\cge 39.3 &\cge 40.0 &\cge 40.4 & 1.51* &\cge 0.06* \\[-2pt]
 300     & $\sim$230 BH &  690  & 15.8 &\cle 2.0$\times$10$^6$ &\cge --11.0 & --0.2 & --1.0 & --1.3 &\cge 38.0 &\cge 38.6 &\cge 39.1 &  3.1* &\cge 0.02* \\[-2pt]
1000     & $\sim$720 BH & 2160  & 29.8 &\cle 6.6$\times$10$^6$ &\cge --12.3 & --0.2 & --0.9 & --1.3 &\cge 36.7 &\cge 37.4 &\cge 37.8 &  5.8* &\cge 0.01* \\[0.05in]
\enddata
\mn
\tablenotetext{a}{\footnotesize Pop III ZAMS stellar mass in \Mo\ from
Table~\ref{tab:tab1}.}

\tablenotetext{b}{\footnotesize Resulting Pop III Stellar Black Hole mass in
\Mo, following \citet{woosley_2002}. Note that for Pop III stellar masses of
100\cle M\cle 200 \Mo\ there are likely no BH leftovers (see \S \ref{sec51}),
which the weighting in \S \ref{sec62} includes.}

\tablenotetext{c}{\footnotesize Resulting Pop III BH Schwarzschild radius \Rs\
in km.}

\tablenotetext{d}{\footnotesize Adopted Pop III BH restframe UV-accretion disk
half-light radius \rhl\ in \Ro. The top tier of BH UV-accretion radii (and
bolometric luminosities) is inferred by scaling from observed microlensed
quasars \citep{blackburne_2011}, the bottom tier was estimated from the 
multi-color thin-disk model discussed in \S \ref{sec552}. For a standard
multi-color accretion disk around a black hole of mass M, we get about \RUV
\cle 40,000 \Rs.}

\tablenotetext{e}{\footnotesize Adopted Pop III BH accretion disk bolometric 
luminosity in \Lo. The quoted luminosities and resulting restframe
UV-magnitudes are upper limits, since they assume that the BH accretion disk
is constantly feeding at the stated luminosities for maximum lifetimes
discussed in \S \ref{sec53}--\ref{sec54}. Therefore, the resulting caustic BH
accretion disk transit rates in Col. 14 are lower limits.}

\tablenotetext{f}{\footnotesize Resulting Pop III BH accretion disk absolute
bolometric AB-magnitude M$_{bol}$.}

\tablenotetext{g}{\footnotesize Combined bolometric+IGM+K-correction to Pop
III star \Mbol\ at z=7, z=12, and z=17, respectively, calculated as in \S 
\ref{sec33} and \ref{sec552}.}

\tablenotetext{h}{\footnotesize BH accretion disk apparent AB-magnitudes at
z=7, z=12, and z=17 in 2016 Planck cosmology \citep{planck_XIII_2016_a}, using
the NIRCam filters that sample restframe UV 1500 \AA, assuming K-corrections
as in Cols. 7--9 and no dust (see \S \ref{sec61}). Distance moduli used are DM
= 49.24, 50.58, and 51.42 mag at z=7, z=12, and z=17, respectively.}

\tablenotetext{i}{\footnotesize Pop III BH accretion disk caustic transit
rise-time $t_{rise}$ at z=12 as estimated in \S \ref{sec44}. Asterisks (*)
indicate BH masses M\cge 24--65 \Mo. For their accretion disks, caustic
transit events are possibly observable to the detection limits of JWST
medium-deep to deep survey epochs reaching AB\cle 28.5--29 mag, assuming
caustic transit magnifications of $\mu$$\simeq$10$^{4}$--10$^{5}$ can elevate
Pop III stellar-mass BH accretion disks with AB\cle 41.5 mag temporarily above
these JWST detection limits. Details are in \S \ref{sec62}. } 

\tablenotetext{j}{\footnotesize The cluster caustic transit rate of
stellar-mass BH UV-accretion disks as estimated in \S \ref{sec44} and
\ref{sec62}, but directly applying Eq.~\ref{eq:NM} and \ref{eq:dAdt}, rather
than the general expression in Eq.~\ref{eq:dNlensdt}, which is only valid for
the ML-relation in Eq.~\ref{eq:masslumrelations} and Table~\ref{tab:tab2} for
Pop III stars.}
\end{deluxetable*}


\mn \subsection{Evolution in Metallicity and Massive Star Duplicity}
\label{sec52}

\sn Since we do not know the duplicity nor the separation distribution of Pop
III stars, nor of the first polluted O-stars in mini-halos, we need to consider
a range of possibilities. Pop III BHs with 5\cle M\cle 720 \Mo\ may accrete more
steadily via Roche-lobe overflow from a (slightly polluted) Pop II.5 companion
star of lower ZAMS mass, as discussed in \S \ref{sec32}. 

The second scenario is much common for O-stars nearby, given their very high
multiplicity, but may not be common for Pop III stars at z\cge 7.
\citet{trenti_2009} suggest that as soon as a massive Pop III star first forms
in a mini-halo, its powerful Lyman-Werner UV-radiation field may prevent lower
mass Pop III stars to form in its immediate surroundings. Self-shielding by
very dense surrounding hydrogen gas against this UV radiation may allow some
neighboring lower mass Pop III stars to still form. \citet{trenti_2009}
therefore also discuss mini-halos that may have more than one Pop III star. 
In their models, Pop III stars generally start forming at z \cle 30--40 (cosmic
ages \cge 99--65 Myr, respectively), followed by slightly polluted Pop II.5
stars that quickly ramps up at z\cle 28--35 (cosmic ages \cge 109--79 Myr,
respectively), or about $\sim$10--15 Myr later in cosmic time.
\citet{sarmento_2018} present hydrodynamical simulations that narrow the Pop
III star redshift range from z$\simeq$20 to z$\simeq$7. In their models,
pristine Pop III stars are still the dominant population at z$\simeq$20, while
at z$\simeq$7, slightly polluted (Z\cle 10$^{-4}$ \Zo) ``Pop II.5'' stars
outnumber Pop III stars by $\sim$10:1. In other words, Pop III stars may have
polluted their surroundings quickly enough that within 10--15 Myr, many lower
mass stars that formed in their neighborhood already have somewhat non-zero
metallicities. 

Comparing the estimated pre-MS lifetimes (\taupreMS) to the \texttt{MESA}
ZAMS--AGB lifetimes in Table~\ref{tab:tab1}, we found in \S \ref{sec31} that
Pop III stars with M$\simeq$20--1000 \Mo\ live short enough (\cle 8 Myr) that
they may have polluted the material from which stars with M$\simeq$1--1.5 \Mo\
formed at z\cge 7, since their pre-MS lifetimes are longer than 6--9 Myr.
Hence, it is possible that most early low-mass stars (M$\simeq$1--1.5 \Mo) may
have been polluted by massive Pop III stars as early as z\cle 20, and
certainly at lower redshifts down to z$\simeq$7. This then also means that it
is possible that very low metallicity Pop II.5 stars may have formed at z\cle 
20 in the vicinity of Pop III stars, perhaps some close enough to form binaries
or multiple star systems with those Pop III stars. In any case, the first
polluted O-stars likely also appeared at z\cle 20. For the latter, the
duplicity fraction may have quickly increased from the very low duplicity
values expected for true zero metallicity Pop III stars --- with lower-mass
companions not forming due to their significant LW radiation
\citep{trenti_2009} --- to the much higher duplicity fraction seen in O-stars
today (see \S \ref{sec32}). 

The metallicity evolution of stellar populations is not well known at high
redshifts \citep[z\cge 4;][]{maiolino_2008, kim_2017}. \citet{trenti_2007} and
\citet{sarmento_2018} suggest that mini-halos and the IGM can get quickly
enriched (to Z$\simeq$$10^{-4}$ \Zo) by a progenitor Pop III SN. The
hydrodynamical simulations of \citet{sarmento_2018} use Adaptive Mesh
Refinement (AMR) to sample the mass range of M$\simeq$10$^{5.5}$--10$^8$ \Mo\
over the redshift range of z$\simeq$8 to z$\simeq$16, where their predicted
metallicities range from Z$\simeq$0.1 \Zo\ at M$\simeq$10$^8$ \Mo\ to
Z$\simeq$0.003 \Zo\ at M$\simeq$10$^{5.5}$ \Mo. Over this mass and redshift
range, their mass-metallicity relation has a slope of
$\Delta$log(Z/\Zo)/log(M/\Mo)$\simeq$0.5--0.6. At masses below 
M$\sim$10$^{5.5}$ \Mo, their AMR simulations have insufficient mass resolution,
but {\it if} the mass-metallicity were to continue with this slope to single
stellar masses as low as M\cle 10$^3$ \Mo, then the non-pristine stars at z\cge
7 could indeed have metallicities as high as Z$\simeq$10$^{-3.5}$ \Zo.
\citet{madau_2017} suggest that at z$\simeq$7--10 the metallicity of massive 
(M$\sim$$10^8$ \Mo) star-forming objects may be as high as 0.03--0.1 \Zo. For
the low mass environments in which slightly polluted Pop II.5 stars form, a
metallicity of Z\cge 10$^{-4}$\Zo\ is thus plausible. 

Recent observational \citep[\eg][]{badenes_2017} has shown that metal-poor
(Z\cle 0.3 \Zo) stars have a multiplicity fraction 2--3$\times$ higher than
metal-rich (Z$\sim$\Zo) stars. Theoretical work on star-formation
\citep[\eg][]{machida_2009} suggested a higher binary frequency in lower
metallicity gas, and that a majority of stars are born as members of
binary/multiple systems for Z\cle 10$^{-4}$\Zo. Hence, for non-zero
metallicities at least the binary fraction increases with decreasing 
metallicity. Physically, this occurs because metal-line cooling becomes
significant above a threshold of Z\cge 10$^{-4}$\Zo, which decreases
fragmentation of the gas clouds that form the stars. We do not know if this
trend continues to hold for truly zero metallicity Pop III stars at z\cge 7. But
it seems possible that any non-zero metallicity massive star will form and
evolve in an environment with a significant binary fraction \citep[see \eg][for
a discussion]{adams_2006, adams_2010}. 

What matters for the current work is that, while some massive stars with zero or
very low metallicity may still exist at z$\simeq$7, at the same time a
sufficient fraction of polluted stars (Z\cge 10$^{-4}$ \Zo) already exists at
z$\simeq$7--17. The latter are critical, since they likely formed with a 
significant fraction of binaries, and so play an essential role in BH
accretion disk feeding via Roche-lobe overflow during its post-main sequence
evolution. 

Mass transfer is not currently considered in the \texttt{MESA} code. Future
work needs to include detailed star formation scenarios with full metallicity
evolution in the ISM at z\cge 7, their subsequent evolutionary tracks at very
low metallicities, and include scenarios of binary evolution that incorporate
mass transfer, and address how mass transfer affects the BH-feeding timescales.

\mn \subsection{Range in Stellar Mass Black Hole Accretion Lifetimes}
\label{sec53} 

\sn We will consider here that any BHs left over after a massive Pop III star's
death may accrete from a surrounding lower-mass, low-metallicity star filling its
Roche lobe during its post-main sequence evolution, causing a UV-bright accretion
disk. The accretion time scales onto these BHs in stellar binaries are not well
known, but may have plausibly lasted as long as the GB lifetimes of the less
massive star in a binary when it fills its Roche lobe. Following the Pop III MS
lifetimes from \S \ref{sec31} and Table~\ref{tab:tab1}, this can happen within
\cle 12\% of their MS-ages, or within 0.3--60 Myr after the first SN of the more
massive star in the pair has occurred. The question then arises: how often can
this scenario have happened for Pop III stars at z\cge 7, whose stellar-mass
BH-remnants would still be around today as leftovers from the First Light epoch? 

If a fraction $(1-\epsilon)$ of the matter is accreted at the Eddington rate,
where $\epsilon$ denotes the radiative efficiency, then the mass of the black
hole will increase exponentially with a characteristic timescale of $t_E =
4\pi G\mu m_p/(\sigma_e c \epsilon)\simeq 45$ Myr. If all remnants of Pop III
stars accreted at the Eddington rate for \cge $10^8$ years, then this would
increase the black hole mass by orders of magnitude, which would increase the
mass density of black holes to values that are excluded by constraints on the
present-day mass density of black holes \citep[see \eg][]{tanaka_2012}. Steady
BH feeding from accretion disks for \cge $10^8$ years would have likely given
rise to black holes that will grow catastrophically to $>$$>$$10^{2}$ \Mo, and
may quickly produce {\it massive} black holes with M\cge 10$^3$--10$^5$ \Mo\ or
more, and become Ultra-Luminous X-ray sources (ULX). While UV-bright accretion
disks around such massive BHs would be more easy to detect by JWST during
caustic transits (see \S \ref{sec44} \& \ref{sec62}), they will likely also be
much more rare. 

For Pop III stellar-mass BH accretion disks, we will therefore consider
lifetimes of at least \cge 0.3 Myr from the massive binary argument in \S
\ref{sec322}. In \S \ref{sec55}, we will assume that the BH accretion disks
are constantly feeding at the luminosities predicted for {\it maximum}
lifetimes of \cle 60 Myr, during which the lowest-mass (M\cge 2.0 \Mo)
companion AGB stars would fill their Roche lobes before reionization is
complete at z$\simeq$7 (\S \ref{sec321}). Given the uncertain accretion times,
the BH accretion disk UV-luminosities derived in \S \ref{sec55} are upper
limits, so their caustic transit rates in \S \ref{sec62} are lower limits. 

\mn \subsection{Efficiency of Massive Pop III Star Black Hole Accretion Disks} 
\label{sec54}

\sn Following the arguments of \S \ref{sec35}, if N\cle 10$^3$ massive Pop III
stars per \arcsecsq\ contribute to the near-IR sky-SB of AB\cge 31 \magarc\ at
2.0 \mum, then a large fraction ($f_{\rm BH}$) of them will leave behind BHs.
Accretion onto these BHs will give rise to additional flux in the IRB. The
ratio of the Pop III to Pop III remnant contribution can be estimated from: 

\begin{equation} 
\frac{S_{\rm Pop III}}{S_{\rm BH}}=\frac{1}{f_{\rm BH}}\frac{t_{\rm Pop
III}}{t_{\rm acc}}\frac{L_{\rm Pop III}}{L_{\rm BH}}.
\label{eq:Lbol}
\end{equation} 

\n Fig.~\ref{fig:fig2} shows that a 300 \Mo\ Pop III star has a luminosity of 
\Lbol$\simeq$2.5$\times 10^{40}$ erg/sec, which agrees quite closely with the
Eddington luminosity associated with an almost equal mass BH, which is
$\sim$10$^{40}$ erg/sec following Eq.~\ref{eq:Lbol}. If we assume that the
fraction $f_{\rm BH}$ of Pop III stars that collapses into BHs produces a black
hole of $\sim$15--70\% of the original stellar ZAMS mass, then we expect
$L_{\rm Pop III}\simeq L_{\rm BH}$ at least at early times. We then obtain: 

\begin{equation}
\frac{S_{\rm Pop III}}{S_{\rm BH}}=\frac{1}{f_{\rm BH}}\frac{t_{\rm Pop
III}}{t_{\rm acc}}.
\end{equation} 

\n The efficiency of gas accretion onto stellar-mass BHs formed by Pop III
stars is discussed by \citet[][]{milosavljevic_2009}. It is possible that
radiative feedback seriously limits the efficiency of gas accretion.
Time-averaged Eddington ratios of $\sim 1\%$ have been reported by, \eg\ 
\citet{park_2012}, although this ratio could be smaller. If accretion occurs 
during the typical $\sim$0.3--60 Myr adopted for early massive binaries, then
these accretion times are less than 1--10\% of the available Hubble time at
z\cge 7. Hence, BHs may have been feeding with a duration of \cle 1--10\% of
the total available time. 

If we take into account that the mass of the BH grows with time, then it is
plausible that $(S_{\rm Pop III}/S_{\rm BH})$\cle 1, \ie\ the remnants of Pop
III stars may contribute more to the near-IR sky-SB than the Pop III stars
themselves. If Pop III remnants form the seeds for supermassive black holes,
including the {\it rare} $M_{\rm BH}\simeq 10^9$ \Mo\ black holes that are seen
in quasars at z\cge 6 \citep{willott_2003, jiang_2007, kurk_2007}, then at
least a small fraction of them must accrete at practically the Eddington rate
with a duty cycle of $\sim$$100\%$ \citep[\eg][]{willott_2010}. If a small
fraction of the remnants accrete so efficiently, then it is not unexpected that
a much larger fraction will accrete with duty cycles intermediate between $1\%$
and $100\%$. Depending on how large the fraction of more slowly accreting BHs
is, this population could contribute significantly more to the near-IR sky-SB
and to caustic transits than the Pop III stars themselves. 

\subsection{Stellar Mass Black Hole Accretion Disk Radii and Luminosities}
\label{sec55} 

\sn Pop III stars with masses M$\simeq$30--1000 \Mo\ can leave BHs behind with
M$\simeq$5--720 \Mo\ (\S \ref{sec51}), except for the mass range around
100--200 \Mo\ where they seem to produce no BHs \citep{woosley_2017}. The
Schwarzschild radii of these Pop III BHs will thus be in the range of
\Rs$\simeq$15--2200 km, as listed in Col. 3 of Table~\ref{tab:tab5}. Using this
range of BH masses and Schwarzschild radii, this section summarizes available
constraints on the resulting sizes and luminosities of stellar-mass BH
accretion disks. Since these parameters are more uncertain than those of Pop
III stars, we will estimate them in two independent ways to permit a
consistency check. The resulting UV-accretion disk radii, bolometric
luminosities, and corresponding \MAB-magnitudes are listed in Cols. 3--6 of
Table~\ref{tab:tab5}, which are described for both methods in the next two
sub-sections. 

\sn \subsubsection{Estimates by Scaling from Observed Microlensed Quasar
Results}
\label{sec551}

\sn A first estimate of \Raccr\ and \Laccr\ can be made from observed
microlensing results on strongly-lensed quasars at z$\simeq$1--2 by
\citet{blackburne_2011}. These authors present accretion disk sizes,
temperatures, and luminosities from their quasar images that were monitored
extensively with ground-based telescopes and through Chandra X-ray fluxes.
Their Eq.~(2) gives a simple relationship between accretion disk half-light
radius (\rhl\ or \Raccr), the quasar SMBH mass, and the observed wavelength,
which in their case is the observed optical that samples restframe $\sim$2500
\AA: 

\begin{equation}
R_{\rm accr} \simeq 1.68 \times 10^{14}\ (\frac{M_{\rm BH}}{10^9\
M_\odot})^{2/3}\ (\frac{\lambda}{{\mu}m})^{4/3}\ \ m.
\label{eq:Raccr}
\end{equation}

\n We rescale this for the JWST NIRCam near-IR filters F115W--F277W, which
sample Pop III objects at z$\simeq$7--17 approximately in the restframe UV at
$\lambda$$\simeq$1500 \AA. From their multi-color microlensing photometry,
\citet{blackburne_2011} derive SMBH masses of order (0.04--2)$\times$10$^9$
\Mo\ and bolometric luminosities in the range of \Lbol$\simeq$(0.1--4)
$\times$10$^{46}$ erg/sec. Their Table~8 suggests that for all 12 quasars the
bolometric accretion-disk luminosity scales with SMBH mass approximately as:

\begin{equation} 
L_{\rm bol} \simeq 3.2 \times 10^{46}\ (\frac{M_{BH}}{10^9\ M_\odot})\ erg/sec.
\label{eq:Lbolaccr}
\end{equation}

\n For a solar luminosity of 3.828$\times$10$^{33}$ erg/sec, this corresponds
to quasar accretion disk luminosities of $\sim$(0.3--10)$\times$10$^{12}$\ \Lo. 

These are remarkable direct constraints to quasar restframe UV-accretion disk
sizes and their luminosities. We do not know if we may scale these values down
from their observed mass range to our range of BH masses of $\sim$5--720 \Mo\
adopted in Table~\ref{tab:tab5}. \citet{blackburne_2011} suggest from the data
over their mass range, their half-light radii scales as: 

\begin{equation}
r_{\rm hl} \propto M_{\rm BH}\ ^{\rho}, 
\label{eq:RUVvsM}
\end{equation}

\n with a best fit of $\rho$$\simeq$0.27$\pm$0.17. This is flatter than the
$\rho$=2/3 slope implied by multi-color accretion disk theory in
Eq.~\ref{eq:Raccr}. If we scale our UV accretion-disk radii down with
$\rho$$\simeq$0.27 from their SMBH mass range, then we obtain very large radii
(\RUV\cge 10$^3$ \Ro) and luminosities for Pop III stellar-mass BH accretion
disks. This suggests that the flat $\rho$-slope derived from their quasar
sample may not hold down to Pop III BH masses, as may be caused by the strong
dependence on black hole mass of the tidal forces around each BH. We therefore
adopt a slope between these values of $\rho$$\simeq$0.5, which is consistent
with the \citet{blackburne_2011} value within their errors, and still provides
a good fit to their data given the small dynamic range in \MBH\ in their
sample. In \S \ref{sec552} we suggest that $\rho$$\simeq$0.5 produces more
consistent overall results for the multi-color thin-disk accretion model. When
we scale the \citet{blackburne_2011} UV accretion-disk radii down with
$\rho$$\simeq$0.5, then we obtain the BH UV half-light radii listed in the top
tier of Table~\ref{tab:tab5}. These are in the range of \RUV$\simeq$1--16 \Ro\
for \MBH$\simeq$5--720 \Mo. The bolometric luminosities listed in the top tier
of Table~\ref{tab:tab5} were scaled down directly with Eq.~\ref{eq:Lbolaccr},
and are in the range of 4$\times$10$^4$--6$\times$10$^6$ \Lo\ for
\MBH$\simeq$5--720 \Mo, respectively. 

\sn \subsubsection{Estimates from Multi-Color Accretion Disk Theory}
\label{sec552} 

\sn In this section, we compare the stellar-mass BH accretion disk sizes and
luminosities as scaled down from the quasar observations in \S \ref{sec551} to
theoretical estimates. 

In the simplest form, accretion disks around black holes are assumed to be
``multi-color'' thin disks, which consist of a series of concentric shells each
of which emit black body radiation characterized by its radial dependent
temperature \citep[\eg][]{shakura_sunyaev_1973, remillard_2006,
blackburne_2011}. In the restframe UV-optical (at $\nu_{Ly\alpha}$\cle 
2.466$\times$10$^{15}$ Hz), the spectrum of the accreting BH is dominated by
the thermal disk component. In the very inner part of the accretion disk, other
radiation mechanisms will likely produce significant X-ray emission, such as
synchrotron radiation in the presence of strong central magnetic fields,
inverse Compton radiation, or thermal bremsstrahlung
\citep{shakura_sunyaev_1973, shakura_sunyaev_1976}. Only the harder part of
this redshifted X-ray emission will make it past the IGM and potentially be
detected by Chandra, but what matters for any JWST detections is the amount of 
associated restframe UV-emission that makes it past the IGM at $\lambda$\cge 
1216 \AA. In the multi-color thin accretion-disk model, the temperature
increases with radius as:

\begin{equation}
T \propto r^{-3/4}. 
\label{eq:TvsR}
\end{equation} 

\n Using Eq.~\ref{eq:RUVvsM}, gas on the inner most stable orbit at
R$\simeq$3\Rs\ has a maximum temperature of about: 

\begin{equation}
T_{max}\ \simeq\ 10\ (\frac{M_{BH}}{100\ M_\odot})^{-\tau}\ keV. 
\label{eq:TmaxvsM}
\end{equation} 

\n Standard thin-disk accretion theory suggests a slope of $\tau$=1/4, but
since we adopted $\rho$$\simeq$1/2 in Eq.~\ref{eq:RUVvsM}, we need to use
$\tau$=3/8 here to maintain consistency with Eq.~\ref{eq:TvsR}. The multi-color
accretion disk models predict similar UV half-light radii \RUV for either slope
$\tau$, since the largest SED differences occur well below restframe 1216 \AA,
and this part of the SED does not make it past the IGM at z\cge 7. 

We will assume here that the maximum temperature of the inner accretion disk in
Eq.~\ref{eq:TmaxvsM} for a 100 \Mo\ BH needs to be at least 10 keV, or
\Tmax$\simeq$3.87$\times$10$^7$ K. This is so that their hard X-ray photons 
can make it past the neutral hydrogen at z\cge 7 \citep{haardt_2012}, {\it and}
when redshifted from z$\simeq$7--17, still be in principle observable in the
Chandra soft X-ray band, which covers 0.5--2.0 keV. This argument is based on
the following: if part of the Spitzer--Chandra cross-correlation power-spectrum
signal \citep{cappelluti_2013, mitchell-wynne_2016} came from redshift z\cge 7,
then the sources that cause it must be {\it both} Spitzer 3--4 \mum\ {\it and}
Chandra X-ray sources, as discussed in \S \ref{sec232}. Both these papers
discussed PBHs as possible candidates for the Spitzer--Chandra
cross-correlation signal. As discussed in \S \ref{sec31}, Pop III stars alone
cannot cause this Spitzer--Chandra cross-correlation signal, since they reach
only a maximum temperature of \Teff\cle $10^5$ K. 

As we move out in radius, the temperature drops as in Eq.~\ref{eq:TvsR}. We
assume that each concentric radius interval in the multi-color accretion disk
emits as a black body with its own temperature. The largest radius that will
contribute to the UV-emission is the one where the black body curve peaks in
the UV longwards of Ly$\alpha$. For our 1500 \AA\ restframe UV-reference, this
largest ring needs to have a temperature of T\cge 3.2$\times$10$^4$ K. This
suggests that we need to go out in radius where the temperature is a factor of
\cge 1200$\times$ lower than in the inner ring. Hence, we need to integrate
out to r$\simeq$13,000 $R_{min}$, where \Rmin$\simeq$3 \Rs\ is the radius of
the innermost stable orbit around the BH. Plugging in the numbers above then
yields \Raccr(UV)$\simeq$1.6$\times$10$^7$ km for the maximum radius of the
UV-emitting region, or $\sim$17 \Ro\ for \MBH= 100 \Mo. Integration of the
{\it actual} multi-color thin-disk light-profiles for a 100 \Mo\ BH yields a
{\it half-light radius} \rhl that is about 1.7$\times$ smaller than this, as
shown below.

We use the multi-color accretion disk model in
Eq.~\ref{eq:TvsR}--\ref{eq:TmaxvsM} for the Pop III BH mass range of 5--720
\Mo\ in Table~\ref{tab:tab5} to predict their UV half-light radii r$_{hl}$, 
their bolometric, and their UV-luminosities. Their UV half-light radii \rhl\
are then simply integrated from the part of the radially-dependent 
UV-accretion disk SED that makes it past the IGM at z\cge 7. These results are
listed in the bottom tier of Table~\ref{tab:tab5}, and shows UV half-light radii
in the range of \RUV$\simeq$2--30\ \Ro. 

At these r$_{hl}$-values, the multi-color accretion disks have an effective
temperature of \Teff$\simeq$47,500--48,000 K for M$\simeq$5--720 \Mo.
Bolometric+IGM+K-corrections were applied to the bolometric luminosities in
Table~\ref{tab:tab5}, as for Pop III stars in \S \ref{sec33}. For multi-color
accretion disks with \Teff$\simeq$47,700 K, these combined BIK-corrections are
--0.3, --1.1, and --1.5 mag at z=7, z=12, and z=17, respectively. These are
comparable to the values in Tables \ref{tab:tab3}--\ref{tab:tab4} for Pop III
RGB and AGB stars with {\it monochromatic} black body disks of similar
temperatures. The bolometric luminosities predicted for the multi-color thin
accretion disks are in the range of 3$\times$10$^4$--7$\times$10$^6$ \Lo\ for
\MBH$\simeq$5--720 \Mo, and are listed in the bottom tier of
Table~\ref{tab:tab5}. 

To check our multi-color accretion disk models for consistency, we first
verified that they reproduce the \citet{blackburne_2011} UV half-light radii 
obtained for accretion disks of z$\simeq$1--2 quasars from Eq.~\ref{eq:Raccr}.
Second, we apply our multi-color accretion disk model to M$\simeq$10$^9$ \Mo\
SMBHs known to be present in quasars at z\cge 6 \citep[\eg][]{willott_2003,
jiang_2007, kurk_2007}. The above equations imply an UV accretion-disk {\it
diameter} of 2\RUV$\simeq$2$\times$10$^5$ \Ro\ for a 10$^9$ \Mo\ SMBH, which is
$\sim$1000 AU or 0.05 pc across, corresponding to a light travel time of \cle 2
days in the restframe. This is comparable to the accretion-disk sizes of QSOs
inferred from variability studies, where the somewhat larger Broad Line Region
can be lightdays--weeks across \citep[\eg][]{butler_2011, kozlowski_2010}. Our
multi-color thin accretion disk model also predicts the unobscured restframe
UV-luminosity for the rare quasars with a 10$^9$ \Mo\ SMBH at z\cge 6, which is
\MUV $\simeq$--27 AB-mag \citep{fan_2001, fan_2003}. Hence, their BIK-corrected
near-IR fluxes are predicted to be \mAB\cge 21 mag at z\cge 7, which is
comparable to what is observed for the highest redshift quasars 
\citep[\eg][]{mortlock_2011}. This extrapolation to QSOs at z\cge 6 then
justifies the slightly modified choices of $\rho$$\simeq$1/2 (instead of 2/3)
and $\tau$$\simeq$3/8 (instead of 1/4) above. 

In summary, Table~\ref{tab:tab5} shows that both estimates of \RUV and \Lbol\
of Pop III stellar-mass BH accretion disks in \S \ref{sec551} and in this
section are within a factor of two or less. We will therefore adopt the two
scaling methods in the equations above, and assume that the resulting range of
properties in Table~\ref{tab:tab5} capture the properties of Pop III BH
UV-accretion disks sufficiently well to make an order-of-magnitude estimate of
the cluster caustic transit rates for Pop III BH accretion disks. 

Given the unknown accretion efficiencies compared to Eddington, or the unknown
accretion lifetimes compared to the maximum accretion lifetimes possible (\S 
\ref{sec31}), the values in Table~\ref{tab:tab5} are upper limits to the Pop
III BH UV-accretion disk luminosities. That is, the luminosities and
resulting \Mbol\ and \mAB-values in Table~\ref{tab:tab5} assume that BH
accretion disks radiate at {\it steady-state} levels inferred by the
multi-color accretion-disk model for maximum lifetimes as discussed in \S 
\ref{sec53}. 

In conclusion, the inner stellar-mass BH accretion disks may be significantly
hotter than the typical T$\simeq$10$^5$ K temperatures of Pop III stars,
plausibly reaching X-ray temperatures at the innermost radii, and reaching
$\sim$30,000 K at the outermost radii. Their UV-bright accretion disks --- if
unobscured by surrounding dust --- have SEDs that can make it in part through the
neutral IGM at z\cge 7 with UV radii \cle 40,000 \Rs. Their restframe UV-radii
are \RUV$\simeq$1--30\ \Ro, and their UV-luminosities are at most
3$\times$10$^4$--7$\times$10$^6$ \Lo\ for M$_{BH}$$\simeq$5--720 \Mo,
respectively. Pop III stellar-mass BH accretion disk radii may thus be similar
to, or somewhat larger than the 1--13 \Ro\ radii of the ZAMS Pop III stars in
Tables~\ref{tab:tab1}--\ref{tab:tab2}, but no larger than the Pop III RGB- or
AGB-star radii in Tables~\ref{tab:tab3}--\ref{tab:tab4}. They would fit well
within the $\sim$7--55 \Ro\ Roche lobe sizes seen in massive binaries discussed
in \S \ref{sec322}, and so are eligible for feeding from a less massive RGB/AGB
star in the binary that is filling its Roche lobe. This assumes that subsequent
generations of (slightly) polluted massive stars at z\cge 7 already have high
enough metallicity to form binaries. The predicted stellar-mass BH accretion disk
UV-radii and maximum luminosities are similar to those of Pop III RGB--AGB stars
in the 10--300 \Mo\ range. We use this to estimate BH accretion disk caustic
transit time and rates in \S \ref{sec62}. 

\subsection{White Dwarfs and Neutron Stars from Low-Mass Pop III Stars}
\label{sec56}

For completeness, we will briefly consider here the potential impacts of White
Dwarfs (WDs) that likely result from low-mass stars (M\cle 5 \Mo) at z\cge 7
\cite[see \eg\ the Z=5$\times$10$^{-3}$ \Zo\ sample of][]{romero_2015}, and of
Neutron Stars (NS) that likely result from Pop III stars at ZAMS masses M\cle
20 \Mo, since both will be far more common than Pop III stellar-mass BHs (see
Table~\ref{tab:tab5} and the IMF-slopes in Fig.~\ref{fig:fig3}). 

Table~\ref{tab:tab1} implies that NSs would not appear until 8--70 Myrs after 
their progenitor stars with 5\cle M\cle 20 \Mo\ form, while WDs would appear at
least \cge 230 Myr after their progenitor stars with M\cle 5 \Mo\ form at
z$\simeq$7--17. If the first stars form at z$\simeq$35--40, then the first NSs
would thus appear soon thereafter, but the first WDs not until z\cle 14. In
either case, the first NS or WD mergers at z\cge 7 would have only 500--700 Myr
to occur. Hence, we will not consider NS--NS mergers such as recently found by
LIGO \citep{abbott_2017_b, abbott_2017_c} and identified by ground-based
follow-up campaigns \citep[\eg][]{chornock_2017,cowperthwaite_2017} in a nearby
galaxy, nor potential NS--WD or WD--WD mergers, as these are far more rare than
regular accretion onto either a WD or a NS. 

The duration of regular accretion onto WDs or NS before they undergo a nuclear
explosion on their surfaces depends mainly on their accretion rates. These in
turn depend on the binary separation, masses of the two components,
evolutionary state of the companion, and the nature of the explosion. For
white dwarfs (novae and super-soft X-ray sources), the recurrence timescales
are $\sim$20 to $\sim$10,000 years \citep{shara_1986, cannizzo_1988, wolf_2013,
henze_2015, shafter_2015, shafter_2017}, and are likely too rare to average out
to a flux that could be detected during a cluster caustic transit. For neutron
stars (X-ray bursters), the recurrence timescales can be hours to weeks
\citep{tanaka_1996, watts_2012}. Their luminosities when averaged over \cge 0.1
year at z\cge 7 would determine if such objects could be seen via caustic
transits by JWST. In all cases, their surface layers explode, after which they
may resume accretion, and may approach their previous steady-state luminosity.
A proper description of WD and NS accretion will thus not only require the
multi-color thin-disk models that we use for BH accretion disks in \S
\ref{sec552}, but also a quantitative modeling of these episodic nuclear
detonation events, which is beyond the scope of the current paper. Future work
will need to consider if accretion onto Pop III NS or WDs can be steady enough
and luminous enough to be a source of caustic transits that is potentially
observable by JWST. 

\bn \section{Estimates of Caustic Transits for Pop III Star Black Hole
Accretion Disks} 
\label{sec6}

\sn In this section, we discuss the possible effects of dust produced by Pop
III stars, and then present our estimates of the cluster caustic transit rates
resulting from stellar-mass BH accretion disks as described in \S \ref{sec5}. 

\mn \subsection{Possible Effects from Dust generated by Pop III Stars}
\label{sec61}

\sn True zero metallicity massive stars, by all modeling investigations to date,
have significantly reduced mass loss. The normal driver of massive-star winds
--- radiation pressure from scattering off metals, is not present. 
Alternatives such as rotational mixing, some dredge-up scenario to bring core
material to the photosphere, or (epsilon- and kappa-) pulsation mechanisms, are
too weak to cause much mass loss \citep{castor_1975, gotberg_2017, renzo_2017}.
Thus, zero metallicity massive stars may not be shrouded by dusty
circumstellar material. For metallicities of Z\cle 10$^{-4}$ \Zo\ (or even
\cle 10$^{-5}$ \Zo) the winds (hence dust) will be at levels more common for
massive stars seen nearby, although still significantly reduced.

If Pop III stars --- or the slightly polluted Pop II.5 stars --- did manage to
produce stellar winds during their main sequence and Blue-Red Supergiant
(BSG-RSG/AGB) phases, this could have deposited dust into the surrounding
medium. When a fraction of Pop III stars goes off as Pair Instability
SuperNovae (PISN), they would deposit additional metals into their immediate
surroundings. Dust formation in the circumstellar material of initially zero
metallicity Pop III stars could thus have added a non-trivial
extinction/reddening factor, especially in their late stellar evolution and
subsequent BH accretion disk stages. Hence, we should consider possible cases
where either Pop III stars or their stellar-mass BH accretion disks are
significantly reddened by dust, or both. 

For non-rotating Pop III stars, this dust could be distributed rather uniformly
and obscure most of the Pop III stars and their BH accretion disks, but for
rotating stars, the situation may be quite different. We do know that Gamma Ray
Bursters (GRBs) are quite visible from $\gamma$--ray to radio waves when viewed
from the right direction. The same is true for unobscured vs. obscured AGN ---
much of their visibility is viewing-angle dependent with respect to the dust
torus. We therefore must consider that at least a fraction of Pop III stars
with significant stellar rotation produced BH accretion disks that are visible
under certain viewing angles, and produce an equal amount of UV-continuum
radiation as the Pop III stars themselves, or perhaps more. Evolving rotating
Pop III star models, dust production, and their likely non-uniform
dust-expulsion mechanism are currently too uncertain to take into account in
the model calculations, and will require more detailed numerical modeling in
future work. 

If both Pop III stars and their stellar-mass BH accretion disks were fully
unobscured, then the average \cge 2$\times$10$^6$ year MS lifetime of Pop III
stars (\S \ref{sec31}) and maximum BH accretion disk lifetimes --- as visible in
the restframe UV --- of \cle 60 Myr would determine their visible ratio. Some
fraction of Pop III BH accretion disks may not be fully obscured, as would be
implied by the Spitzer--Chandra power spectrum results in \S \ref{sec232},
{\it if} some of this signal came from z\cge 7 \citep{cappelluti_2013,
cappelluti_2017}. In reality, nature may have well produced some observable
combination of obscured and unobscured Pop III stars and their BH accretion
disks, as it has for the iEBL from spheroids, disks, and unobscured AGN at
lower redshifts in Fig.~\ref{fig:fig1}. For that reason, we allowed the maximum
sky-SB of \cge 31.0 \magarc\ of \S \ref{sec23} to be {\it either} fully caused
by Pop III stars {\it or} by their BH accretion disks, or by a combination of
the two not exceeding this SB-level. JWST may be able to distinguish between
the two through {\it chromatic} effects of caustic transits, as discussed in \S
\ref{sec72}. 

\mn \subsection{Implied Estimates of Cluster Caustic Transits for Pop III Star
Black Hole Accretion Disks without Microlensing}
\label{sec62}

\sn In this section, we present estimates of the cluster caustic transit rates
resulting from stellar-mass BH accretion disks as described in \S \ref{sec5}.
To first order, for Pop III stellar-mass BHs, the same principles apply as
above, so unless stated otherwise, we use the equations in \S \ref{sec44}. 

As discussed in \S \ref{sec552}, the expected BH accretion disk radii are
similar to, or somewhat larger than, the 1--13 \Ro\ radii of the ZAMS Pop III
stars in Tables~\ref{tab:tab1}--\ref{tab:tab2}, but no larger than Pop III RGB
or AGB star radii in Tables~\ref{tab:tab3}--\ref{tab:tab4}. The maximum BH
accretion disk luminosities are in general similar to those of Pop III RGB--AGB
stars in the 10--100 \Mo\ range, or $\sim$10$^4$--10$^7$ \Lo. 

Pop III stellar-mass BH accretion disks --- when lensed through cluster caustic
transits --- thus also have rise times of order one to several hours. For their
similar luminosities, the decline times will then be also of the order of a
year, as discussed in \S \ref{sec44}. These, together with their transit rates
predicted for the Pop III BH radii and luminosities in \S \ref{sec55} are
listed in Table~\ref{tab:tab5}. 

For a Pop III ZAMS mass function slope of $\alpha$$\simeq$2, the weights for
each of the mass bins for BH accretion disks in Table~\ref{tab:tab5} are very
similar, following Eq.~\ref{eq:dNlensdtalpha}. The resulting {\it total} transit
rates for Pop III stellar-mass BH accretion disks with \MBH\cge 24--720 \Mo\
that are in principle observable with JWST to AB\cle 28.5--29 mag across the
caustics are predicted to be \cge 0.18 per cluster per year for the top tier
in Table~\ref{tab:tab5}, and \cge 0.24 per cluster per year for the bottom tier,
respectively. 

Because the luminosities and the resulting \Mbol\ and \mAB-values in
Table~\ref{tab:tab5} are upper limits (\S \ref{sec55}), the inferred BH
accretion-disk transit-rates are lower limits, as indicated in
Table~\ref{tab:tab5}. I.e., if the actual accretion efficiencies were 10$\times$
lower, then the BH accretion luminosities would be $\sim$2.5 mag fainter, and
the caustic transit rates could be several times higher. This is because 
there would be 10$\times$ as many faint objects per mass bin that make-up the
near-IR SB adopted in \S \ref{sec23} that contribute to caustic transits above 
the detection limit, but there would also be fewer mass bins contributing above
the JWST detection limit. 

The limits to the caustic transit rates of stellar-mass BH accretion disks of
$\sim$0.2 per cluster per year are similar to caustic transit rate of
$\sim$0.32 per cluster per year obtained for Pop III ZAMS+RGB+AGB stars (\S
\ref{sec44}). For BHs, they could be several times higher, depending on their
actual accretion efficiency. 

We briefly discuss this in the context of the lifetime differences between Pop
III stars and their stellar-mass BH accretion disks that could affect the mix
of caustic transits JWST may observe. In Table~\ref{tab:tab5}, BHs with
UV-accretion disks bright enough to be detected by JWST during caustic transits
have M$_{BH}$$\simeq$24--720 \Mo\ and AB$\simeq$37--42 mag, respectively. Pop
III stars with 30\cle M\cle 1000 \Mo\ that produce BHs have ZAMS ages of
5.6--2.1 Myr (Table~\ref{tab:tab1}) with an average of $\sim$3 Myr. Pop III
stars of masses M$\simeq$2--20 \Mo\ live considerably longer than this during
their AGB stage, where they could fill their Roche lobes for up to 0.6--60 Myr,
with an IMF-weighted average GB age of $\sim$6 Myr. Hence, during their AGB
stage 2--20 \Mo\ stars could feed the BH that is leftover from a 30--1000 \Mo\
star for a maximum duration that is significantly longer than the ZAMS lifetime
of the massive Pop III star that produced this BH. 

In summary, depending on how steady and efficient BH feeding by a lower mass
AGB star in its Roche lobe is, stellar-mass BH accretion disks may be about as
likely as Pop III stars at z\cge 7 to cause cluster caustic transits that could
be observed by JWST, and possibly more likely. Stellar-mass BH accretion
disks with a SB$\simeq$31.0 \magarc\ (or $\sim$1 \nWsqmsr) could produce about
one caustic transit per 5 clusters per year, and perhaps as many as one event
per 2 clusters per year. As for the Pop III stars in \S \ref{sec44}, a
dedicated JWST program that monitors 3 clusters per year for a number of years
could possibly detect several caustic transits for stellar-mass BH accretion
disks. If their SB were to be as dim as $\sim$36.0 \magarc, which corresponds
to $\sim$10 Pop III BH accretion disks per \arcsecsq\ (see
Fig.~\ref{fig:fig1}), then 30 clusters would have to be monitored for up to 10
years to detect any caustic transits from BH accretion disks at z\cge 7.

Appendix D discusses the uncertainty estimates in the main parameters that
determine the caustic {\it transit rates} and {\it rise times} of Pop III 
stellar-mass black hole accretion disks. As in \S \ref{sec45} and Appendix C,
the combined uncertainty in their caustic transit rates follows from the
adopted effective caustic length $L_{caust}$ (with $\sim$0.3 dex uncertainty),
the cluster transverse velocity \vT\ ($\sim$0.3 dex), and the uncertainty from
the presence of microlensing in the ICL (\cge 0.5 dex). The uncertainty in the
Pop III stellar-mass black hole accretion disk luminosity $L$ is larger than
for Pop III stars. This is due to their uncertain accretion efficiency, or
accretion duration, which we assume is uncertain by at least 0.5 dex, as
discussed in \S \ref{sec54} and Appendix D. On the other hand, the uncertainty
in the 1--4 \mum\ sky-SB from Pop III stellar-mass black hole accretion disks
may be smaller than that of Pop III stars, since the Spitzer--Chandra power
spectrum results (\S \ref{sec232}) hint at a possible contribution from
(stellar-mass) black holes at z\cge 7. As discussed in Appendix D, the error in
their power spectrum {\it signal} is estimated at \cge 0.15 dex. 

Since these parameters are independent, the combined uncertainty for the caustic
transit rates of stellar-mass black holes is thus at least $\sim$0.7 dex, but for
somewhat different reasons than for Pop III stars. This is indicated by the
vertical black error range in Fig.~\ref{fig:fig1}. Given the Spitzer--Chandra
power-spectrum signal discussed in \S \ref{sec232} and Appendix D, and the
possibility that their non steady-state luminosities may increase their caustic
transit rates for a given near-IR sky-SB, as discussed in \S \ref{sec54}, the
caustic transit rates for stellar-mass black hole accretion disks may be closer
to the upper value indicated in black in Fig.~\ref{fig:fig1}. Within the
uncertainties detailed in Appendices C--D, it is thus possible that stellar-mass
black hole accretion disks at z\cge 7 may outshine the sky-SB from Pop III stars
in the observed near-IR, and that they may produce correspondingly more caustic
transits. Only a long-term, dedicated observing program may be able to tell the
difference between these two possible sources of caustic transits at z\cge 7, as
discussed in \S \ref{sec7}. 

In conclusion, Pop III star rotation, the way dust is produced and expelled
during and after the Pop III star evolutionary sequence, the massive star binary
fraction, and the subsequent stellar-mass BH accretion-timescales and
accretion-efficiency may well in the end determine which of the two has the
best chance to be detected by JWST via cluster caustic transits. 

\bn \section{Possible Observing Programs to Detect Pop III Caustic Transits}
\label{sec7}

\sn JWST's lifetime requirement is 5 years and its lifetime goal is 10 years
\citep{gardner_2006}. JWST's actual mass is currently about 200 kg under its
allotted 6500 kg launch-mass, so its propellant tank has been completely filled,
enabling a maximum possible lifetime of 11--14 years with proper angular
momentum management if no hardware components {\it and their spares} fail
before that time. JWST carries a number of HST/SM4 heritage parts, and the 
HST/WFC3 hardware is operating just fine three years past its design lifetime.
Hence, contemplating a compelling time-domain science case for a JWST mission
with a 5--10 year baseline is possible. 

\mn \subsection{Characteristics of a JWST Survey to Find Pop III Caustic
Transits at z\cge 7}
\label{sec71}

\sn To observe caustic transits from First Light objects, a dedicated JWST
observing program will be required of at least several, and up to 30 clusters
for a duration of 1--10 years (see Fig.~\ref{fig:fig1}). Depending on their
exact contribution to the {\it diffuse} 1--4 \mum\ sky-SB (\cle 0.01--0.1
\nWsqmsr), such a JWST observing program to detect individual Pop III stars
and/or their stellar-mass BH accretion disks at z\cge 7 may well require to
monitor --- in the optimistic case that {\it most} of the NIR power-spectrum
signal comes from z\cge 7 --- a few suitable galaxy clusters during a year. In
the most pessimistic case that there exist really only a few Pop III objects per
square arcsecond and/or that most of them are shrouded by dust --- JWST may
need to monitor 30 clusters twice a year for a good fraction of JWST's 5--10
years lifetime to detect a few Pop III caustic transits. All of these cluster
observations would require coeval images in four NIRCam filter-pairs and/or
four NIRISS filters to constrain the spectral signature and redshift of a Pop
III caustic transit candidate. These would appear as z\cge 7 dropout candidates
that vary with time, either increasing rapidly and then slowly fading, or vice
versa. Their rise time are of the order of hours, while their fading times are
a good fraction of a year. 

Both cases pose interesting challenges to JWST IR-array data reduction
techniques: great care must be taken that a sudden increase in magnified object
flux during a caustic transit does not get rejected as an artifact or a cosmic
ray in the series of images taken that day. Also, care must be taken that a
slow increase in magnified flux of an object that approaches the caustic {\it 
from the other side} does not get misinterpreted as a slowly variable faint
Galactic brown dwarf star or a weak variable AGN \citep[\eg][]{cohen_2006}. The
nature of such ``reverse transits'' may therefore not be obvious when first
identified observationally by JWST. 

Could Pop III caustic transits cause a real difference in the luminosity
function at z\cge 7 in the field \citep[\eg][]{bouwens_2017} compared to
clusters \citep{livermore_2017}? {\it If} in the most optimistic case, several
Pop III objects at z\cge 7 were {\it always} seen transiting a cluster caustic
in any given year, then this could artificially boost the number of z\cge 7
objects seen behind clusters. This may not be obvious if such Pop III objects
resided in small star-forming objects that are well below the HST or JWST
detection limits, so one would not know in advance to expect caustic transits
at these locations. This is unlike the caustic transiting objects detected by
\citet{kelly_2017_b} and \cite{rodney_2017}, where there was a {\it known} faint
galaxy at a given location on the cluster caustic. While such caustic transit
detections of Pop III objects at z\cge 7 behind clusters could be real, they
may need additional lensing magnification-corrections in order to represent the
unlensed background universe at z\cge 7, and so could affect the derived
steady-state LF. Detailed JWST studies of high quality LFs at z\cge 7 that are
well sampled behind different clusters may reveal cluster-to-cluster
differences in caustic properties. Cosmic variance of the z\cge 7 population
will also require to average over a significant number of line-of-sights, by
observing a number of clusters with JWST. 

Microlensing by faint stars in the lensing cluster ICL may decrease the
magnifications from $\sim$10$^4$--10$^5$ to \cge 10$^3$, but greatly lengthen
the visibility time of the caustic transit, where a transiting microlensed
object may be visible for many decades or longer. We outlined an observing
strategy that JWST may use to observe these objects. To minimize the effects
from microlensing in the modeling of caustic transits, one could also target
some compact galaxy clusters at 0.3\cle z\cle 0.5 that have a smaller fraction
of ICL compared to their total galaxy light at the z\cge 7 lensing contours,
but that --- due to their compactness --- have excellent lensing properties
\citep[\eg][]{griffiths_2018}. 

\mn \subsection{Possible Spectral Differences between Pop III Star and Stellar
Mass BH Caustic Transits}
\label{sec72}

\sn A dedicated multi-band JWST monitoring program of well studied lensing
clusters may be able to detect the {\it chromatic differences} expected
between caustic transits of stellar-mass BH accretion disks and those of Pop III
stars, perhaps including spectroscopic confirmation. 

The one significant difference between Pop III stellar-mass BH accretion disks
and Pop III stars is likely the presence of a hard X-ray component that
contributes very significantly at the inner accretion disk radii, and that will
{\it also} have a significant energy tail longwards of \Lya 1216 \AA. No such
X-ray component would exist for the Pop III stars themselves, since ignoring
their limb-darkening and any star-spots, their stellar photospheres have nearly
uniform temperatures of T$\simeq$10$^5$ K (\S \ref{sec31}). Hence, Pop III
stars will not show significant chromatic behavior that may be traced during a
caustic transit, but BH accretion disks could show such chromaticity if they
were detected close to the actual caustic transit. 

Any differences in the effective UV-radii between BH accretion disks and Pop
III stars are important, since the maximum magnification obtained during a
caustic transit increases strongly for objects with smaller effective UV-radii
(Eq.~\ref{eq:muvsd} in \S \ref{sec44}). A very hot BH accretion disk crossing
a caustic could have much larger magnifications at its smallest intrinsic
X-ray--UV-bright radii, since these radii contribute a {\it larger fraction of
the total energy longwards of Ly$\alpha$} than they do for Pop III stars. Since
the maximum magnification scales as 1/$\sqrt{R_{UV}}$ (Eqs.~\ref{eq:muvsd1} \& 
\ref{eq:muvsd}), their smallest UV-bright radii (\S \ref{sec55}) could undergo
a maximum magnification, $\mu_{max}$, that is considerably larger during a
caustic crossing, which could boost their observed rates accordingly compared
to Pop III stars. 

Specifically, the inner (bluer) part of the BH accretion disk would be
magnified much more than its outer (redder) part. The ratio in magnifications
should follow $\sqrt{(r_{out}/r_{in})}$, where $r_{out} > r_{in}$ are the
largest and smallest BH UV-accretion disk radii discussed in \S \ref{sec552},
respectively. This will result in chromaticity due to lensing, where the {\it
shape} of the BH light-curve peaks during a caustic crossing would depend more
strongly on restframe UV-wavelength, unlike that of the Pop III stars. For
JWST, there could be a $\sim$1 dex difference in magnification between the
bluer and redder filters for an object undergoing a caustic transit at z\cge 7.
If a caustic transit maximum is observed almost simultaneously in different
JWST filters, we could then constrain the BH mass using Eq.~\ref{eq:RUVvsM},
assuming its scaling holds with slope $\rho$$\simeq$1/2 to stellar BH masses
(see \S \ref{sec55}). This would be an indirect way of confirming that part of
the light observed from a z\cge 7 object undergoing a caustic transit
originates in accretion disks around stellar-mass BHs. 

Pop III stars may also be detected or confirmed by JWST in other ways. For
instance, \citet{macpherson_2013} consider the prospect of finding a Pop III
hyper-nova ``in flagrante'', and suggest a detection rate of
2.78$\times$10$^{-6}$ per JWST field-of-view (FOV) and a probability of 37\%
that JWST will serendipitously image an afterglow during its lifetime. What
JWST truly will find from the Pop III epoch may include these and other
unexpected surprises. It is therefore critical that JWST First Light surveys
are well designed to optimize the possible detection of Pop III objects
directly. 

\mn \subsection{Role of the Next Generation Ground-Based Optical--Near-IR
Telescopes in Caustic Transits}
\label{sec73}

\sn JWST will be able to detect and monitor caustic transits during its 5--10
year lifetime. It is therefore useful to consider which other facilities can
observe caustic transits on longer timescales. JWST's unique advantage is its
very dark Zodiacal sky in L2 (AB\cge 23--24 \magarc\ at
$\lambda$$\simeq$2.0--3.5 \mum; Fig.~\ref{fig:fig1}), and it stable PSF over a
relative wide FOV \citep[2\arcmpt 2$\times$4\arcmpt 4]{rieke_2005}. Together
with its 25 m$^2$ collecting area, JWST should be able to reach AB\cge 28.5 mag
routinely \citep{windhorst_2008}. The next generation 25--40 m ground-based
telescopes --- the European Extremely Large Telescope (E-ELT), the Giant
Magellan Telescope (GMT), and the Thirty Meter Telescope (TMT)\
\footnote{\url{http://www.eso.org/sci/facilities/eelt/},
\url{http://www.gmto.org/resources/}, and \url{http://www.tmt.org/}.} --- will
have much larger collecting area, and narrower PSFs when using Multi-Conjugate
(laser-assisted) Adaptive Optics, although perhaps not as stable as JWST's
PSFs, and they will have lower Strehl ratios. They will also have a 1--2 \mum\ 
sky foreground that is \cge 7 mag brighter than JWST's in L2. As a
consequence, the next generation ground-based telescopes may be able to reach
AB\cle 29 mag in integrations of hours at 1--2 \mum, but --- given their
adaptive optics --- only over a smaller FOV (\cle 
20\arcs$\times$20\arcs---1\arcm$\times$1\arcm). Ground-based telescopes will
have reduced sensitivity at wavelengths $\lambda$\cge 2--2.2 \mum\ because of
the strongly increasing thermal foreground. For that reason, JWST will be able
to better address any chromatic differences between caustic transits of Pop III
stars and their stellar-mass BH accretion-disks (\S \ref{sec72}), especially
those at z \cge 12 that require several very sensitive filters at $\lambda$\cge
2 \mum, where ground-based telescopes cannot reach AB$\sim$29 mag due to the 
much brighter thermal foreground. 

Confirming spectra of caustic transits by Pop III stars or their stellar-mass
BH accretion disks could be taken with the JWST NIRISS and NIRSpec
spectrographs, and also with the next generation near-IR spectrographs on the
ELT, GMT, and TMT telescopes. Of particular interest would be to detect the
1640 \AA\ He line, which is expected to be present in the ionized regions around
Pop III stars, or their BH accretion disks, with T\cge 10$^5$ K
\citep{schaerer_2002, sobral_2015}.

We do not need to catch a caustic transit event at the precise moment of
crossing the caustic. It may be sufficient if a Pop III star is seen \cle 10
years before or after a caustic crossing, when the typical magnification may
well be of order 10$^4$, which can make a Pop III star with AB$\simeq$38 mag
visible to JWST. In five years time, the observed flux would increase (or
decrease) steadily by a factor $\sim$$\sqrt{2}$, which could be identified as a
star heading towards (or away from) a caustic. Perhaps the caustic transit of
such stars will not be observed during JWST's lifetime, but the next-generation
ground-based telescopes will be able to continue to monitor such stars for a much
longer period, when a given star appears to be heading towards a caustic in
several years time.

In summary, the next generation ground-based telescopes can monitor at 1--2 \mum\
--- over a much longer period than JWST --- individual Pop III caustic transits
that JWST will have detected at 1--4 \mum\ during its lifetime, and also discover
new ones on timescales longer than JWST's lifetime. This capability would be
particularly useful to follow-up on caustic transits that may be affected by
microlensing, and so may stretch out over many decades. Because of its much
wider 1--4.5 \mum\ wavelength range over which it can reach AB$\simeq$29 mag,
JWST will be essential to distinguish between possible {\it chromatic}
differences between Pop III stars and BH caustic transits. 

\bn \section{Summary and Conclusions}
\label{sec8}

\mn The following are the main conclusions of our paper:

\sn 1) The panchromatic (0.1--500 \mum) discrete galaxy counts
\citep{driver_2016} converge well at almost all wavelengths, resulting in iEBL
values from discrete objects that are well determined (to within 20\%) and
similar to those obtained at 1--4 \mum\ from $\gamma$-ray blazar spectral
distortions. Therefore, limits to the {\it diffuse} 1--4 \mum\ EBL are likely 
below \cle 1--2 \nWsqmsr, which we consider as ``hard'' upper limit for any Pop
III contribution to the EBL. 

\sn 2) Based on recent near-IR \citep{kashlinsky_2012, kashlinsky_2015,
mitchell-wynne_2016} and near-IR--X-ray power-spectrum \citep{cappelluti_2013} 
results and theoretical estimates, we adopt tighter constraints to the sky-SB
from Pop III BH accretion disks of \cle 0.11 \nWsqmsr\ (\ie\ sky-SB\cge 31
AB-\magarc\ at 2.0 \mum). From observational and theoretical considerations of
the cosmic SFH, we adopt similar upper limits to the 2.0 \mum\ SB for Pop III
stars themselves.

\sn 3) These adopted near-IR Pop III sky-SB values lead to a predicted rate of
\cle 0.32 Pop III star caustic transits per cluster per year that may be
observable with JWST to AB\cle 28.5 mag, with rise-times of less than a few
hours and decay timescales of less than a year, or vice versa, depending on
from which direction the Pop III object approaches the caustic: starting at the
``sharp edge'' of the caustic, or starting at the other side that declines
smoothly as $1/\sqrt{d}$. Microlensing by intracluster medium objects can
reduce transit magnifications, but lengthen visibility times. 

\sn 4) For Pop III stellar-mass BH accretion disks and their anticipated
accretion times of 0.3--60 Myr, we suggest cluster caustic transit rates that
are similar to those of Pop III stars, amounting to \cge 0.2 Pop III BH
accretion disk caustic transits per massive cluster per year. The BH feeding
timescales compared to the Pop III star lifetimes --- and the amount and
distribution of self-produced dust around the Pop III stars and their
subsequent BH accretion disks --- will determine which one of the two compact
UV sources will yield the most frequent cause of cluster caustic transits that
could be observed by JWST or the next generation 25--40 m ground-based
telescopes. 

\sn 5) In the case that the actual caustic transit rates from Pop III stars or
their stellar-mass BH accretion-disks are much lower than our suggested
predictions, the actual detection rate --- or upper limits thereto --- by JWST
over its 5--10 year lifetime will significantly constrain the Pop III objects
that our universe contains. Any firmly detected Pop III caustic transit would
be one of the most exciting First Light discoveries with JWST. 

If no Pop III caustic transits are seen with JWST by monitoring $\sim$30
clusters over 5--10 years, despite a long dedicated campaign, then the SB of
Pop III stars and their stellar-mass BH accretion disks may truly be fainter
than SB\cge 36--37 \magarc\ at 2.0 \mum. In other words, the true Pop III star
density would be very low indeed, with only a few Pop III stars per square
arcsec in the 1--4 \mum\ sky. While not as exciting as a number of significant
caustic transit detections at z\cge 7, such a null experiment would be
interesting in itself, as it would significantly constrain the sky-SB of Pop
III objects at z\cge 7 that may contribute to the diffuse EBL. Either way, the
experiment would allow JWST to directly constrain the First Light epoch. 

In summary, unlensed Pop III stars or their stellar-mass BH accretion disks may
have fluxes of AB$\simeq$35--41.5 mag at z$\simeq$7--17, and so will {\it not}
be directly detectable by JWST. However, cluster caustic transits with
magnifications of $\mu$$\simeq$10$^4$--10$^5$ may well render them temporarily
detectable to JWST in medium-deep to deep observations (AB\cle 28.5--29 mag) on
timescales of months to a year, with rise-times less than a few hours. Deep
and well time-sequenced observations of the best-lensing clusters carried out
throughout JWST's lifetime would fulfill its promise to the US Congress and
citizens as NASA's ``First Light'' telescope. 

\acknowledgements We dedicate this paper to Phil Sabelhaus, who during his life
heroically fought every day to manage the JWST project during its first decade:
Phil is our hero --- we are certain that without Phil, JWST could not have
succeeded. 

We thank Drs. Fred Adams, Mia Bovill, Michele Cirasuolo, Seth Cohen, Tim de
Zeeuw, Brenda Frye, Pat McCarthy, and Steve Rodney for helpful discussions. We
thank Teresa Ashcraft, Mia Bovill, Harrison Bradley, Seth Cohen, Robert Groess,
Victoria Jones, Bhavin Joshi, Brent Smith, Cameron White, and the referee for
very useful comments on the manuscript, Rolf Jansen for help with AASTeX. We
thank the following scientists for their hospitality during working visits when
part of this work was completed: Prof. John Peacock, James Dunlop and Gillian
Wright in Edinburgh, Leon Koopmans in Groningen, and Dr. Huub R\"ottgering in
Leiden.

This work was funded by NASA JWST Interdisciplinary Scientist grants NAG5-12460,
NNX14AN10G, and 80NSSC18K0200 to RAW from GSFC. FXT acknowledges support from
NASA under the Theoretical and Computational Astrophysics Networks (TCAN) grant
NNX14AB53G, by NSF under the Software Infrastructure for Sustained Innovation
(SI$^2$) grant 1339600 and grant PHY-1430152 for the Physics Frontier Center
``Joint Institute for Nuclear Astrophysics --- Center for the Evolution of the
Elements'' (JINA-CEE). JSBW acknowledges the support of the Australian Research
Council. JMD acknowledges support of projects AYA2015-64508-P (MINECO/FEDER, UE),
AYA2012-39475-C02-01, and Consolider Project CSD2010-00064 funded by the
Ministerio de Economia y Competitividad of Spain. 

\sn \software{
\texttt{APT} Astronomer's Proposal Tool (STScI)
\url{http://www.stsci.edu/hst/proposing/apt},
\texttt{MESA} \citep{paxton_2011, paxton_2013, paxton_2015},
\texttt{Python} \url{https://www.python.org}, 
\texttt{matplotlib} \citep{hunter_2007}, 
\texttt{NumPy} \citep{der_walt_2011},
\texttt{Source Extractor} \citep{bertin_2006}
}

\facilities{
\texttt{James Webb Space Telescope} \citep{gardner_2006}
}

\bibliographystyle{aasjournal}

\begin{thebibliography}{}

\expandafter\ifx\csname natexlab\endcsname\relax\def\natexlab#1{#1}\fi

\providecommand{\url}[1]{\href{#1}{#1}}

\bibitem[{{Abbott} {et~al.}(2016{\natexlab{a}}){Abbott}, {Abbott}, {Abbott},
{Abernathy}, {Acernese}, {Ackley}, {Adams}, {Adams}, {Addesso}, {Adhikari}, \&
et~al.}]{abbott_2016_a} {Abbott}, B.~P., {Abbott}, R., {Abbott}, T.~D., {et~al.}
2016{\natexlab{a}}, Physical Review Letters, 116, 061102

\bibitem[{{Abbott} {et~al.}(2016{\natexlab{b}}){Abbott}, {Abbott}, {Abbott},
{Abernathy}, {Acernese}, {Ackley}, {Adams}, {Adams}, {Addesso}, {Adhikari}, \&
et~al.}]{abbott_2016_b} ---. 2016{\natexlab{b}}, Physical Review Letters, 116,
241103

\bibitem[{{Abbott} {et~al.}(2016{\natexlab{c}}){Abbott}, {Abbott}, {Abbott},
{Abernathy}, {Acernese}, {Ackley}, {Adams}, {Adams}, {Addesso}, {Adhikari}, \&
et~al.}]{abbott_2016_c} ---. 2016{\natexlab{c}}, \apjl, 818, L22

\bibitem[{{Abbott} {et~al.}(2016{\natexlab{d}}){Abbott}, {Abbott}, {Abbott},
{Abernathy}, {Acernese}, {Ackley}, {Adams}, {Adams}, {Addesso}, {Adhikari}, \&
et~al.}]{abbott_2016_d} ---. 2016{\natexlab{d}}, \apjl, 832, L21

\bibitem[{{Abbott} {et~al.}(2016{\natexlab{e}}){Abbott}, {Abbott}, {Abbott},
{Abernathy}, {Acernese}, {Ackley}, {Adams}, {Adams}, {Addesso}, {Adhikari}, \&
et~al.}]{abbott_2016_e} ---. 2016{\natexlab{e}}, \apjl, 833, L1

\bibitem[Abbott et al.(2017{\natexlab{a}})]{abbott_2017_a} Abbott, B.~P.,
Abbott, R., Abbott, T.~D., et al.\ 2017{\natexlab{a}}, Physical Review Letters,
118, 221101

\bibitem[Abbott et al.(2017{\natexlab{b}})]{abbott_2017_b} Abbott, B.~P.,
Abbott, R., Abbott, T.~D., et al.\ 2017{\natexlab{b}}, Physical Review Letters,
119, 161101 

\bibitem[Abbott et al.(2017{\natexlab{c}})]{abbott_2017_c} Abbott, B.~P.,
Abbott, R., Abbott, T.~D., et al.\ 2017{\natexlab{c}}, \apjl, 848, L13 

\bibitem[Abel et al.(2002)]{abel_2002} Abel, T., Bryan, G.~L., \& Norman,
M.~L.\ 2002, Science, 295, 93

\bibitem[Acebron et al.(2017)]{acebron_2017} Acebron, A., Jullo, E., Limousin,
M., et al.\ 2017, \mnras, 470, 1809 

\bibitem[Adams et al.(2006)]{adams_2006} Adams, F.~C., Proszkow, E.~M.,
Fatuzzo, M., \& Myers, P.~C.\ 2006, \apj, 641, 504 

\bibitem[Adams(2010)]{adams_2010} Adams, F.~C.\ 2010, \araa, 48, 47 

\bibitem[Ahnen et al.(2016)]{ahnen_2016} Ahnen, M.~L., Ansoldi, S., Antonelli,
L.~A., et al.\ 2016, \aap, 590, A24 

\bibitem[Alpaslan et al.(2012)]{alpaslan_2012} Alpaslan, M., Robotham, A.~S.~G.,
Driver, S., et al.\ 2012, \mnras, 426, 2832 

\bibitem[Andrews et al.(2017{\natexlab{a}})]{andrews_2017_a} Andrews, S.~K.,
Driver, S.~P., Davies, L.~J.~M., et al.\ 2017{\natexlab{a}}, \mnras, 464, 1569 

\bibitem[Andrews et al.(2017{\natexlab{b}})]{andrews_2017_b} Andrews, S.~K.,
Driver, S.~P., Davies, L.~J.~M., et al.\ 2017{\natexlab{b}}, \mnras, 470, 1342 

\bibitem[Angus \& McGaugh(2008)]{angus_2008} Angus, G.~W., \& McGaugh, S.~S.\
2008, \mnras, 383, 417 

\bibitem[{{Arendt} {et~al.}(2016){Arendt}, {Kashlinsky}, {Moseley}, \&
{Mather}}]{arendt_2016} {Arendt}, R.~G., {Kashlinsky}, A., {Moseley}, S.~H., \&
{Mather}, J. 2016, \apj, 824, 26

\bibitem[Ashcraft et al.(2017)]{ashcraft_2017} Ashcraft, T. A., Windhorst, R.
A., Jansen, R. A., et al. 2017, \pasp, resubmitted (astro-ph/1703.09874)

\bibitem[Badenes et al.(2017)]{badenes_2017} Badenes, C., Mazzola, C.,
Thompson, T.~A., et al.\ 2017, \apj, submitted (astro-ph/1711.00660)

\bibitem[Bahcall \& Oh(1996)]{bahcall_1996} Bahcall, N.~A., \& Oh, S.~P.\
1996, \apjl, 462, L49 

\bibitem[Barkana \& Loeb(2001)]{barkana_2001} Barkana, R., \& Loeb, A.\
2001, \physrep, 349, 125

\bibitem[{{Barkana} \& {Loeb}(2002)}]{barkana_2002} {Barkana}, R., \& {Loeb}, A.
2002, \apj, 578, 1

\bibitem[{{Barkat} {et~al.}(1967){Barkat}, {Rakavy}, \& {Sack}}]{barkat_1967}
{Barkat}, Z., {Rakavy}, G., \& {Sack}, N. 1967, Physical Review Letters, 18, 379

\bibitem[Bastian et al.(2010)]{bastian_2010} Bastian, N., Covey, K.~R., \&
Meyer, M.~R.\ 2010, \araa, 48, 339 

\bibitem[{{Beichman} {et~al.}(2012){Beichman}, {Rieke}, {Eisenstein}, {Greene},
{Krist}, {McCarthy}, {Meyer}, \& {Stansberry}}]{beichman_2012} {Beichman},
C.~A., {Rieke}, M., {Eisenstein}, D., {et~al.} 2012, \procspie, 8442, Space
Telescopes and Instrumentation: Optical, Infrared, \& Millimeter Wave, 84422N 

\bibitem[{{Belczynski} {et~al.}(2016){Belczynski}, {Heger}, {Gladysz}, {Ruiter},
{Woosley}, {Wiktorowicz}, {Chen}, {Bulik}, {O'Shaughnessy}, {Holz}, {Fryer}, \&
{Berti}}]{belczynski_2016} {Belczynski}, K., {Heger}, A., {Gladysz}, W.,
{et~al.} 2016, \aap, 594, A97

\bibitem[Bertin \& Arnouts(1996)]{bertin_2006} Bertin, E., \& Arnouts, S.\ 1996,
\aaps, 117, 393 

\bibitem[{{Bessell} {et~al.}(1998){Bessell}, {Castelli}, \&
{Plez}}]{bessell_1998} {Bessell}, M.~S., {Castelli}, F., \& {Plez}, B. 1998,
\aap, 333, 231

\bibitem[Biteau \& Williams(2015)]{biteau_2015} Biteau, J., \& Williams, D.~A.\
2015, \apj, 812, 60 

\bibitem[Blackburne et al.(2011)]{blackburne_2011} Blackburne, J.~A., Pooley,
D., Rappaport, S., \& Schechter, P.~L.\ 2011, \apj, 729, 34 

\bibitem[{{Bond} {et~al.}(1984){Bond}, {Arnett}, \& {Carr}}]{bond_1984} {Bond},
J.~R., {Arnett}, W.~D., \& {Carr}, B.~J. 1984, \apj, 280, 825

\bibitem[Bouwens et al.(2015)]{bouwens_2015} Bouwens, R.~J., Illingworth,
G.~D., Oesch, P.~A., et al.\ 2015, \apj, 803, 34 

\bibitem[Bouwens et al.(2017)]{bouwens_2017} Bouwens, R.~J., Illingworth,
G.~D., Oesch, P.~A., et al.\ 2017, \apj, 843, 41 

\bibitem[Bovill(2016)]{bovill_2016} Bovill, M. S.\ 2016, presentation at the
October 2016 Montreal JWST Workshop
\url{http://craq-astro.ca/jwst2016/agenda_en.php/}


\bibitem[Bromm, Kudritzki, \& Loeb(2001)]{bromm_2001} Bromm, V., Kudritzki,
R.P., \& Loeb, A.\ 2001, \apj, 552, 464

\bibitem[Butler \& Bloom(2011)]{butler_2011} Butler, N.~R., \& Bloom,
J.~S.\ 2011, \aj, 141, 93 

\bibitem[Caminha et al.(2017)]{caminha_2017} Caminha, G.~B., Grillo, C., Rosati,
P., et al.\ 2017, \aap, 600, A90 

\bibitem[Calzetti et al.(1994)]{calzetti_1994} Calzetti, D., Kinney, A.~L., \&
Storchi-Bergmann, T.\ 1994, \apj, 429, 582 

\bibitem[Cannizzo et al.(1988)]{cannizzo_1988} Cannizzo, J.~K., Shafter,
A.~W., \& Wheeler, J.~C.\ 1988, \apj, 333, 227 

\bibitem[{{Cappelluti} {et~al.}(2013){Cappelluti}, {Kashlinsky}, {Arendt},
{Comastri}, {Fazio}, {Finoguenov}, {Hasinger}, {Mather}, {Miyaji}, \&
{Moseley}}]{cappelluti_2013} {Cappelluti}, N., {Kashlinsky}, A., {Arendt},
R.~G., {et~al.} 2013, \apj, 769, 68

\bibitem[Cappelluti et al.(2017)]{cappelluti_2017} Cappelluti, N., Li, Y.,
Ricarte, A., et al.\ 2017, \apj, 837, 19 

\bibitem[{{Casagrande} {et~al.}(2006){Casagrande}, {Portinari}, \&
{Flynn}}]{casagrande_2006} {Casagrande}, L., {Portinari}, L., \& {Flynn}, C.
2006, \mnras, 373, 13

\bibitem[Castor et al.(1975)]{castor_1975} Castor, J.~I., Abbott, D.~C., \&
Klein, R.~I.\ 1975, \apj, 195, 157 

\bibitem[{{Chatzopoulos} {et~al.}(2013){Chatzopoulos}, {Wheeler}, \&
{Couch}}]{chatzopoulos_2013} {Chatzopoulos}, E., {Wheeler}, J.~C., \& {Couch},
S.~M. 2013, \apj, 776, 129

\bibitem[Choi et al.(2016)]{choi_2016} Choi, J., Dotter, A., Conroy,
C., et al.\ 2016, \apj, 823, 102 

\bibitem[Chornock et al.(2017)]{chornock_2017} Chornock, R., Berger, E.,
Kasen, D., et al.\ 2017, \apjl, 848, L19 

\bibitem[Clowe et al.(2006)]{clowe_2006} Clowe, D., Brada{\v c}, M., Gonzalez,
A.~H., et al.\ 2006, \apjl, 648, L109 

\bibitem[Cohen et al.(2006)]{cohen_2006} Cohen, S.~H., Ryan, R.~E., Jr.,
Straughn, A.~N., et al.\ 2006, \apj, 639, 731 

\bibitem[Conroy(2013)]{conroy_2013} Conroy, C.\ 2013, \araa, 51, 393 

\bibitem[{{Cooray} {et~al.}(2012){Cooray}, {Gong}, {Smidt}, \&
{Santos}}]{cooray_2012} {Cooray}, A., {Gong}, Y., {Smidt}, J., \& {Santos},
M.~G. 2012, \apj, 756, 92

\bibitem[Cowperthwaite et al.(2017)]{cowperthwaite_2017} Cowperthwaite, P.~S.,
Berger, E., Villar, V.~A., et al.\ 2017, \apjl, 848, L17 

\bibitem[Coulter et al.(2017)]{coulter_2017} Coulter, D.~A., Lehmer, B.~D., 
Eufrasio, R.~T., et al.\ 2017, \apj, 835, 183 

\bibitem[{{de Mink} \& {Mandel}(2016)}]{de-mink_2016} {de Mink}, S.~E., \&
{Mandel}, I. 2016, \mnras, 460, 3545

\bibitem[Diaferio(1999)]{diaferio_1999} Diaferio, A.\ 1999, \mnras, 309, 610 

\bibitem[Diego et al.(2015{\natexlab{a}})]{diego_2015_a} Diego, J.~M.,
Broadhurst, T., Molnar, S.~M., Lam, D., \& Lim, J.\ 2015{\natexlab{a}}, \mnras,
447, 3130 

\bibitem[Diego et al.(2015{\natexlab{b}})]{diego_2015_b} Diego, J.~M.,
Broadhurst, T., Zitrin, A., et al.\ 2015{\natexlab{b}}, \mnras, 451, 3920 

\bibitem[Diego et al.(2016{\natexlab{a}})]{diego_2016_a} Diego, J.~M.,
Broadhurst, T., Chen, C., et al.\ 2016{\natexlab{a}}, \mnras, 456, 356 

\bibitem[Diego et al.(2016{\natexlab{b}})]{diego_2016_b} Diego, J.~M.,
Broadhurst, T., Wong, J., et al.\ 2016{\natexlab{b}}, \mnras, 459, 3447 

\bibitem[Diego et al.(2017)]{diego_2017} Diego, J.~M., Kaiser, N., Broadhurst,
T., et al.\ 2017, \apj, resubmitted (astro-ph/1706.10281) 

\bibitem[Dressler(1991)]{dressler_1991} Dressler, A.\ 1991, \nat, 350, 391 

\bibitem[{{Driver} {et~al.}(2016){Driver}, {Andrews}, {Davies}, {Robotham},
{Wright}, {Windhorst}, {Cohen}, {Emig}, {Jansen}, \& {Dunne}}]{driver_2016}
{Driver}, S.~P., {Andrews}, S.~K., {Davies}, L.~J., {Robotham}, A.~S.~G.,
{Wright}, A.~H., {Windhorst}, R.~A., {Cohen}, S.~H., {Emig}, K., {Jansen}, R.~A.
\& {Dunne}, L. 2016, \apj, 827, 108 (D16)

\bibitem[{{Duch{\^e}ne} \& {Kraus}(2013)}]{duchene_2013} {Duch{\^e}ne}, G., \&
{Kraus}, A. 2013, \araa, 51, 269

\bibitem[Dwek \& Krennrich(2013)]{dwek_2013} Dwek, E., \& Krennrich, F.\ 2013,
Astroparticle Physics, 43, 112 

\bibitem[Ebeling et al.(2014)]{ebeling_2014} Ebeling, H., Ma, C.-J., \& Barrett,
E.\ 2014, \apjs, 211, 21 

\bibitem[Emilio et al.(2012)]{emilio_2012} Emilio, M., Kuhn, J.~R., Bush, R.~I.,
\& Scholl, I.~F.\ 2012, \apj, 750, 135 

\bibitem[Fan et al.(2001)]{fan_2001} Fan, X., Narayanan, V.~K., Lupton, R.~H.,
et al.\ 2001, \aj, 122, 2833 

\bibitem[Fan et al.(2003)]{fan_2003} Fan, X., Strauss, M.~A., Schneider, D.~P.,
et al.\ 2003, \aj, 125, 1649 

\bibitem[{{Farmer} {et~al.}(2015){Farmer}, {Fields}, \& {Timmes}}]{farmer_2015}
{Farmer}, R., {Fields}, C.~E., \& {Timmes}, F.~X. 2015, \apj, 807, 184

\bibitem[{{Farmer} {et~al.}(2016){Farmer}, {Fields}, {Petermann}, {Dessart},
{Cantiello}, {Paxton}, \& {Timmes}}]{farmer_2016} {Farmer}, R., {Fields}, C.~E.,
{Petermann}, I., {et~al.} 2016, \apjs, 227, 22

\bibitem[Faulkner(1967)]{faulkner_1967} Faulkner, J.\ 1967, \apj, 147, 617

\bibitem[{{Fields} {et~al.}(2016){Fields}, {Farmer}, {Petermann}, {Iliadis}, \&
{Timmes}}]{fields_2016} {Fields}, C.~E., {Farmer}, R., {Petermann}, I.,
{Iliadis}, C., \& {Timmes}, F.~X. 2016, \apj, 823, 46

\bibitem[Fields et al.(2017)]{fields_2017} Fields, C.~E., Timmes, F.~X., Farmer,
R., et al.\ 2017, \apj, submitted (astro-ph/1712.06057)

\bibitem[Finkelstein et al.(2015)]{finkelstein_2015} Finkelstein, S.~L., Ryan,
R.~E., Jr., Papovich, C., et al.\ 2015, \apj, 810, 71 

\bibitem[Finkelstein(2016)]{finkelstein_2016} Finkelstein, S.~L.\ 2016, \pasa,
33, e037 

\bibitem[Fixsen et al.(1996)]{fixsen_1996} Fixsen, D.~J., Cheng, E.~S., Gales,
J.~M., et al.\ 1996, \apj, 473, 576 

\bibitem[Fixsen(2009)]{fixsen_2009} Fixsen, D.~J.\ 2009, \apj, 707, 916 

\bibitem[Flower(1996)]{flower_1996} Flower, P.~J.\ 1996, \apj, 469, 355 

\bibitem[{{Fraley}(1968)}]{fraley_1968} {Fraley}, G.~S. 1968, \apss, 2, 96

\bibitem[Frank et al.(2002)]{frank_2002} Frank, J., King, A., \& Raine, D.~J.\
2002, Accretion Power in Astrophysics, pp.~398.~ISBN 0521620538 Cambridge
University Press, (Cambridge, UK)

\bibitem[{{Fryer} {et~al.}(2001){Fryer}, {Woosley}, \& {Heger}}]{fryer_2001}
{Fryer}, C.~L., {Woosley}, S.~E., \& {Heger}, A. 2001, \apj, 550, 372

\bibitem[{{Gardner} {et~al.}(2006){Gardner}, {Mather}, {Clampin}, {Doyon},
{Greenhouse}, {Hammel}, {Hutchings}, {Jakobsen}, {Lilly}, {Long}, {Lunine},
{McCaughrean}, {Mountain}, {Nella}, {Rieke}, {Rieke}, {Rix}, {Smith},
{Sonneborn}, {Stiavelli}, {Stockman}, {Windhorst}, \& {Wright}}]{gardner_2006}
{Gardner}, J.~P., {Mather}, J.~C., {Clampin}, M., {et~al.} 2006, \ssr, 123, 485

\bibitem[Giavalisco et al.(2004)]{giavalisco_2004} Giavalisco, M., Ferguson,
H.~C., Koekemoer, A.~M., et al.\ 2004, \apjl, 600, L93 

\bibitem[G{\"o}tberg et al.(2017)]{gotberg_2017} G{\"o}tberg, Y., de
Mink, S.~E., \& Groh, J.~H.\ 2017, \aap, 608, A11 

\bibitem[Greif et al.(2011)]{greif_2011} Greif, T.~H., Springel, V., White,
S.~D.~M., et al.\ 2011, \apj, 737, 75 

\bibitem[Griffiths et al.(2018)]{griffiths_2018} Griffiths, A., Conselice, C.
Conselice, C.~J., Alpaslan, M., et al.\ 2018, \mnras, in press
(astro-ph/1801.01140)

\bibitem[Grogin et al.(2011)]{grogin_2011} Grogin, N.~A., Kocevski, D.~D.,
Faber, S.~M., et al.\ 2011, \apjs, 197, 35 

\bibitem[Guszejnov et al.(2016)]{guszejnov_2016} Guszejnov, D., Krumholz, M.~R.,
\& Hopkins, P.~F.\ 2016, \mnras, 458, 673 

\bibitem[Haardt \& Madau(2012)]{haardt_2012} Haardt, F., \& Madau, P.\ 2012,
\apj, 746, 125 

\bibitem[Madau \& Haardt(2015)]{haardt_2015} Madau, P., \& Haardt, F.\ 2015,
\apjl, 813, L8 

\bibitem[Hathi et al.(2008)]{hathi_2008} Hathi, N.~P., Jansen, R.~A., Windhorst,
R.~A., et al.\ 2008, \aj, 135, 156 

\bibitem[{{Helgason} {et~al.}(2016){Helgason}, {Ricotti}, {Kashlinsky}, \&
{Bromm}}]{helgason_2016} {Helgason}, K., {Ricotti}, M., {Kashlinsky}, A., \&
{Bromm}, V. 2016, \mnras, 455, 282

\bibitem[Henze et al.(2015)]{henze_2015} Henze, M., Ness, J.-U., Darnley,
M.~J., et al.\ 2015, \aap, 580, A46 

\bibitem[HESS~Collaboration(2013)]{abramowski_2013} HESS~Collaboration,
Abramowski, A., Acero, F., et al.\ 2013, \aap, 550, A4

\bibitem[H.~E.~S.~S.~Collaboration et al.(2017)]{abdalla_2017} 
H.~E.~S.~S.~Collaboration, Abdalla, H., Abramowski, A., et al.\ 2017, \aap, 606,
A59 

\bibitem[Hinshaw et al.(2009)]{hinshaw_2009} Hinshaw, G., Weiland, J.~L., Hill,
R.~S., et al.\ 2009, \apjs, 180, 225 

\bibitem[{{Hirschi}(2007)}]{hirschi_2007} {Hirschi}, R. 2007, \aap, 461, 571

\bibitem[Hoffman et al.(2015)]{hoffman_2015} Hoffman, Y., Courtois, H.~M., \&
Tully, R.~B.\ 2015, \mnras, 449, 4494 

\bibitem[Hoffman et al.(2017)]{hoffman_2017} Hoffman, Y., Pomar{\`e}de, D.,
Tully, R.~B., \& Courtois, H.~M.\ 2017, Nature Astronomy, 1, 0036 

\bibitem[Hogg(1999)]{hogg_1999} Hogg, D.~W.\ 1999, astro-ph/9905116 

\bibitem[Hogg et al.(2002)]{hogg_2002} Hogg, D.~W., Baldry, I.~K., Blanton,
M.~R., \& Eisenstein, D.~J.\ 2002, astro-ph/0210394 

\bibitem[Hosokawa et al.(2016)]{hosokawa_2016} Hosokawa, T., Hirano, S.,
Kuiper, R., et al.\ 2016, \apj, 824, 119 

\bibitem[Hoyle \& Lyttleton(1942)]{hoyle_1942} Hoyle, F., Lyttleton, R.A. \
1942, \mnras, 102, 177

\bibitem[{Hunter(2007)}]{hunter_2007} Hunter, J.~D. 2007, Computing In Science
\& Engineering, 9, 90

\bibitem[Ishiyama et al.(2016)]{ishiyama_2016} Ishiyama, T., Sudo, K., Yokoi,
S., et al.\ 2016, \apj, 826, 9 

\bibitem[Jansen et al.(2017)]{jansen_2017} Jansen, R.~A., \& Webb Medium Deep
Fields IDS GTO team, American Astronomical Society Meeting Abstracts, 230,
\#216.02 

\bibitem[Jauzac et al.(2014)]{jauzac_2014} Jauzac, M., Cl{\'e}ment, B.,
Limousin, M., et al.\ 2014, \mnras, 443, 1549 

\bibitem[Jauzac et al.(2015)]{jauzac_2015} Jauzac, M., Richard, J., Jullo, E.,
et al.\ 2015, \mnras, 452, 1437 

\bibitem[Jiang et al.(2007)]{jiang_2007} Jiang, L., Fan, X., Vestergaard, M.,
et al.\ 2007, \aj, 134, 1150 

\bibitem[Kashlinsky et al.(2012)]{kashlinsky_2012} Kashlinsky, A., Arendt,
R.~G., Ashby, M.~L.~N., et al.\ 2012, \apj, 753, 63 

\bibitem[Kashlinsky et al.(2015)]{kashlinsky_2015} Kashlinsky, A., Mather,
J.~C., Helgason, K., et al.\ 2015, \apj, 804, 99 

\bibitem[Kayser et al.(1986)]{kayser_1986} Kayser, R., Refsdal, S., \& Stabell,
R.\ 1986, \aap, 166, 36 

\bibitem[Kawamata et al.(2016)]{kawamata_2016} Kawamata, R., Oguri, M.,
Ishigaki, M., Shimasaku, K., \& Ouchi, M.\ 2016, \apj, 819, 114 

\bibitem[Kelly et al.(2017{\natexlab{a}})]{kelly_2017_a} Kelly, P.~L., Diego,
J.~M., Nonino, M., et al.\ 2017{\natexlab{a}}, The Astronomer's Telegram,
No.~10005, 5, 

\bibitem[Kelly et al.(2017{\natexlab{b}})]{kelly_2017_b} Kelly, P.~L., Diego,
J.~M., Rodney, S., et al.\ 2017{\natexlab{b}}, Nature Astr., resubmitted 
(astro-ph/1706.10279)

\bibitem[Kelsall et al.(1998)]{kelsall_1998} Kelsall, T., Weiland, J.~L., Franz,
B.~A., et al.\ 1998, \apj, 508, 44 

\bibitem[Kennicutt(1998)]{kennicutt_1998} Kennicutt, R.~C., Jr.\ 1998, \apj,
498, 541 

\bibitem[Kim et al.(2017)]{kim_2017} Kim, D., Jansen, R.~A., \& Windhorst,
R.~A.\ 2017, \apj, 804, 28 

\bibitem[{{Kiminki} \& {Kobulnicky}(2012)}]{kiminki_2012} {Kiminki}, D.~C., \&
{Kobulnicky}, H.~A. 2012, \apj, 751, 4

\bibitem[Koekemoer et al.(2011)]{koekemoer_2011} Koekemoer, A.~M., Faber,
S.~M., Ferguson, H.~C., et al.\ 2011, \apjs, 197, 36 

\bibitem[Koekemoer et al.(2013)]{koekemoer_2013} Koekemoer, A.~M., Ellis, R.~S.,
McLure, R.~J., et al.\ 2013, \apjs, 209, 3 

\bibitem[Kohri et al.(2014)]{kohri_2014} Kohri, K., Nakama, T., \& Suyama, T.\
2014, \prd, 90, 083514 

\bibitem[Koz{\l}owski et al.(2010)]{kozlowski_2010} Koz{\l}owski, S.,
Kochanek, C.~S., Udalski, A., et al.\ 2010, \apj, 708, 927 

\bibitem[{{Kozyreva} \& {Blinnikov}(2015)}]{kozyreva_2015} {Kozyreva}, A., \&
{Blinnikov}, S. 2015, \mnras, 454, 4357

\bibitem[{{Kozyreva} {et~al.}(2017){Kozyreva}, {Gilmer}, {Hirschi},
{Fr{\"o}hlich}, {Blinnikov}, {Wollaeger}, {Noebauer}, {van Rossum}, {Heger},
{Even}, {Waldman}, {Tolstov}, {Chatzopoulos}, \& {Sorokina}}]{kozyreva_2017}
{Kozyreva}, A., {Gilmer}, M., {Hirschi}, R., {et~al.} 2017, \mnras, 464, 2854

\bibitem[Kurk et al.(2007)]{kurk_2007} Kurk, J.~D., Walter, F., Fan,
X., et al.\ 2007, \apj, 669, 32 

\bibitem[Kurucz(2005)]{kurucz_2005} Kurucz, R.~L.\ 2005, Mem. S.A.It. Suppl.,
8, 189 \url{http://kurucz.harvard.edu/sun.html}

\bibitem[Lagattuta et al.(2017)]{lagattuta_2017} Lagattuta, D.~J., Richard,
J., Cl{\'e}ment, B., et al.\ 2017, \mnras, 469, 3946 

\bibitem[Lam et al.(2014)]{lam_2014} Lam, D., Broadhurst, T., Diego, J.~M., et
al.\ 2014, \apj, 797, 98 

\bibitem[{{Lewis} {et~al.}(2000){Lewis}, {Ibata}, \& {Wyithe}}]{lewis_2000}
{Lewis}, G.~F., {Ibata}, R.~A., \& {Wyithe}, J.~S.~B. 2000, \apjl, 542, L9

\bibitem[Livermore et al.(2017)]{livermore_2017} Livermore, R.~C., Finkelstein,
S.~L., \& Lotz, J.~M.\ 2017, \apj, 835, 113 

\bibitem[Lorentz et al.(2015)]{lorentz_2015} Lorentz, M., Brun, P., \&
Sanchez, D.\ 2015, 34th International Cosmic Ray Conference (ICRC2015), 34, 777 

\bibitem[Lotz et al.(2017)]{lotz_2017} Lotz, J.~M., Koekemoer, A., Coe, D., et
al.\ 2017, \apj, 837, 97 

\bibitem[Machida et al.(2009)]{machida_2009} Machida, M.~N., Omukai, K.,
Matsumoto, T., \& Inutsuka, S.-I.\ 2009, \mnras, 399, 1255 

\bibitem[{{Macpherson} {et~al.}(2013){Macpherson}, {Coward}, \&
{Zadnik}}]{macpherson_2013} {Macpherson}, D., {Coward}, D.~M., \& {Zadnik},
M.~G. 2013, \apj, 779, 73

\bibitem[{{Madau} \& {Silk}(2005)}]{madau_2005} {Madau}, P., \& {Silk}, J. 2005,
\mnras, 359, L37

\bibitem[Madau \& Dickinson(2014)]{madau_2014} Madau, P., \& Dickinson, M.\
2014, \araa, 52, 415 

\bibitem[Madau \& Fragos(2017)]{madau_2017} Madau, P., \& Fragos, T.\ 2017,
\apj, 840, 39 

\bibitem[Mahler et al.(2018)]{mahler_2018} Mahler, G., Richard, J.,
Cl{\'e}ment, B., et al.\ 2018, \mnras, 473, 663 

\bibitem[Maiolino et al.(2008)]{maiolino_2008} Maiolino, R., Nagao, T., Grazian,
A., et al.\ 2008, \aap, 488, 463 

\bibitem[Mamajek et al.(2015)]{mamajek_2015} Mamajek, E.~E., Prsa, A., Torres,
G., et al.\ 2015, astro-ph/1510.07674 

\bibitem[Mas-Ribas et al.(2016)]{mas-ribas_2016} Mas-Ribas, L., Dijkstra, M., \&
Forero-Romero, J.~E.\ 2016, \apj, 833, 65 

\bibitem[Matsuoka et al.(2011)]{matsuoka_2011} Matsuoka, Y., Ienaka, N., Kawara,
K., \& Oyabu, S.\ 2011, \apj, 736, 119 

\bibitem[Mattila et al.(2017)]{mattila_2017} Mattila, K., V{\"a}is{\"a}nen, P.,
Lehtinen, K., von Appen-Schnur, G., \& Leinert, C.\ 2017, \mnras, 470, 2152 

\bibitem[Mayer et al.(2017)]{mayer_2017} Mayer, P., Harmanec, P., Chini, R., et
al.\ 2017, \aap, 600, A33 

\bibitem[Meneghetti et al.(2017)]{meneghetti_2017} Meneghetti, M., Natarajan,
P., Coe, D., et al.\ 2017, \mnras, 472, 3177 

\bibitem[Milosavljevi{\'c} et al.(2009)]{milosavljevic_2009} Milosavljevi{\'c},
M., Bromm, V., Couch, S.~M., \& Oh, S.~P.\ 2009, \apj, 698, 766 

\bibitem[Miralda-Escude(1991)]{miralda_escude_1991} Miralda-Escude, J.\ 1991,
\apj, 379, 94 

\bibitem[{{Mitchell-Wynne} {et~al.}(2016){Mitchell-Wynne}, {Cooray}, {Xue},
{Luo}, {Brandt}, \& {Koekemoer}}]{mitchell-wynne_2016} {Mitchell-Wynne}, K.,
{Cooray}, A., {Xue}, Y., {et~al.} 2016, \apj, 832, 104

\bibitem[Molnar et al.(2013)]{molnar_2013} Molnar, S.~M., Broadhurst, T.,
Umetsu, K., et al.\ 2013, \apj, 774, 70 

\bibitem[Morishita et al.(2017)]{morishita_2017} Morishita, T., Abramson,
L.~E., Treu, T., et al.\ 2017, \apj, 846, 139 

\bibitem[Morgan et al.(2015)]{morgan_2015} Morgan, R.~J., Windhorst, R.~A.,
Scannapieco, E., \& Thacker, R.~J.\ 2015, \pasp, 127, 803 

\bibitem[Mortlock et al.(2011)]{mortlock_2011} Mortlock, D.~J., Warren, S.~J.,
Venemans, B.~P., et al.\ 2011, \nat, 474, 616 

\bibitem[Natarajan et al.(2017)]{natarajan_2017} Natarajan, P., Chadayammuri,
U., Jauzac, M., et al.\ 2017, \mnras, 468, 1962 

\bibitem[Negrello et al.(2017)]{negrello_2017} Negrello, M., Amber, S.,
Amvrosiadis, A., et al.\ 2017, \mnras, 465, 3558 

\bibitem[Oguri et al.(2017)]{oguri_2017} Oguri, M., Diego, J.~M., Kaiser, N.,
Kelly, P.~L., \& Broadhurst, T.\ 2017, \apj, submitted (astro-ph/1710.00148)

\bibitem[{{Ohkubo} {et~al.}(2009){Ohkubo}, {Nomoto}, {Umeda}, {Yoshida}, \&
{Tsuruta}}]{ohkubo_2009} {Ohkubo}, T., {Nomoto}, K., {Umeda}, H., {Yoshida}, N.,
\& {Tsuruta}, S. 2009, \apj, 706, 1184

\bibitem[Oke \& Gunn(1983)]{okegunn_1983} Oke, J.~B., \& Gunn, J.~E.\ 1983,
\apj, 266, 713 

\bibitem[Owers et al.(2011)]{owers_2011} Owers, M.~S., Randall, S.~W., Nulsen,
P.~E.~J., et al.\ 2011, \apj, 728, 27 

\bibitem[Pagel \& Portinari(1998)]{pagel_1998} Pagel, B.E.J., \& Portinari, L.\
1998, \mnras, 298, 747

\bibitem[Park \& Ricotti(2012)]{park_2012} Park, K., \& Ricotti, M.\ 2012, \apj,
747, 9 

\bibitem[{{Paxton} {et~al.}(2011){Paxton}, {Bildsten}, {Dotter}, {Herwig},
{Lesaffre}, \& {Timmes}}]{paxton_2011} {Paxton}, B., {Bildsten}, L., {Dotter},
A., {et~al.} 2011, \apjs, 192, 3

\bibitem[{{Paxton} {et~al.}(2013){Paxton}, {Cantiello}, {Arras}, {Bildsten},
{Brown}, {Dotter}, {Mankovich}, {Montgomery}, {Stello}, {Timmes}, \&
{Townsend}}]{paxton_2013} {Paxton}, B., {Cantiello}, M., {Arras}, P., {et~al.}
2013, \apjs, 208, 4

\bibitem[{{Paxton} {et~al.}(2015){Paxton}, {Marchant}, {Schwab}, {Bauer},
{Bildsten}, {Cantiello}, {Dessart}, {Farmer}, {Hu}, {Langer}, {Townsend},
{Townsley}, \& {Timmes}}]{paxton_2015} {Paxton}, B., {Marchant}, P., {Schwab},
J., {et~al.} 2015, \apjs, 220, 15

\bibitem[Petermann \& Timmes(2018)]{petermann_2018} Petermann, I., \& Timmes, F.
X. 2018, private communication

\bibitem[Planck Collaboration et al.(2014)]{planck_XXVII_2014} Planck
Collaboration, Aghanim, N., Armitage-Caplan, C., et al.\ 2014, \aap, 571, A27 

\bibitem[Planck Collaboration et al.(2016{\natexlab{a}})]{planck_XIII_2016_a}
Planck Collaboration, Ade, P.~A.~R., Aghanim, N., et al.\ 2016{\natexlab{a}},
\aap, 594, A13 

\bibitem[Planck Collaboration et al.(2016{\natexlab{b}})]{planck_XLVI_2016_b}
Planck Collaboration, Aghanim, N., Ashdown, M., et al.\ 2016{\natexlab{b}},
\aap, 596, A107 

\bibitem[Planck Collaboration et al.(2016{\natexlab{c}})]{planck_II_2016_c}
Planck Collaboration, Ade, P.~A.~R., Aghanim, N., et al.\ 2016{\natexlab{c}},
\aap, 594, A2 

\bibitem[Planck Collaboration et al.(2016{\natexlab{d}})]{planck_XLVII_2016_d}
Planck Collaboration, Adam, R., Aghanim, N., et al.\ 2016{\natexlab{d}}, \aap,
596, A108 

\bibitem[Portinari et al.(2010)]{portinari_2010} Portinari, L., Casagrande, L.,
\& Flynn, C.\ 2010, \mnras, 406, 1570 

\bibitem[Postman et al.(2012)]{postman_2012} Postman, M., Coe, D.,
Ben{\'{\i}}tez, N., et al.\ 2012, \apjs, 199, 25 

\bibitem[Pr{\v s}a et al.(2016)]{prsa_2016} Pr{\v s}a, A., Harmanec, P., Torres,
G., et al.\ 2016, \aj, 152, 41 

\bibitem[Remillard \& McClintock(2006)]{remillard_2006} Remillard, R.~A., \&
McClintock, J.~E.\ 2006, \araa, 44, 49 

\bibitem[Renzo et al.(2017)]{renzo_2017} Renzo, M., Ott, C.~D., Shore, S.~N.,
\& de Mink, S.~E.\ 2017, \aap, 603, A118 

\bibitem[Rieke et al.(2005)]{rieke_2005} Rieke, M.~J., Kelly, D., \& Horner,
S.\ 2005, \procspie, 5904, 1 

\bibitem[Robotham et al.(2011)]{robotham_2011} Robotham, A.~S.~G., Norberg, P.,
Driver, S.~P., et al.\ 2011, \mnras, 416, 2640 

\bibitem[Rodney et al.(2017)]{rodney_2017} Rodney, S.~A., Balestra, I.,
Bradac, M., et al.\ 2017, Nature Astr., in press (astro-ph/1707.02434)

\bibitem[Romero et al.(2015)]{romero_2015} Romero, A.~D., Campos, F., \&
Kepler, S.~O.\ 2015, \mnras, 450, 3708 

\bibitem[{{Rydberg} {et~al.}(2013){Rydberg}, {Zackrisson}, {Lundqvist}, \&
{Scott}}]{rydberg_2013} {Rydberg}, C.-E., {Zackrisson}, E., {Lundqvist}, P., \&
{Scott}, P. 2013, \mnras, 429, 3658

\bibitem[Rydberg et al.(2015)]{rydberg_2015} Rydberg, C.-E., Zackrisson, E.,
Zitrin, A., et al.\ 2015, \apj, 804, 13 

\bibitem[Salpeter(1955)]{salpeter_1955} Salpeter, E.~E.\ 1955, \apj, 121, 161 

\bibitem[{{Sana} {et~al.}(2012){Sana}, {de Mink}, {de Koter}, {Langer}, {Evans},
{Gieles}, {Gosset}, {Izzard}, {Le Bouquin}, \& {Schneider}}]{sana_2012} {Sana},
H., {de Mink}, S.~E., {de Koter}, A., {et~al.} 2012, Science, 337, 444

\bibitem[{{Sarmento} {et~al.}(2017){Sarmento}, {Scannapieco}, \&
{Pan}}]{sarmento_2017} {Sarmento}, R., {Scannapieco}, E., \& {Pan}, L. 2017,
\apj, 834, 23

\bibitem[{{Sarmento} {et~al.}(2018){Sarmento}, {Scannapieco}, \&
{Cohen}}]{sarmento_2018} {Sarmento}, R., {Scannapieco}, E., \& {Cohen}, S.
2018, \apj, in press (astro-ph/1710.09878) 

\bibitem[Scalo(1986)]{scalo_1986} Scalo, J.~M.\ 1986, \fcp, 11, 1 

\bibitem[Schaerer(2002)]{schaerer_2002} Schaerer, D.\ 2002, \aap, 382, 28 

\bibitem[Shafter et al.(2015)]{shafter_2015} Shafter, A.~W., Henze, M.,
Rector, T.~A., et al.\ 2015, \apjs, 216, 34 

\bibitem[Shafter(2017)]{shafter_2017} Shafter, A.~W.\ 2017, \apj, 834, 196 

\bibitem[Shakura \& Sunyaev(1973)]{shakura_sunyaev_1973} Shakura, N.~I., \&
Sunyaev, R.~A.\ 1973, \aap, 24, 337 

\bibitem[Shakura \& Sunyaev(1976)]{shakura_sunyaev_1976} Shakura, N.~I., \&
Sunyaev, R.~A.\ 1976, \mnras, 175, 613 

\bibitem[Shara et al.(1986)]{shara_1986} Shara, M.~M., Livio, M., Moffat,
A.~F.~J., \& Orio, M.\ 1986, \apj, 311, 163 

\bibitem[{{Smith} {et~al.}(2007){Smith}, {Li}, {Foley}, {Wheeler}, {Pooley},
{Chornock}, {Filippenko}, {Silverman}, {Quimby}, {Bloom}, \&
{Hansen}}]{smith_2007} {Smith}, N., {Li}, W., {Foley}, R.~J., {et~al.} 2007,
\apj, 666, 1116

\bibitem[Smith et al.(2018)]{smith_2018} Smith, B.~M., Windhorst, R.~A.,
Jansen, R.~A., et al.\ 2018, \apj, in press (astro-ph/1602.01555v2)

\bibitem[Sobral et al.(2015)]{sobral_2015} Sobral, D., Matthee, J., Darvish, B.,
et al.\ 2015, \apj, 808, 139

\bibitem[Springel \& Farrar(2007)]{springel_2007} Springel, V., \& Farrar,
G.~R.\ 2007, \mnras, 380, 911 

\bibitem[Stacy et al.(2016)]{stacy_2016} Stacy, A., Bromm, V., \& Lee, A.~T.\
2016, \mnras, 462, 1307 

\bibitem[Stanway et al.(2016)]{stanway_2016} Stanway, E.~R., Eldridge, J.~J., \&
Becker, G.~D.\ 2016, \mnras, 456, 485 

\bibitem[{{Sugimoto} \& {Nomoto}(1980)}]{sugimoto_1980} {Sugimoto}, D., \&
{Nomoto}, K. 1980, \ssr, 25, 155

\bibitem[Sukhbold \& Woosley(2014)]{sukhbold_2014} Sukhbold, T., \& Woosley,
S.~E.\ 2014, \apj, 783, 10 

\bibitem[Sukhbold \& Woosley(2016)]{sukhbold_2016} Sukhbold, T., \& Woosley,
S.~E.\ 2016, \apjl, 820, L38 

\bibitem[Susa et al.(2014)]{susa_2014} Susa, H., Hasegawa, K., \& Tominaga, N.\
2014, \apj, 792, 32 

\bibitem[Tanaka et al.(2012)]{tanaka_2012} Tanaka, T., Perna, R., \& Haiman, Z.\
2012, \mnras, 425, 2974 

\bibitem[Tanaka \& Shibazaki(1996)]{tanaka_1996} Tanaka, Y., \& Shibazaki, N.\
1996, \araa, 34, 607 

\bibitem[Thompson \& Nagamine(2012)]{thompson_2012} Thompson, R., \& Nagamine,
K.\ 2012, \mnras, 419, 3560 

\bibitem[{{Trenti} \& {Stiavelli}(2007)}]{trenti_2007} {Trenti}, M., \&
{Stiavelli}, M. 2007, \apj, 667, 38

\bibitem[{{Trenti} \& {Stiavelli}(2009)}]{trenti_2009} ---. 2009, \apj, 694, 879

\bibitem[Trujillo \& Fliri(2016)]{trujillo_2016} Trujillo, I., \& Fliri, J.\
2016, \apj, 823, 123 

\bibitem[Tucker et al.(1998)]{tucker_1998} Tucker, W., Blanco, P., Rappaport,
S., et al.\ 1998, \apjl, 496, L5 

\bibitem[Turk et al.(2009)]{turk_2009} Turk, M.~J., Abel, T., \& O'Shea, B.\
2009, Science, 325, 601 

\bibitem[{van~der Walt {et~al.}(2011)van~der Walt, Colbert, \&
Varoquaux}]{der_walt_2011} van~der Walt, S., Colbert, S.~C., \& Varoquaux, G.
2011, Computing in Science Engineering, 13, 22

\bibitem[Watkins \& Feldman(2015{\natexlab{a}})]{watkins_2015_a} Watkins, R., \&
Feldman, H.~A.\ 2015{\natexlab{a}}, \mnras, 447, 132 

\bibitem[Watkins \& Feldman(2015{\natexlab{b}})]{watkins_2015_b} Watkins, R., \&
Feldman, H.~A.\ 2015{\natexlab{b}}, \mnras, 450, 1868 

\bibitem[Watts(2012)]{watts_2012} Watts, A.~L.\ 2012, \araa, 50, 609 

\bibitem[Watson et al.(2014)]{watson_2014} Watson, W.~A., Iliev, I.~T., Diego,
J.~M., et al.\ 2014, \mnras, 437, 3776 

\bibitem[{{Wheeler}(1977)}]{wheeler_1977} {Wheeler}, J.~C. 1977, \apss, 50, 125

\bibitem[Willott et al.(2003)]{willott_2003} Willott, C.~J., McLure,
R.~J., \& Jarvis, M.~J.\ 2003, \apjl, 587, L15 

\bibitem[Willott et al.(2010)]{willott_2010} Willott, C.~J., Albert, L.,
Arzoumanian, D., et al.\ 2010, \aj, 140, 546 

\bibitem[{{Windhorst} {et~al.}(2008){Windhorst}, {Hathi}, {Cohen}, {Jansen},
{Kawata}, {Driver}, \& {Gibson}}]{windhorst_2008} {Windhorst}, R.~A., {Hathi},
N.~P., {Cohen}, S.~H., {et~al.} 2008, Advances in Space Research, 41, 1965 

\bibitem[{{Windhorst} {et~al.}(2011){Windhorst}, {Cohen}, {Hathi}, {McCarthy},
{Ryan}, {Yan}, {Baldry}, {Driver}, {Frogel}, {Hill}, {Kelvin}, {Koekemoer},
{Mechtley}, {O'Connell}, {Robotham}, {Rutkowski}, {Seibert}, {Straughn},
{Tuffs}, {Balick}, {Bond}, {Bushouse}, {Calzetti}, {Crockett}, {Disney},
{Dopita}, {Hall}, {Holtzman}, {Kaviraj}, {Kimble}, {MacKenty}, {Mutchler},
{Paresce}, {Saha}, {Silk}, {Trauger}, {Walker}, {Whitmore}, \&
{Young}}]{windhorst_2011} {Windhorst}, R.~A., {Cohen}, S.~H., {Hathi}, N.~P.,
{et~al.} 2011, \apjs, 193, 27 (W11)

\bibitem[Wolf et al.(2013)]{wolf_2013} Wolf, W.~M., Bildsten, L., Brooks, J.,
\& Paxton, B.\ 2013, \apj, 777, 136 

\bibitem[{Woosley {et~al.}(2002)Woosley, Heger, \& Weaver}]{woosley_2002}
Woosley, S.~E., Heger, A., \& Weaver, T.~A. 2002, Rev. Mod. Phys., 74, 1015

\bibitem[Woosley(2017)]{woosley_2017} Woosley, S.~E.\ 2017, \apj, 836, 244 

\bibitem[{{Yoon} {et~al.}(2008){Yoon}, {Cantiello}, \& {Langer}}]{yoon_2008}
{Yoon}, S.-C., {Cantiello}, M., \& {Langer}, N. 2008, in American Institute of
Physics Conference Series, Vol. 990, First Stars III, ed. B.~W. {O'Shea} \&
A.~{Heger}, 225--229

\bibitem[Yue et al.(2013)]{yue_2013} Yue, B., Ferrara, A., Salvaterra, R., Xu,
Y., \& Chen, X.\ 2013, \mnras, 433, 1556 

\bibitem[{{Yusof} {et~al.}(2013){Yusof}, {Hirschi}, {Meynet}, {Crowther},
{Ekstr{\"o}m}, {Frischknecht}, {Georgy}, {Abu Kassim}, \&
{Schnurr}}]{yusof_2013} {Yusof}, N., {Hirschi}, R., {Meynet}, G., {et~al.}
2013, \mnras, 433, 1114

\bibitem[{{Zackrisson} {et~al.}(2015){Zackrisson}, {Gonz{\'a}lez}, {Eriksson},
{Asadi}, {Safranek-Shrader}, {Trenti}, \& {Inoue}}]{zackrisson_2015}
{Zackrisson}, E., {Gonz{\'a}lez}, J., {Eriksson}, S., {et~al.} 2015, \mnras,
449, 3057

\bibitem[Zemcov et al.(2017)]{zemcov_2017} Zemcov, M., Immel, P., Nguyen, C.,
et al.\ 2017, Nature Communications, 8, 15003 

\bibitem[Zhang et al.(2010)]{zhang_2010} Zhang, F., Han, Z., Li, L., Guo, J., \&
Zhang, Y.\ 2010, \apss, 329, 249 

\bibitem[Zitrin et al.(2013)]{zitrin_2013} Zitrin, A., Menanteau, F., Hughes,
J.~P., et al.\ 2013, \apjl, 770, L15 

\end{thebibliography}

\ve 


\vspace*{-0.600cm}
\n\cl{Appendix A.\ Perturbing the Cluster Velocity Distribution to Constrain
the Maximum \vT-Value}
\label{secAppA}

\begin{figure*}[!hptb]
\hspace*{-0.10cm}
\n\cl{
\includegraphics[width=1.000\txw,angle=0]{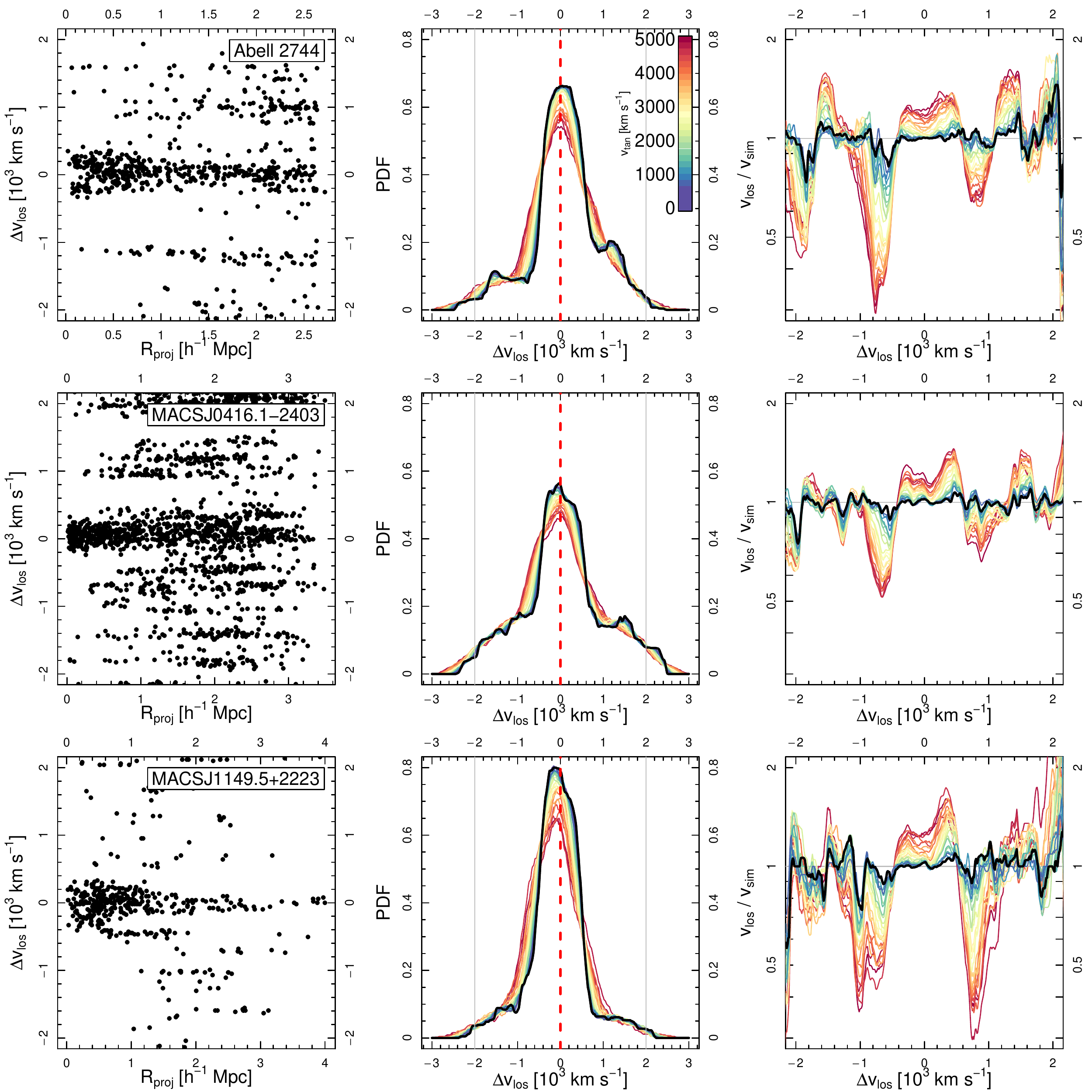}
}
\centering{}
\vspace*{-0.300cm}
\caption{
The redshift space distribution of galaxies in three lensing clusters (from the
HFF program) suitable for lensing of First Light objects and caustic transit
studies at z\cge 7 with JWST. The left-hand panels show the projected radial
distance from the cluster center of each galaxy as a function of its
line-of-sight velocity. The middle panels show the distribution of line-of-sight
velocities for each cluster, where the thick black lines indicate the real
velocity data from the left-hand panels. The colored lines show the modified
line-of-sight velocity distributions, after a random {\it space} velocity has
been added to the whole cluster, affecting both its \vlos\ and its \vT-vectors.
The resulting added transverse velocity increments range between
\vT$\simeq$500 to 5000 \kms, as indicated by the color scale-bar in the top
middle panel. The right-hand panels display the residuals from the middle
panels for better visibility between the models. We only detect a significant
deviation from the observed \vlos\ distribution when \vT\cge 1700--2300 \kms.
We adopt \vT\cle 1000 \kms\ as an upper limit to the transverse velocity of
these clusters in the plane of the sky when calculating the possible caustic
transit rate of First Light objects that may be seen by JWST at z\cge 7. At 
lower \vT-values, differences between the perturbed $N(v)$ model and the 
actual redshift data cannot be distinguished. 
}\label{fig:fig5}
\end{figure*}

\n To test the maximum values of \vT\ for galaxy clusters likely to be observed
by the community for caustic transit observations, we examined the available
redshift space distribution of the galaxies in three well studied HFF
clusters. For a circularized cluster in virial equilibrium, the {\it central}
distribution of cluster galaxies in redshift space (\ie\ projected distance,
$R_{\mathrm{proj}}$, from the cluster center as a function of line-of-sight
velocity $v_{\mathrm{los}}$) is expected to resemble a ``trumpet''
\citep{diaferio_1999, alpaslan_2012}. This is clearly visible in the left-hand
panels of Fig.~\ref{fig:fig5}, where we show the redshift space distribution of
galaxies in the HFF clusters Abell 2744, MACS J0416-2403 and MACS J1149.5+2223.
Redshift information for Abell 2744 was taken from \citet{owers_2011}, and for
MACS J0416-2403 plus MACS J1149.5+2223 from \citet{ebeling_2014}. In the middle
panels of Fig.~\ref{fig:fig5}, we display the observed velocity distribution of
\vlos\ from the left-hand panels as thick black lines. We expect the
distribution of the \vlos\ for the central cluster to peak around 0 \kms\ with
respect to the cluster redshift. But all three HFF clusters are embedded in
significant large scale structures in velocity space, with typical separations
between different structures along the line-of-sight between $\sim$500 and
$\sim$2000 \kms. MACS J0416-2403 is composed of two merging clusters, with a
small difference in redshift between them. The black lines suggests that the
central core of all three clusters appears to have significant sub-structure in
velocity space, especially for MACS J0416-2403 and MACS J1149.5+2223, whose
central cores are significantly non-Gaussian in their redshift distribution
$N(v)$. Each line-of-sight may have a number of virialized sub-structures ---
including the main cluster itself --- each with approximately a Gaussian
velocity distribution $N(v)$. One could think of the velocity distribution in
the {\it central core} of the black lines as some combination of virialized
Gaussians with a broader non-virialized component of galaxies, a fraction of
which are falling into the main cluster. 

The question then arises: by how much can we perturb the {\it space velocity} of
the cluster itself before the \vlos-distribution is noticeably changed from the
observed redshift distribution, which then also provides a limit to the {\it
maximum transverse velocity component} that can be randomly added? To implement
this, we take the line-of-sight velocities of all galaxies within the central
cluster itself and perturb them with a space velocity vector that is aligned at
45\degree\ with respect to the line-of-sight. We ensure that each galaxy is
perturbed by a similar value to mimic the effect of a true space velocity on the
observed redshift distribution of the central cluster. Galaxies whose projected
distance is between the cluster center and the median projected distance for
that cluster are perturbed by a vector whose magnitude is 500 \kms\ {\it less}
than galaxies at the outskirts of the cluster, and whose projected distance is
greater than that of the median. This simulates the effects of a differential
velocity disturbance, since the galaxies closer to the cluster potential-well
likely experience a velocity perturbation resulting from the cluster
space-velocity that is smaller in magnitude compared to the local velocity
dispersion. The magnitude of the perturbing velocity was drawn from a normal
distribution centered on the number given in the colored legend of
Fig.~\ref{fig:fig5}, with a standard deviation of 500 \kms. The angle was
drawn from a uniform distribution ranging from 40\degree--50\degree. Finally,
we extract the y-component of this resulting velocity vector, and assign {\it
that} to be the new line-of-sight velocity for each galaxy. The colored lines
in the middle and right panels of Fig.~\ref{fig:fig5} show these modified
line-of-sight velocity distributions {\it after} the space velocity has been
added to the cluster, and decomposed into the \vlos\ and the \vT-vector
component for each galaxy. The added transverse velocity increments range
between \vT$\simeq$500 to 5000 \kms, as indicated by the color-bar. 

In the right-hand panels of Fig.~\ref{fig:fig5}, we display the residuals from
the middle panels for better visibility between the models. We only detect a
significant deviation from the measured \vlos\ distribution if we increase the
space velocity such that the components added to the transverse velocity exceed
\vT\cge 1000 \kms, where the excess becomes clearly visible in
Fig.~\ref{fig:fig5} at \vT\cge 2000 \kms\ (green--red curves). A simple
normalized $\chi^2$-estimate for each fit shows that the reduced $\chi$-square
\vlos-values start to exceed unity when space velocities have been added with
transverse velocity components considerably higher than 1000 \kms. For Abell
2744, the reduced $\chi$-square exceed unity at \vT\cge 1900 \kms, for MACS
J0416-2403 at \vT\cge 2300 \kms, and for MACS J1149.5+2223 at \vT\cge 1715
\kms. We obtain similar results when we add space velocities to each galaxy
that are more randomized in angle, or when we add more randomized values of the
space velocity to each of the cluster sub-clumps. 

In conclusion, Fig.~\ref{fig:fig5} thus shows that adding space velocities
with projected {\it transverse} components much larger than \vT $\simeq$1000
\kms\ imply projected components of this space velocity added {\it along the
line-of-sight} that are not consistent with the available redshift data in the
cluster core. We will thus adopt an upper limit of \vT\cle 1000 \kms\ for the
maximum transverse velocity of these clusters at 0.3\cle z\cle 0.5 in the
plane of the sky when calculating the possible Pop III caustic transit rates of
First Light objects that may be seen by JWST at z\cge 7. For some
substructures in each cluster, the \vT-values may well be as high as 1000 \kms,
or perhaps somewhat higher.


\vspace*{-0.000cm}
\n


\ve 

\n\cl{Appendix B1.\ Caustic Transit Rates in the Presence of Microlensing: I.
The General Case}
\label{secAppB1}

\mn When microlenses are present, the rate of caustic transit events is
sensitive to the mass function of low-mass stars in the cluster ICL 
\cite[\eg][]{miralda_escude_1991}. If brighter Pop III stars are more common,
relatively modest microlensing peaks with $\mu_{max}$$\simeq$$5\times 10^3$ can
momentarily amplify a bright star (AB\cle 37 mag) to above the detection limit
of JWST. The rate of events will then be dominated by the microlens-peaks of the
brightest stars, and the rate of events will be proportional to the optical
depth of microlenses. Significantly fainter stars may be sufficiently magnified
only if the disruption of the cluster caustic by microlenses is moderate. In
this case, most events will be produced by relatively faint Pop III stars
crossing the lightly disrupted caustic, which can have maximum magnification of
$\mu_{max}\simeq 10^5$ (see \S \ref{sec4}). The rate of events will be dominated
in this case by the anticipated, much more numerous fainter stars
(AB$\simeq$40--42 mag) when crossing the caustic, and will be proportional to
the surface mass density of stars in the ICL. A rough estimate of the expected
rate of events, $R$, in the presence of microlenses can be obtained based on
the predictions from \citet{kelly_2017_b} \& \citet{diego_2017}, who modeled
the HFF clusters with \vT$\simeq$1000 \kms: 

\begin{equation}
R({\rm yr}^{-1}) = A(>\mu)\,\rho_{\ast}\,r(\Sigma,v_T) =
\frac{B_o}{\mu^2}\,\rho_{\ast}\,r(\Sigma,v_T). 
\end{equation}

\n Here, $A(>\mu)$ is the area in the source plane above a given magnification
$\mu$, which scales as $B_o/\mu^2$ in the presence of microlenses and at high
magnification. Also, $\rho_{\ast}$ is the surface mass density of Pop III stars
above $z=7$, so that $A(>\mu).\rho_{\ast}$ is the number of Pop III stars
undergoing microlensing at a particular moment. Last, r($\Sigma$,\vT) is the
rate of microlens caustic (hereafter ``microcaustic'') events a moving object
in the background ($z>7$) would encounter if the surface mass density of
microlenses is $\Sigma$, and the background object is moving with a transverse
velocity \vT\ with respect to the network of microcaustics.
\citet{kelly_2017_b} estimated that r($\Sigma$,\vT) is of order 0.1 yr$^{-1}$
for an event like Icarus. Using the equations of \citet{diego_2017} and
assuming the total length of the caustics to be L$\simeq$100\arcs, we estimate
that $B_o$$\simeq$1.8$\times$10$^{-4}$ \arcsecsq, or A($\mu$\cge
3$\times\,10^3$) $\simeq$0.002 \arcsecsq\ for an HFF-like cluster in the
presence of microlenses. The value $\mu\simeq 3 \times 10^3$ is adopted to
select regions in the source plane associated with microlensing peaks that will
reach $\mu_{max}$\cge 10$^4$. Then, if there are $\rho_{\ast}$$\simeq$100 Pop
III objects/\arcsecsq\ brighter than AB$\simeq$38 mag, we would expect for each
HFF-like cluster $R({\rm yr}^{-1})\simeq 0.2\times\,r(\Sigma, v_T) \simeq$ 0.02
caustic transits yr$^{-1}$ if we extrapolate the results of
\citet{kelly_2017_b} to the entire caustic region, \ie\ one event when
monitoring 5 HFF-like clusters for 10 years. 

We note that both the estimates with microlensing in this Appendix and those in
\S \ref{sec43} based on the adopted transverse velocity are in good agreement.
These numbers should be compared with the expected caustic transit rate if we
assume there are no microlenses. In this case, the magnification would be
described by $\mu\simeq 20/\sqrt{d}$ which implies $d$=1.6$\times$10$^{-5}$
arcsec for $\mu$$\simeq$10$^4$. Note that the above expression would give $\mu =
(20/\sqrt{1.6\times 10^{-5}}$)$\simeq$5000, but the total magnification would
be $\mu$=10$^4$ when we account for the double image produced in the image
plane. If the perimeter of the caustic region is L$\simeq$100\arcs, then the
area over which a magnification larger than $\mu$$\simeq$10$^4$ can be attained
is $\simeq$0.0016 \arcsecsq. This number is comparable to the value estimated
above when microlenses are included ($\sim$0.002 \arcsecsq). 

The similarity of results obtained with and without microlenses could have been
anticipated from basic principles. Owing to flux conservation, the number of
photons collected after integrating for a long period (tens to hundreds of
years) should be the same independent of the presence (and number) of
microlenses. The distribution of microlenses determines how this magnification
is redistributed. A lens plane without microlenses results in large
magnifications concentrated in a unique narrow region around the caustic (\ie\ 
a single very bright peak; see \S \ref{sec44}), while a lens-plane populated
with microlenses will break apart the single caustic into multiple (smaller)
microcaustics. Thus, the rate of high magnification events ($\mu_{max}$\cge
10$^4$) would be similar whether there are microlenses or not, but in the case
without microlenses we would see a single very bright peak when the Pop III star
crosses the caustic, while in the case with microlenses we would see many
(smaller) peaks hundreds of years before (or after) the star crosses the
position of the cluster caustic \citep[\eg][]{diego_2017}. Extreme
magnification ($\mu_{max}$ $\simeq$ 10$^6$) can be attained {\it only} when
microlenses are not included, and may only occur for the lower-mass BH
accretion disks whose inner X-ray--UV bright core-radii may be much smaller
than those of Pop III stars (\S \ref{sec552}), so that all large magnifications
are concentrated in a single peak around the cluster caustic. When microlenses
are included, the caustic region is expanded in size, as shown in
\citet{diego_2017}, so there is higher probability of a star in the source
plane to align with a microcaustic. However, the magnified peaks will be
correspondingly fainter, so only the more rare, brighter Pop III stars, or the
brighter stellar-mass BH accretion disks, may produce caustic transits that can
be observed by JWST. 

\bn\cl{Appendix B2.\ Caustic Transit Rates in the Presence of Microlensing: II.
The Case of Relatively Bright Sources}
\label{secAppB2}

\mn If a background source at z\cge 7 were to be relatively ``bright'' (AB\cle
37 mag for the unlensed source), virtually all microlensing peaks --- with
magnifications of about one to several thousand \citep{kelly_2017_b} --- can
be observed if the star is sufficiently close to the critical curve (\ie\
$\mu$\cge $10^3$). The number of events in that case will be approximately
equal to the number of microcaustics that the background source encounters, as
it moves across the web of microcaustics. In this section, we therefore present
an estimate of the case that JWST may observe when the background object is
relatively bright. Instead of computing the probability of an event based on
the area above a given magnification as in Appendix B1, we can simply estimate
the number of times a microcaustic is crossed, since a rare but very bright Pop
III star (AB\cle 35--37.5 mag, see Table~\ref{tab:tab2}--\ref{tab:tab4}) may be
directly observed, if the microlensing magnification from the cluster ICL at
its location is at least $\mu$$\simeq$10$^3$. In this case, even modest
microlenses with sub-solar masses can produce changes in flux of 0.5 magnitudes
or more. Having a bright star undergoing such frequent encounters with
microcaustics is possible, as discussed by \citet[\eg][]{kelly_2017_b}. These
variations in flux may be observed with JWST when the star crosses the network
of microlensing caustics. 

The probability of having a microlensing event at a given distance, $\theta$,
from the critical curve is given by the effective optical depth to
microlensing, which is defined as the fractional area at that location that is
being affected by microlensing. Using the model in \citet{diego_2017}, the
effective optical depth from microlenses is given by:

\begin{equation}
\tau(\theta) \simeq 21\times 10^{-2}\ \Sigma(M_{\odot}/pc^2)\ /\ \theta(\arcs)
\end{equation}

\n For sources at high redshift, the critical curve moves to distances of order
1\arcm\, from the center of the Brightest Cluster Galaxy (BCG), where the impact
of microlenses is expected to be small, but still not necessarily negligible.
At these angular distances, \citet{diego_2017} estimated for MACS1149 that the
surface mass-density of microlenses decreases by approximately two orders of
magnitude with respect to the one estimated at the position of Icarus. This
corresponds to a mass surface density of $\Sigma \simeq 0.1\ M_{\odot}/pc^2$.
Hence, if we restrict our analysis to the region where the effective optical
depth of microlenses reaches the saturation level (\ie\ $\tau\simeq1$), then
this implies an angular distance of $\theta\simeq 20$ milliarcsec (mas). That
is, the two counter-images of the lensed background object would appear on
either side of the critical curve and be separated by $\sim$40 mas. At these
separations, both counter images would form a single unresolved --- or at best
a slightly resolved --- object in the JWST mosaics. At 20 mas distance from the
critical curve, the model in \citet{diego_2017} predicts that the magnification
from the cluster is approximately $\mu$$\simeq$5000, so that any star brighter
than AB$\simeq$38 mag transiting the caustic at this location could be detected
by JWST to AB\cle\ 29 mag.

In principle, we could see twice the caustic transit rate in this case, since
this unresolved image would contain fluctuations form both sides of the critical
curve (\ie\ from its positive and negative parity). In reality, the rate on the
side with negative parity is expected to be a factor $\sqrt{2}$ times smaller
than the rate on the side with positive parity. This difference in rate can be
obtained from \citet{oguri_2017}, who estimated that the maximum dimension (or
cross-section ${\rm C_S}$) of the microcaustic on the side with positive parity
is ${\rm C_S}(M)=\theta_e(M)\sqrt{\mu_t}/\mu_r$, while its shape is that of a
stretched diamond. The Einstein radius of the microlens, $\theta_e$, depends on
the mass of the microlens and the angular-diameter distances from the observer
to the lens, from the lens to the background object, and from the observer to
the background object. On the side with negative parity this extension is
smaller by a factor $\sqrt{8}$, but the caustic is divided into two
semi-diamond shapes, so that a star crossing the microcaustic would cross
caustic lines 4$\times$ instead of twice. Hence, the effective length (or
cross-section) of the caustic on the side with negative parity is smaller by a
factor $\sqrt{8}/2=\sqrt{2}$. 

With the above ingredients it is possible to estimate the expected number of
caustic crossings for relatively bright sources (AB\cle 37 mag), which are now
the microcaustics formed by the local microlenses. Each microcaustic is shaped
as a diamond, or a double semi-diamond, on the sides with positive and negative
parity respectively \citep[for details, see \eg][]{diego_2017, oguri_2017}. For
simplicity, we assume that the microcaustic crossing events take place within
the region where the saturation level to lensing is reached (\ie\ the 40 mas
region surrounding the critical curve mentioned above). In the regime where the
saturation level to lensing has not been reached, one can still use the
relation between the position in the source plane, $\beta$, and the position in
the image plane, $\theta$, given by standard lensing theory,
$\beta=\theta^2/C$, where the constant $C$ depends on the lens strength (\ie\ 
the gradient of the lensing potential), and both $\beta$ and $\theta$ are
given with respect to the caustic and critical curves, respectively. For a
cluster like MACS1149, \cite{diego_2017} estimated $C$$\simeq$68\arcs. For
this particular value of $C$, one obtains $\beta$$\simeq$6 micro-arcsec, or $2.5
\times 10^{-2}$ pc at $z\simeq 10$. The distance traveled by a moving
background star with respect to the caustic network is $d(v)=1 \times 10^{-4}
(v_T/1000\ km/s)$ pc/yr. During this time, the star may encounter multiple
microcaustics, depending on the surface density of microlenses. For simplicity,
we assume that the background source is moving perpendicular to the maximum
extension of the diamond-shaped macro-caustics. This is a reasonable
assumption, since the microlens caustics are typically stretched by very large
factors, so to first order they can be approximated by straight parallel
lines. Since each microcaustic has a cross section ${\rm C_S}(M)$ that scales
with $\theta_e(M)$ \citep[see][or the expressions above]{oguri_2017}, the
yearly rate of intersections with a microcaustic of mass $M$ is then given by: 

\begin{equation}
r(M)=2(1+\sqrt{2})\ .\ n(M)\mu\ .\ d(v)\ .\ 2{\rm C_S}(M).
\end{equation}

\n Here, the first term, $2(1+\sqrt{2})$, accounts for the events produced on
either side of the critical curve and the fact that a microcaustic is crossed
twice (4$\times$ for the side with negative parity). The second term, $n(M)\mu$,
is the number density of microlenses with mass $M$ in the lens plane, $n(M)$,
which is increased by a factor $\mu$ in the source plane. The third term,
$d(v)$, is the distance traveled by the background object in one year. The 
last term, ${\rm C_S}(M)$ is the cross-section of a microcaustic of mass M.
For realistic distributions of $n(M)$, one should integrate $r(M)$ to compute
the caustic transit rate, but for our purposes we adopt the simple scenario
where all microlenses have similar masses of $M\simeq 1$\,\Mo. In that case, we
get: $n(M)=\Sigma(M)=0.1/pc^2$ and $r(m)=4.7 \times 10^{-7}\mu_t^{3/2}$ per
year. Here we assumed that \vT\cle 1000\ \kms, $z_{lens}$=0.5, and 
$z_{source}$=10, for which one obtains $\theta_e$=2.3$\times 10^{-6}$ arcsec, or
0.0097 pc at z=10. Also, $\mu = \mu_t\mu_r$, so that the dependency with
$\mu_r$ cancels out and the rate depends only on $\mu_t^{3/2}$. To estimate the
value of $\mu_t$, we adopt the model in \cite{diego_2017}, according to which
$\mu_t = \mu/5$ and $\mu\simeq 100/\theta$, where $\theta$ is in arcseconds. 
At the point where the effective optical depth of microlenses reaches the
saturation level, $\theta=0.02''$, we get $\mu_t \simeq 1000$, and the caustic
transit rate becomes $r(M)\simeq 0.015$ per year. 

The above rate is the expected rate per year for one background star
intersecting $n(M)=\Sigma \simeq 0.1/pc^2$ stars per $pc^2$ with mass 
M$\simeq$1 \Mo. During the time the star is moving across the saturation region
and towards the main caustic of the cluster, it will intersect many micro
caustics until it reaches the main caustic, after which it fades away forever
from our vantage point. We ignore the equally likely case where the star
approaches from the main caustic from other the direction, which is discussed
in \S \ref{sec44} \& \ref{sec71}, but is observationally much harder to
recognize. We can then estimate the time it takes to cross the saturation
region as 0.0097 (pc)/10$^{-4}$ (pc/year)= 97 years, so in this time the
background star would cross 1.5 microcaustics before reaching the main cluster
caustic. The final boosting factor is expected to be modest in the regions of
the critical curve with a very small density of microlenses, which would amount
to $\sim$2.5 caustic crossings instead of just one. 

Finally, we note that the approximations made above assumed a very conservative
low density of microlenses, about two orders of magnitude smaller than in the
outskirts of the BCG region. There may be certain regions along the critical 
curve where the density of microlenses increases very significantly, for
instance near an area with a larger fraction of ICL or near a cluster member
galaxy. If in these areas the number density of microlenses, n(M), increases
substantially, the rate would increase by a similar amount. Assuming one could
estimate the luminosity function of Pop III stars at z\cge 7 in the future from
a long term monitoring program of cluster caustic transits, one would expect
their LF to show a significant excess at the highest luminosities, as a
consequence of the caustic/microcaustic crossings boosting their observed
luminosities the most. This would be akin to the lensing tail observed in the
bright-end of the high redshift sub-mm galaxy luminosity function, like for
instance has been seen for the lensed Herschel sample
\citep[\eg][]{negrello_2017}. 

\bn\cl{Appendix C.\ Uncertainty Estimates for Caustic Transit Rates of Pop
III Stars at z\cge 7}
\label{secAppC}

\sn Here we estimate the uncertainties in the caustic {\it transit rates} and
{\it rise times} of Pop III stars at z\cge 7. The combined uncertainty in their
caustic transit rates follows from the multiplicative sources of error in 
Eqs.~\ref{eq:NM}, \ref{eq:dAdt}, and \ref{eq:dNlensdt1}. These are the
adopted effective caustic length $L_{caust}$ (in \arcs), the cluster transverse
velocity \vT\ (in \kms), the stellar luminosity $L$ (in $L_{100}$), and the
1--4 \mum\ sky-SB from Pop III stars (in \magarc). 

The error on the effective cluster caustic length $L_{caust}$ is estimated to be
$\sim$0.3 dex (Fig. \ref{fig:fig4}b), which incorporates the measurement errors
in tracing $L$ along the caustics, and the differences in caustic lengths
between current lensing models (see \S \ref{sec43}). The error in the cluster
transverse velocity \vT\ is estimated to be at least 0.3 dex, following the
discussion in \S \ref{sec422} and Appendix A. This includes the uncertainty in
the \vT\ values as constrained in Fig.~\ref{fig:fig5}, and their \vT-values as
projected onto the plane of the sky that assumed an average foreshortening of
$<$sin(i)$>$$\simeq$1/2, as discussed in \S \ref{sec431}. 

\citet{choi_2016} predict the stellar luminosities, radii, and \Teff-values over
a wide range of masses (0.2\cle M\cle 30 \Mo) and metallicities (--4.0 \cle
Z/\Zo \cle +0.5) using the same \texttt{MESA} models as in \S \ref{sec31}, and
compare these to data from a large number of detached eclipsing binaries in our
Galaxy. They find that the stellar luminosities predicted by the \texttt{MESA}
models in general follow the detached eclipsing binary data to within 0.2 dex.
While not anchored yet in data for (nearly) zero metallicity stars with M\cge 30
\Mo\ as needed for Pop III stars at z\cge 7, we will adopt the 0.2 dex error in
the predicted luminosities and radii for lower mass stars to be representative
for more massive, (nearly) zero metallicity Pop III stars at z\cge 7. Future
work will need to Monte Carlo model the shape of the error distribution in
these parameters \citep[\eg][]{fields_2017}. 

The uncertainties in the caustic rise times follow for each mass from the two
parameters in Eq.~\ref{eq:tRMb}: $R_{100}$ and \vT. \citet{choi_2016} also find
that the predicted stellar radii follow the detached eclipsing binary data to
within $\sim$0.2 dex. This is less than the 0.3 dex uncertainty in the \vT\
values. Since both parameters are independent, the resulting uncertainties in
the caustic {\it rise times} are thus \cle 0.4 dex. 

For Pop III binary or multiple stars the situation may be more complex, as
discussed in \S \ref{sec32} and \ref{sec52}, but can be approximated as
following. Unless a lot of mass exchange happens continuously in Pop III
binaries, to first order a binary --- which will generally be of unequal mass
(see Eq.~\ref{eq:Nq}) --- consists of two stellar photospheres with luminosities
and radii that are determined by their ZAMS mass (\S \ref{sec31}). Without mass
exchange, their radii will be in the range 2--13 \Ro\ during the ZAMS stage (see
Table~\ref{tab:tab2}), while their typical binary separations are expected to be
$\sim$10--100 \Ro\ (\S \ref{sec323}). The more common lower mass Pop III
stars with M\cle 20 \Mo\ in a binary are simply too faint to be seen during
caustic transits (Tables \ref{tab:tab2}--\ref{tab:tab4}). However, when both
Pop III stars in a binary have masses M\cge 30--50 \Mo, caustic transits of Pop
III binaries will to first order consist of multiple peaks, each with a transit
rise-time less than a few hours as specified in Eq.~\ref{eq:tRM}, while for
constant \vT\ these events will thus be separated in time by hours to days.
Therefore, to first order Pop III star multiplicity does not affect the
calculated caustic transit rates, other than producing two successive caustic
transits that are likely separated by hours to days, such as potentially
already observed at z$\simeq$1.5 by \citet{kelly_2017_b}. JWST epochs observed
hours--days apart may thus observe multiple caustic transits for Pop III binary
stars and identify each by its different SED colors, as discussed in \S
\ref{sec7}. In any case, massive Pop III binary stars will likely lead to a
double caustic transit, and as long as both caustic transit events are
observationally recognized as coming from stars with different radii
(rise-times) and SEDs, they will not lead to a significant overcounting of Pop
III binary star caustic transits. 

Fast rotating massive stars will evolve more towards the blue than their
non-rotating counterparts, since their mass-loss is not driven by the classic
line-driven winds \citep{castor_1975}, but by wave transport, which is not
incorporated in our current models. The effects of this on the radii and
luminosities of massive stars will need to be addressed in future work. 

The presence of microlensing in the foreground cluster ICL will require to
adjust these calculations, as discussed in \S \ref{sec432} and Appendix
B1--B2. To first order, microlensing may reduce the magnification of each
caustic transit event from values of $\mu$\cge 10$^4$ to several 1000
\citep{diego_2017}, but multiply the number of caustic transit events seen per
unit time accordingly, while preserving the total lensed flux during the
crossing of all (micro-)caustics by this object. Therefore, depending on the
IMF slope, as discussed in \S \ref{sec34} and Eq.~\ref{eq:dNlensdtalpha}, the
more rare, but more massive Pop III stars with M\cge 100--50 \Mo\ may be
magnified more often than without microlensing at the expense of the more
common, lower mass Pop III stars (M\cle 50 \Mo), which may now become more
often invisible due to their smaller (microlensed) magnification. Because the
same star could be seen microlensed several times over a decade or longer
\citep{diego_2017}, this could lead to overcounting of the caustic crossings,
unless the observations allow us to recognize that these caustic crossings all
came from a star with the same radius (rise-time) and SED. Given this possible
overcounting from microlensing, we take the uncertainty in the caustic transit
rates induced by microlensing to be at least 0.5 dex. 

The last major uncertainty in the caustic transit calculations is the value of
the 1--4 \mum\ sky-SB that comes from Pop III stars at z\cge 7, which according
to the discussion in \S \ref{sec231} is \cge 31.4 \magarc. Given that recent
hierarchical models yield values of the SFR at z$\simeq$7--10
\citep{sarmento_2018} that are within a factor of 2--3 from the
\citet{madau_2014} SFR at z$\simeq$7--8 (Eq.~\ref{eq:sfrd}), we will adopt the
uncertainty in the 1--4 \mum\ Pop III star sky-SB to be $\sim$0.3--0.5 dex. It
is not likely that the true 1--4 \mum\ sky-SB is much higher than this amount
by many times this uncertainty, because the fitted values to the SFH data at
7\cle z\cle 10 by \citet{madau_2014}, \citet{madau_2017}, and
\citet{finkelstein_2016} do not permit this, and because at least \cge\ 75\% of
the near-IR sky-SB that comes from 7\cle z\cle 17 is already produced in the
redshift bins at z$\simeq$7--8 (\S \ref{sec231}). But it is possible that the
1--4 \mum\ sky-SB from Pop III stars is significantly {\it lower} by factors of
10--100 or more, as discussed in \ref{sec35}, \ref{sec45}, and as indicated by
the (light orange) range in sky-SB levels in Fig. \ref{fig:fig1} that may come
from Pop III stars. The resulting Pop III star caustic transit rates discussed in
\S \ref{sec44} can therefore be regarded as upper limits, and the consequences of
this for the JWST observing strategy are discussed in \S \ref{sec45} and
\ref{sec7}. 

\bn\cl{Appendix D.\ Uncertainty Estimates for Caustic Transit Rates of 
Stellar-Mass BH Accretion Disks at z\cge 7} 
\label{secAppD}

\sn Here we estimate the uncertainties in the caustic transit rates of Pop III
stellar-mass black hole accretion disks at z\cge 7. We will follow the same
reasoning as in Appendix C, with some important differences. The error
estimates for the total caustic length $L_{caust}$ and the cluster transverse
motion \vT\ are $\sim$0.3 dex for each, as discussed in Appendix C. The
combined uncertainty from overcounting due to microlensing also remains at 0.5
dex (Appendix C). 

The main differences with uncertainties in the caustic transit rates of Pop III
stars are twofold. First, the uncertainty in the adopted 3--4 \mum\ sky-SB from
stellar-mass black hole accretion disks is significant, like it is for Pop III
stars, but unlike that of Pop III stars, it is not necessarily an upper limit.
Following the discussion in \S \ref{sec232}, the IR and IR--X-ray
power-spectrum results observed in the object-free Spitzer and Spitzer--Chandra
images, respectively \citep{kashlinsky_2012, cappelluti_2013,
mitchell-wynne_2016} have an (amplitude)$^2$ that is at least $\sim$0.3 dex
uncertain between these papers. As discussed in \S \ref{sec232} and
\ref{sec551}, the Spitzer--Chandra power-spectrum results hint at a component
caused by (stellar-mass) black holes, since Pop III stars simply do not get hot
enough to cause this signal. Since this power-spectrum did not come from
discrete objects seen down to either the Spitzer or Chandra detection limits,
it is possible that a significant fraction of the near-IR sky-SB of \cge 31
\magarc\ that we derived in \S \ref{sec232} comes from (stellar-mass) black
hole accretion disks at z\cge 7. We therefore adopt the uncertainty in the
near-IR sky-SB {\it signal itself} for stellar-mass black hole accretion disks
to be at least half this, or \cge 0.15 dex, following the derivation in \S
\ref{sec232}. 

Second, unlike that of Pop III stars, the uncertainty in the predicted
luminosities of stellar-mass black hole accretion disks is no longer smaller 
than the uncertainty in the other parameters. The two methods of \S
\ref{sec551} and \ref{sec552} predicted their luminosities consistently (top
and bottom tiers of Table \ref{tab:tab5}), but this assumed that these BHs
were always accreting. The largest uncertainty in $L$ comes from their accretion
efficiency, or accretion duration, as discussed in \S \ref{sec54}, which we
assume is uncertain by at least 0.5 dex. This could reduce their luminosities
from the steady-state values that we adopted in \S \ref{sec55}, and so {\it
increase} their caustic transit rates for a given sky-SB, as discussed in \S
\ref{sec62}. 

\ve 


\end{document}